\documentclass[12pt]{article}

\pdfoutput=1

\usepackage[utf8]{inputenc}
\usepackage[T1]{fontenc}
\usepackage[compat=1.1.0]{tikz-feynman}
\usepackage{amsmath,amssymb,mathtools}
\usepackage[separate-uncertainty=true,retain-unity-mantissa=false]{siunitx}
\usepackage{slashed}
\usepackage{graphicx}
\usepackage{booktabs,multicol,multirow}
\usepackage[shortlabels]{enumitem}
\usepackage{hyperref}
\usepackage[capitalize]{cleveref}
\usepackage{xspace}
\usepackage[noblocks]{authblk}
\usepackage{color}
\usepackage{soul}
\usepackage[textsize=tiny]{todonotes}
\usepackage{subcaption}
\usepackage{ulem}
\usepackage{bbold}
\usepackage{cite}

\newcommand\citere[1]{Ref.~\cite{#1}}
\newcommand\citeres[1]{Refs.~\cite{#1}}

\newcommand{\cp}{\ensuremath{{\cal CP}}}

\newcommand{\br}{\rm{BR}}

\newcommand{\gev}{\;\text{GeV}\xspace}

\newcommand{\fb}{\;\mathrm{fb}\xspace}
\newcommand{\ifb}{\;\mathrm{fb}^{-1}\xspace}

\newcommand{\cL}{{\cal L}}
\newcommand{\cLint}{\cL_{\mathrm{int}}}

\newcommand{\vw}{\ensuremath{v_{\mathrm{w}}}}

\renewcommand{\rm}{\mathrm}
\newcommand{\nn}{\nonumber}
\newcommand{\HB}{\texttt{HiggsBounds}}

\newcommand{\mhh}{\ensuremath{m_{hh}}}
\newcommand{\mbbbb}{\ensuremath{m_{b\bar b b\bar b}}}
\newcommand{\lahhh}{\lambda_{hhh}}
\newcommand{\lahhH}{\lambda_{hhH}}

\newcommand{\laijk}{\lambda_{ijk}}

\newcommand{\rlahhH}[1]{\hat{\lambda}_{hhH}^{(#1)}}

\newcommand{\rlaijk}[1]{\hat{\lambda}_{ijk}^{(#1)}}

\newcommand{\kala}{\kappa_\lambda}

\newcommand{\sihh}{\sigma_{hh}^{\mathrm{RxSM}}}

\newcommand{\siZhh}{\sigma_{Zhh}^{\text{RxSM}}}

\newcommand{\sivvhh}{\sigma_{\nu \bar{\nu} hh}^{\text{RxSM}}}

\newcommand{\sivvhhSM}{\sigma_{\nu \bar{\nu} hh}^{\text{SM}}}
\newcommand{\eeZhh}{\ensuremath{e^+e^- \to Zhh}}
\newcommand{\eenunuhh}{\ensuremath{e^+e^- \to \nu\bar\nu hh}}

\newcommand{\zthh}{Z_{hh}^{(0)}}
\newcommand{\zohh}{Z_{hh}^{(1)}}

\date{}

\evensidemargin \oddsidemargin
\begin{document}
\thispagestyle{empty}
\def\thefootnote{\fnsymbol{footnote}}

\begin{flushright} \texttt{DESY-25-136}\\
\texttt{IFT–UAM/CSIC-25-110}
\end{flushright}
\vspace{3em}
\begin{center}
{\Large{\bf Complementarity of gravitational wave analyses\\[0.25em] and di-Higgs production in the exploration of the\\[0.5em] Electroweak Phase Transition dynamics in the RxSM}
}
\\
\vspace{3em}
{
Johannes\,Braathen$^{1}$\footnotetext[0]{\hspace{-0.65cm}\href{mailto:johannes.braathen@desy.de}{johannes.braathen@desy.de},\href{mailto:sven.heinemeyer@cern.ch}{sven.heinemeyer@cern.ch},\\ 
\href{mailto:carlos.pulido@estudiante.uam.es}{carlos.pulido@estudiante.uam.es},
\href{mailto:alain.verduras@desy.de}{alain.verduras@desy.de}}
Sven\,Heinemeyer$^{2,3}$, 
Carlos\,Pulido\,Boatella$^{2}$,
Alain\,Verduras\,Schaeidt$^{1}$
}\\[2em]
{\sl $^1$ Deutsches Elektronen-Synchrotron DESY, Notkestr.~85, 22607 Hamburg, Germany}\\[0.2em]
{\sl $^2$ Instituto de F\'isica Te\'orica (UAM/CSIC), Cantoblanco, 28049, Madrid, Spain}\\[0.2em]
{\sl $^3$ Particle Theory and Cosmology Group, Center for Theoretical Physics of the Universe,\\ 
Institute for Basic Science (IBS), Daejeon 34126, South Korea}
\end{center}

\vspace{2ex}

\begin{abstract}\noindent
    The real singlet extension of the Standard Model (SM), RxSM, is one of the simplest Beyond-the-Standard Model (BSM) theories that can accommodate a strong first-order electroweak phase transition (SFOEWPT). We survey the possible thermal histories of the early Universe in the RxSM, and find that a SFOEWPT can occur in this model as a single- or two-step phase transition. We investigate complementary approaches to probe such scenarios experimentally: either via searches for a stochastic background of gravitational waves (GWs) or via searches for di-Higgs production processes at future collider experiments: the HL-LHC, or a possible high-energy $e^+e^-$ collider. For these analyses we consistently include one-loop corrections to the trilinear Higgs couplings. We find that entirely different phenomenological signals are possible, depending on how the SFOEWPT occurs. In scenarios where such a transition is driven by the Higgs doublet direction in field space, BSM deviations in properties of the detected Higgs boson, particularly in the trilinear scalar coupling, typically lead to observable signals at colliders, while the regions of parameter space with detectable GW signals are very narrow. On the other hand, if the SFOEWPT is triggered by the singlet field direction, the detected Higgs boson is very SM-like and no signs of BSM physics would appear in di-Higgs production processes. However, strong GW signals could be produced for significant parts of the RxSM parameter space with singlet-driven SFOEWPT. This work highlights the crucial importance of exploiting complementary experimental directions to determine the dynamics of the electroweak phase transition and access the shape of the Higgs potential realised in Nature. 
\end{abstract}
\def\thefootnote{\arabic{footnote}}
\setcounter{footnote}{0}

\newpage

\tableofcontents

\newpage


\section{Introduction}
\label{sec:intro}

Understanding the dynamics of the electroweak phase transition (EWPT) is one of the most important challenges ahead for High-Energy Physics. While the discovery in 2012 of a scalar boson with a mass of $\sim 125 \gev$ at the LHC~\cite{Aad:2012tfa,Chatrchyan:2012xdj,Khachatryan:2016vau} has confirmed the Brout-Englert-Higgs mechanism~\cite{Englert:1964et,Higgs:1964pj,Guralnik:1964eu} as the origin of electroweak symmetry breaking (EWSB), the actual way in which the corresponding phase transition took place remains unknown. 
Additionally, our best theoretical framework at present to describe fundamental interactions, the Standard Model (SM)~\cite{Glashow:1961tr,Weinberg:1967tq,Salam:1968rm}, remains incomplete and leaves unanswered a number of fundamental issues, such as the origin of the baryon asymmetry of the Universe (BAU), or the nature of dark matter. The Higgs sector is closely connected to many of these questions, and many Beyond-the-Standard-Model (BSM) theories that aim to solve one or more of them feature extended scalar sectors.  

In this context, the issue of the dynamics of the EWPT is of high importance, not only because it is a rare fundamental example of spontaneous symmetry breaking, but also because it relates to the possibility of explaining the BAU through the scenario of electroweak baryogenesis (EWBG)~\cite{Kuzmin:1985mm,Cohen:1993nk}. Among the three Sakharov conditions~\cite{Sakharov:1967dj} that must be fulfilled in order to explain the origin of the BAU dynamically, one condition is departure from thermal equilibrium. In scenarios of EWBG, this is achieved if the EWPT occurs as a strong first-order electroweak phase transition (SFOEWPT). During such a transition, patches of space transition from a ``false'' vacuum to a deeper, and thus more stable, ``true'' vacuum~\cite{Espinosa:1993bs,Ham:2004cf,Profumo:2007wc,Espinosa:2007qk,Barger:2008jx,Espinosa:2011ax,Morrissey:2012db,Curtin:2014jma,Kurup:2017dzf,Ramsey-Musolf:2019lsf}, leading to the nucleation and expansion of bubbles of the deeper vacuum phase. Given the measured mass of the Higgs boson, the EWPT happens in the SM as a smooth cross-over~\cite{Kajantie:1996mn}, meaning that the SM fails to explain the BAU via EWBG. On the other hand, BSM models with extended Higgs sectors can accommodate a SFOEWPT in the early Universe and thus provide the possibility of successful electroweak baryogenesis. 

Various experimental directions offer opportunities to investigate the EWPT, from different, complementary, angles. 
A first direction is to reconstruct the form of the Higgs potential realised in Nature by studying processes at high-energy colliders that probe relevant trilinear (and, in a more distant future) quartic couplings. First and foremost among these are di-Higgs production processes: considering only the production of the detected Higgs boson,~$h$, the dominant channel at the (HL-)LHC is $gg\to hh$~\cite{LHCHiggsCrossSectionWorkingGroup:2016ypw}, while at possible $e^+e^-$ colliders (with sufficiently high energies), the processes $e^+e^-\to Zhh$ and $e^+e^-\to \nu\bar{\nu}hh$ become accessible (see e.g.\ Refs.~\cite{Barklow:2017awn,LinearColliderVision:2025hlt,Altmann:2025feg}). These provide direct access to the trilinear self-coupling of the Higgs boson, denoted $\lambda_{hhh}$, that controls the shape of the Higgs potential away from the EW minimum, along the Higgs direction in field space. In particular, a SFOEWPT taking place along the Higgs field direction would be associated with a significant deviation in $\lambda_{hhh}$ from its SM prediction --- see e.g.\ Refs.~\cite{Grojean:2004xa,Kanemura:2004ch} for early studies and Refs.~\cite{Kakizaki:2015wua,Hashino:2016rvx,Hashino:2016xoj,Basler:2017uxn,Biekotter:2022kgf,Bittar:2025lcr} for more recent examples. In models with extended scalar sectors, BSM deviations in $\lambda_{hhh}$ can arise due to radiative corrections from the BSM scalars, even in scenarios where other properties of the detected Higgs boson are SM-like, like for instance aligned~\cite{Gunion:2002zf} scenarios of multi-Higgs models. These types of effects, which are driven by splittings between the different BSM mass scales of the models, were initially found in Two-Higgs-Doublet Models (2HDM)~\cite{Kanemura:2002vm,Kanemura:2004mg}, but are now known to occur in broad ranges of BSM theories~\cite{Aoki:2012jj,Kanemura:2015fra,Kanemura:2015mxa,Arhrib:2015hoa,Kanemura:2016sos,Kanemura:2016lkz,He:2016sqr,Kanemura:2017wtm,Kanemura:2017gbi,Chiang:2018xpl,Basler:2018cwe,Senaha:2018xek,Braathen:2019pxr,Kanemura:2019slf,Braathen:2019zoh,Braathen:2020vwo,Basler:2020nrq,Bahl:2022jnx,Bahl:2022gqg,Falaki:2023tyd,Bahl:2023eau,Aiko:2023nqj,Cherchiglia:2024abx,Basler:2024aaf,Bahl:2025wzj,Braathen:2025qxf}.

Di-Higgs production has not yet been observed at the LHC. The SM production cross-section turns out to be three orders of magnitude smaller than the one for single Higgs production due to the destructive interference between the diagram involving $\lahhh$ and that without it. This leads to a very suppressed di-Higgs cross-section for SM-like values of the Higgs-boson couplings (including $\lahhh$), but also means that BSM effects in $\lahhh$ can significantly alter the cross-section --- by up to several orders of magnitude.  
A variety of search results have already been obtained by the ATLAS and CMS collaborations at CERN for this process. These have allowed setting bounds on both the di-Higgs cross-section as well as on 
$\lahhh$~\cite{ATLAS:2022jtk,ATLAS:2022kbf,CMS:2022dwd,ATLAS:2024ish,CMS:2024awa,ATLAS:2025lbo}, or equivalently the coupling modifier $\kappa_\lambda\equiv \lahhh/\lahhh^{\text{SM,(0)}}$, where $\lahhh^{\text{SM,(0)}}$ is the SM tree-level prediction for the trilinear Higgs coupling. In Ref.~\cite{Bahl:2022jnx}, it was demonstrated that comparing the experimental limits on $\lahhh$ with precise theoretical predictions already provides a powerful tool to probe otherwise unconstrained parameter space of BSM theories. Moreover, considerable improvements in experimental limits are expected at the HL-LHC~\cite{CMS:2025hfp}.

More generally, models with extended scalar sectors feature several trilinear scalar couplings $\lambda_{ijk}$ (which can also receive large radiative corrections, see e.g.~\citeres{Arco:2025pgx,Braathen:2025qxf,anyHH}), and multiple di-Higgs processes, like $gg\to h_i h_j$, $e^+e^-\to Zh_ih_j$, or $e^+e^-\to \nu\bar\nu h_ih_j$, are needed to reconstruct the form of the potential entirely. However, these processes are even more challenging to access than those with $hh$ final states. On the other hand, $hh$ processes also already offer relevant information about BSM trilinear scalar couplings, see e.g.~\citeres{Arco:2025nii,Braathen:2025qxf}. We will therefore restrict our attention to the processes with $hh$ final states in this paper. 

Another important direction to investigate the possibility of a SFOEWPT is to search for potential cosmological relics of such a 
phase transition. The nucleation, expansion and collisions of bubbles of the true vacuum in the early Universe would have been violent processes, that could have given rise to a stochastic background of gravitational waves (GWs) (see e.g.\ Refs.~\cite{Grojean:2006bp,Ashoorioon:2009nf,No:2011fi,Huber:2015znp,Kakizaki:2015wua,Hashino:2016rvx,Dorsch:2016nrg,Kang:2017mkl,Bruggisser:2018mrt,Chala:2018ari,Morais:2018uou,Hashino:2018zsi,Hashino:2018wee,Goncalves:2021egx,Biekotter:2022kgf,Lewicki:2024ghw}, and the review~\cite{Athron:2023xlk}), to primordial black holes~\cite{Kodama:1982sf,Liu:2021svg,Hashino:2021qoq,Jung:2021mku,Kawana:2022olo,Lewicki:2023ioy,Gouttenoire:2023naa,Baldes:2023rqv,Flores:2024lng,Lewicki:2024ghw,Kanemura:2024pae,Hashino:2025fse,Franciolini:2025ztf,Kierkla:2025vwp}, or even primordial magnetic fields~\cite{Vachaspati:1991nm,Ellis:2019tjf,Olea-Romacho:2023rhh}. We focus here on the first possibility, GWs, which if produced during a SFOEWPT would have a spectrum of frequencies peaking within the sensitivity of future space-based GW interferometers like LISA~\cite{Caprini:2019egz,LISACosmologyWorkingGroup:2022jok}. Other observatories, like DECIGO~\cite{Kawamura:2011zz} or BBO~\cite{Corbin:2005ny} are also being discussed, however, LISA is the only one of these projects that has already been approved. 

In this work, we investigate the general real-singlet extension of the SM, which we denote RxSM (sometimes also referred to as ``xSM'', ``SSM'' or ``HSM''). The scalar sector of this model features two physical CP-even Higgs states $h$ and $H$. In the rest of this work, we consider that $h$ corresponds to the detected Higgs boson with a mass of $\sim 125 \gev$, while the BSM scalar $H$ is assumed to be heavier. Several variants of real singlet models have been devised, differing by the symmetry assignments of the singlet state. The RxSM considered in this work corresponds to the most general version of the model, for which no additional symmetry (in particular no $\mathbb{Z}_2$ symmetry) is imposed.
This adds three (two) additional free parameters to the theory compared to the variants featuring an unbroken (spontaneously broken) $\mathbb{Z}_2$ symmetry. In turn, these can have important phenomenological implications for the trilinear scalar couplings, which can receive large BSM contributions already at the tree level or significant BSM radiative corrections, depending on the considered region of the parameter space.
The RxSM (without a $\mathbb{Z}_2$ symmetry, from now on we will implicitly assume this) is known~\cite{Espinosa:1993bs} to be one of the most minimal Higgs-sector extensions capable of accommodating a SFOEWPT, and the simplest allowing this with a one-step transition. By itself, the RxSM does not allow introducing BSM sources of CP violation, which are known from the Sakharov conditions to be needed for baryogenesis, and can therefore not explain the BAU alone. However, it serves as a very useful theory to investigate the dynamics of the EWPT in presence of a singlet scalar --- and one could in principle devise extensions of the RxSM where some additional sector, for instance a dark sector, could contain the required BSM sources of CP violation (see e.g.\ Refs.~\cite{Shelton:2010ta,Espinosa:2011eu,Azevedo:2018fmj,Hall:2019ank,Biermann:2022meg,Hooper:2025fda,Roy:2025zvo} for examples of this type of model building). 
The possibility of a SFOEWPT and the related phenomenology has already been investigated in the RxSM and other variants of singlet models: for instance the dynamics of the EWPT and/or potential production of GWs were analysed in Refs.~\cite{Espinosa:2007qk,Profumo:2007wc,Huang:2016cjm,Alves:2018jsw,Liu:2021jyc,Ellis:2022lft,Blasi:2023rqi,Ramsey-Musolf:2024ykk,Niemi:2024vzw,Gould:2024jjt,Niemi:2024axp,Ghosh:2022fzp,Roy:2022gop}, while the production of primordial black holes was considered in Ref.~\cite{Goncalves:2024vkj}. Collider signatures of the RxSM have been studied recently in e.g.\ Refs.~\cite{Feuerstake:2024uxs,Lewis:2024yvj,Aboudonia:2024frg}, and in particular scenarios with a SFOEWPT were considered in Refs.~\cite{Li:2019tfd,Liu:2021jyc,Zhang:2023jvh,Palit:2023dvs,Ramsey-Musolf:2024ykk}. 

In this work, we explore the interplay between the Higgs potential and its thermal history, with collider phenomenology at the (HL-)LHC and $e^+e^-$ colliders. For this purpose, we have implemented the RxSM into the public tool \texttt{BSMPTv3.1.1}~\cite{Basler:2018cwe,Basler:2020nrq,Basler:2024aaf}. With this, we investigate the different possible thermal evolutions of the vacuum that can occur in allowed parts of the RxSM parameter space. We ascertain the cases in which GWs could be produced, and those were the signal-to-noise ratio (SNR) at LISA would be sufficient to detect these GWs. In complementarity to this, we study the di-Higgs processes $gg\to hh$ (at the HL-LHC), and $e^+e^-\to Zhh$ and $e^+e^-\to \nu\bar{\nu}hh$ (at a 1-TeV $e^+e^-$ collider). In the RxSM, these processes offer sensitivity to the trilinear scalar couplings $\lahhh$ and $\lahhH$. Given that our analysis of the EWPT dynamics is performed at the one-loop level, for consistency we therefore also take BSM effects into account in di-Higgs production up to the same order. We employ for this the fully on-shell renormalisation scheme and phenomenological setup from our previous article~\cite{Braathen:2025qxf}. We find that two main types of behaviours are possible for RxSM scenarios with a SFOEWPT. On the one hand, if the EWPT is driven by the doublet field direction, noticeable effects on Higgs couplings can be expected, resulting in BSM signals at colliders, although the strength of the transition is only strong enough to produce a detectable signal of GWs in a very limited fraction of the parameter space. On the other hand, if a SFOEWPT is driven by the singlet field direction, markedly stronger signals are possible from GWs, but the properties of the detected Higgs boson are typically very SM-like, so that colliders would not offer much information. Our finding highlight the complementarity of collider searches for di-Higgs productions and searches for cosmological relics of a SFOEWPT to probe the parameter space of the RxSM and determine the dynamics of the EWPT in this model. 

This paper is organised as follows: in \cref{sec:model} we review the RxSM, our choices of notations, as well as the theoretical and experimental constraints included in our analyses. In \cref{sec:effective_potential}, we present our setup to investigate the dynamics of the EWPT and the spectrum of the background of stochastic GWs produced during a SFOEWPT. We apply this setup to the RxSM in \cref{sec:pheno}: we explore the different possible thermal histories, identify regions of the RxSM that would yield a SFOEWPT and a detectable GW signal. In \cref{sec:collider}, we analyse whether such scenarios can be complementarily probed via di-Higgs production searches at high-energy colliders. We provide our conclusions in \cref{sec:conclusions}. 


\section{The RxSM: the general singlet extension of the SM}
\label{sec:model}

\subsection{Definitions and notations}
\label{sec:model_def}
We investigate here the most general real singlet extension of the SM~\cite{Lerner:2009xg,Li:2019tfd,Gonderinger:2009jp,Arco:2025nii}, which we refer to as RxSM. This model extends the Higgs sector of the SM with a real singlet, denoted $S$, and unlike other variants of singlet models, its Lagrangian does not exhibit any $\mathbb{Z}_2$ symmetry under transformations of $S$. After EWSB, the SM-like doublet $\Phi$ and the singlet $S$ can be expanded as
\begin{equation}
{\Phi}=\frac{1}{\sqrt{2}}\left(
 \begin{matrix}
 \sqrt{2}\,G^+  \\
 v+\phi+i\, G^0
 \end{matrix}\right)\,,\quad\text{and}\quad S=s+v_S, \label{eq:fields}
\end{equation}
where $\phi$ is the \cp-even component of the doublet, $G^0$ ($G^\pm$) is the neutral (charged) would-be Goldstone boson, and $v$ is the EW (or SM) VEV. Moreover, $s$ denotes the BSM scalar field and $v_S$ is the singlet VEV. The tree-level scalar potential of the RxSM is given by
\begin{align}
    V^{(0)}({\Phi},{ S})=\mu^2|\Phi|^2+\frac{\lambda}{2}|\Phi|^4+\kappa_{SH}|\Phi|^2 S+\frac{\lambda_{SH}}{2}|\Phi|^2 S^2+\frac{M_S^{2}}{2}{S}^2+\frac{\kappa_S}{3}{ S}^3+\frac{\lambda_S}{2}S^4. \label{eq:potential}
\end{align}
Given the absence of any additional symmetry restricting the allowed interactions, this potential would in principle contain both linear and cubic terms in the singlet field. 
However, in the RxSM 
the singlet VEV $v_S$ can be redefined in order to absorb one of these terms~\cite{Espinosa:2011ax}, and we have used here this freedom to remove the linear tadpole term. Furthermore, we note that because the singlet $S$ is real, \cp-violation cannot be included in the scalar potential of the RxSM (like what is the case for the SM); this means that all the parameters in \cref{eq:potential} are real.

The scalar sector of the RxSM is described in terms of nine parameters, namely 
the two mass parameters, the five trilinear and quartic couplings in the Lagrangian, and the doublet and singlet VEVs. 
The minimisation conditions of the potential (i.e.\ the tadpole equations), which can be written as
\begin{equation}
    \left.\frac{dV^{(0)}}{d\phi}\right|_{\phi=0,s=0}=t_{\phi}=0,\qquad\left.\frac{dV^{(0)}}{ds}\right|_{\phi=0,s=0}=t_S=0,
\end{equation}
yield two relations between the Lagrangian parameters, 
\begin{align}
    \mu^2&=-\frac{\lambda v^2}{2}-\kappa_{SH}v_S -\frac{\lambda_{SH}v_S^2}{2}\,,\nn\\
    M_S^{2}&=-2\lambda_Sv_S^2 -\kappa_Sv_S - \frac{\kappa_{SH}v^2}{2v_S}-\frac{\lambda_{SH}v^2}{2}\,,
\end{align}
thus allowing to reduce the number of free parameters down to seven. We note that from this point on, we set $t_\phi=t_S=0$ in all expressions throughout the paper (if one is interested in the renormalisation of the scalar sector, it is however convenient to keep the parametric dependence on $t_\phi$ and $t_S$; this was done for instance in Ref.~\cite{Braathen:2025qxf}). 
Next, expanding the potential using  \cref{eq:fields}, the \cp-even mass matrix is found to be
\begin{equation}
    \mathcal{M}^2= 
    \begin{pmatrix}
\frac{d^{2}V^{(0)}}{d\phi^{2}} & \frac{d^{2}V^{(0)}}{d\phi ds} \\[0.5em]
 \frac{d^{2}V^{(0)}}{d\phi ds}  & \frac{d^{2}V^{(0)}}{ds^{2}}
\end{pmatrix}\equiv
\begin{pmatrix}
\mathcal{M}_\phi^2 & \mathcal{M}_{\phi s}^2 \\[0.5em]
 \mathcal{M}_{\phi s}^2  &\mathcal{M}_s^2\,,
\end{pmatrix}
\end{equation}
with
\begin{align}
    \mathcal{M}_\phi^2&=\mu^2+\frac{3\lambda v^2}{2}+\kappa_{SH}v_S+\frac{\lambda_{SH}v_S^2}{2}\,, \nn\\ 
    \mathcal{M}_s^2&= M_S^{2}+\frac{\lambda_{SH}v^2}{2}+2\,(\kappa_S+3\lambda_S v_S)\,v_S \,,\nn\\
    \mathcal{M}_{\phi s}^2&=(\kappa_{SH}+\lambda_{SH}v_S)\,v\,. 
\end{align}

\noindent
The gauge eigenstates $\phi$ and $s$ can be rewritten in terms of mass eigenstates $h$ and $H$ via a mixing matrix $R_\alpha$ defined by,
\begin{equation}
    \begin{pmatrix}
  h \\
  H
\end{pmatrix}=
R_{\alpha}^T
  \begin{pmatrix}
  \phi \\
  s
\end{pmatrix}=
\begin{pmatrix}
c_\alpha & s_\alpha \\
-s_\alpha & c_\alpha
\end{pmatrix}
  \begin{pmatrix}
  \phi \\
  s
\end{pmatrix}\,,
\label{mixingmatrix}
\end{equation}
where we have used the shorthand notations $c_x\equiv\cos x$ and $s_x\equiv \sin x$. 
In the following, we assume that $h$ is the detected Higgs boson with a mass around 125 GeV, while $H$ is a BSM Higgs boson, always considered to be heavier than $h$ --- because we want the decay $H\to hh$ to be kinematically allowed. 
After diagonalising the mass matrix, the \cp-even scalar mass eigenvalues and the mixing angle $\alpha$ are found to be
\begin{align}
    m_h^2 &= \mathcal{M}_\phi^2 c^2_\alpha + \mathcal{M}_s^2s^2_\alpha + \mathcal{M}_{\phi s}^2 s_{2\alpha}\,,\nn\\
    m_H^2 &= \mathcal{M}_\phi^2s^2_\alpha+ \mathcal{M}_s^2c^2_\alpha - \mathcal{M}_{\phi s}^2 s_{2\alpha}\,,\nn\\
    \tan2\alpha &= \frac{2\mathcal{M}_{\phi s}^2}{\mathcal{M}_\phi^2  - \mathcal{M}_s^2}\,.
\end{align}

For our numerical setup (as well as for the renormalisation of the RxSM Higgs sector, discussed in Ref.~\cite{Braathen:2025qxf}), it is especially helpful to employ 
as inputs the following seven parameters 
\begin{equation}
   m_h,\,m_H,\,\alpha,\,v,\,v_S,\,\kappa_S,\,\kappa_{SH},
\end{equation}
which we will refer to as ``\textit{mass basis}''. Among these, $m_h\simeq 125\gev$ and $v\simeq 250 \gev$ are fixed, so that five free BSM parameters remain. 

The Lagrangian mass parameters and quartic couplings can be re-expressed in terms of $m_h$, $m_H$, $\alpha$, and the 
Lagrangian trilinear couplings $\kappa_{SH}$ and $\kappa_S$. 
We find 
\begin{align}
\label{eq:rep_lambdas_and_masses}
\lambda &= \frac{c_\alpha^2m_h^2+s_\alpha^2m_H^2}{v^2} \,,\nn\\
\lambda_{SH}&=\frac{(m_h^2-m_H^2)c_{\alpha}s_{\alpha}}{vv_S}-\frac{\kappa_{SH}}{v_S}\,,\nn\\
\lambda_S&=\frac{m_h^2+m_H^2+(m_H^2-m_h^2)c_{2\alpha}}{8v_S^2}-\frac{2\kappa_Sv_S^2-\kappa_{SH}v^2}{8v_S^3}\,,\nn\\
M_S^2&=-\frac{\kappa_{SH}v^2+2\kappa_Sv_S^2+(m_h^2+m_H^2)v_S+(m_h^2-m_H^2)(vs_{2\alpha}-v_Sc_{2\alpha})}{4v_S}\,,\nn\\
\mu^2&=-\frac{1}{2}\kappa_{SH}v_S-\frac{1}{2}\big(m_h^2c_\alpha^2+m_H^2s_\alpha^2\big)-\frac{1}{2}\frac{v_S}{v}\big(m_h^2-m_H^2\big)c_\alpha s_\alpha\,.
\end{align}

Finally, using \cref{eq:rep_lambdas_and_masses}, the following expressions can be obtained for the tree-level expressions of the trilinear Higgs couplings in the mass basis 
{\allowdisplaybreaks
\begin{align}
\label{eq:trilinears_0L}
    \lambda_{hhh}=&\ \frac{1}{4 v v_S^2} \Big\{ -3  v_S \left[ \kappa_{SH} v^2 - 3 m_h^2  v_S \right] c_{\alpha} + 3  v_S \left[ \kappa_{SH} v^2 + m_h^2 v_S \right] c_{3\alpha}  \nn\\
    &\hspace{1.4cm}+2 v \left[ 3 \kappa_{SH} v^2 + 6 m_h^2 v_S - 2 \kappa_S v_S^2 \right] s^3_{\alpha} \Big\}\,,\nn\\
    \lambda_{hhH}=&\ \frac{1}{4 v v_S^2} s_{\alpha} \Big\{ -2 v_S \left[ \kappa_{SH} v^2 +(2 m_h^2 + m_H^2) v_S \right]- 2 v_S [ 3 \kappa_{SH} v^2 +(2 m_h^2 + m_H^2) v_S ] c_{2\alpha} \nn\\
    &\hspace{1.75cm}+ v \left[3 \kappa_{SH} v^2 + 2 v_S \left( 2 m_h^2 + m_H^2 - \kappa_S v_S \right) \right] s_{2\alpha} \Big\}\,,\nn\\
    \lambda_{hHH}=&\ \frac{1}{4 v v_S^2}c_{\alpha} \Big\{ 2 v_S \left[ \kappa_{SH} v^2 + (m_h^2 + 2 m_H^2)  v_S \right]- 2 v_S [ 3 \kappa_{SH} v^2 +(m_h^2 + 2 m_H^2)  v_S ] c_{2\alpha} \nn\\
    &\hspace{1.75cm}+ v \left[3 \kappa_{SH} v^2 + 2 v_S \left( m_h^2 + 2 m_H^2 - \kappa_S v_S \right) \right] s_{2\alpha} \Big\}\,,\nn\\
    \lambda_{HHH}=&\ \frac{1}{8 v_S^2} \Big\{ 3 \left[ 3 \kappa_{SH} v^2 + 6 m_H^2 v_S - 2 \kappa_S v_S^2 \right] c_{\alpha} + [ 3 \kappa_{SH} v^2 + 6 m_H^2 v_S -2 \kappa_S v_S^2 ] c_{3\alpha} \\
    &\hspace{1.25cm}+ 12 v_S \left[ \kappa_{SH} v - m_H^2 \frac{v_S}{v} + \left( \kappa_{SH} v + m_H^2 \frac{v_S}{v} \right) c_{2\alpha} \right] s_{\alpha} \Big\}\,.\nn
\end{align}
}


\subsection{Theoretical and experimental constraints on the RxSM}
\label{sec:constraints}

In this section, we summarise briefly the theoretical and experimental constraints taken into account in our analysis. 

On the theoretical side, we first have to ensure boundedness-from-below of the potential, 
which implies that the quartic couplings must be positive in all field directions. To ensure this, we demand that the determinant of the Hessian matrix of the potential be positive, leading to the conditions
\begin{equation}
    \lambda > 0\,,\quad \lambda_S > 0\,,\quad \text{and}\quad\lambda_{SH} > -2\sqrt{\lambda\;\lambda_S}\,.
    \label{eq:bfb}
\end{equation}
Second, we ensure perturbativity of the couplings, by requiring that
\begin{equation}
    \text{max}\bigg[\frac{\lambda}{2},\,\frac{\lambda_S}{2},\,\frac{|\lambda_{SH}|}{2}\bigg]<4\pi.
    \label{eq:pert}
\end{equation}
Finally, we also want pertubative unitarity to be fulfilled. In order to verify this, we compute the scalar $2 \rightarrow 2$ scattering amplitude matrix in the high-energy limit --- using results from Ref.~\cite{Braathen:2017jvs} --- and we require that the eigenvalues should all be below 1. 

In addition, we check the compatibility of our considered parameter points with experimental measurements. For this purpose, we consider on the one hand constraints from direct searches for extra Higgs bosons at colliders. The exclusion limits at the $95\%$ C.L.\ of relevant BSM Higgs boson searches (including Run~2 data from the LHC) are included in 
the public code \HB~\texttt{v.6}~\cite{Bechtle:2008jh,Bechtle:2011sb,Bechtle:2013wla,Bechtle:2015pma,Bechtle:2020pkv,Bahl:2022igd}, which is itself part of the public code \texttt{HiggsTools}~\cite{Bahl:2022igd}. 
On the other hand, agreement with measurements of the mass and signal strengths of the detected Higgs boson at the LHC is checked via the public tool \texttt{HiggsSignals}~\texttt{v.3}~\cite{Bechtle:2013xfa,Bechtle:2014ewa,Bechtle:2020uwn,Bahl:2022igd} (also included in \texttt{HiggsTools}). 
We note that constraints arising from di-Higgs measurements at the LHC are not checked a priori when generating scan points, but rather they are a core part of our phenomenological investigations in \cref{sec:collider}. 


\section{Thermodynamics of the RxSM}
\label{sec:effective_potential}

In this section, we review the theoretical setup we use to investigate the thermal evolution in the early-Universe of RxSM scenarios. We first introduce the one-loop temperature-dependent effective potential in the RxSM 
before discussing how we compute the thermal dynamics of the electroweak potential. 
Finally, we explain 
our calculation of the stochastic GW background produced in scenarios with a SFOEWPT. For all these steps, we have used the code \texttt{BSMPTv3.1.1}~\cite{Basler:2018cwe,Basler:2020nrq,Basler:2024aaf}, in which we implemented the general RxSM.

\subsection{Effective potential}
The first step in order to study the thermodynamics of an extended Higgs sector is to compute the temperature-dependent effective potential. In this work, we have used a perturbative 4D approach, where the potential is expanded as
\begin{equation}
    V_{\mathrm{eff}}^{(1)}(\phi,T)=V^{(0)}(\phi)+V_{\mathrm{CW}}^{(1)}(\phi)+V_{\mathrm{CT}}^{(1)}(\phi)+V_{\mathrm{T}}^{(1)}(\phi,T)+V_{\mathrm{daisy}}^{(1)}(\phi,T)\,.
    \label{VT}
\end{equation}
Here $V^{(0)}$ is the tree-level potential given in \cref{eq:potential}, $V_{\mathrm{CW}}^{(1)}$ is the one-loop effective (or Coleman-Weinberg~\cite{cw}) potential at $T=0$ (calculated in the $\overline{\mathrm{MS}}$ scheme), $V_{\mathrm{CT}}^{(1)}$ denotes a one-loop counter term (CT) potential (also defined at $T=0$), while the temperature-dependent parts of the potential are given by the one-loop thermal corrections term $V_{\mathrm{T}}^{(1)}$ and a resummation term $V_{\mathrm{daisy}}^{(1)}$ (discussed below). The notation $\phi$ is used to denote collectively the scalar field degrees of freedom of the RxSM.

The Coleman-Weinberg potential can be computed straightforwardly in the $\overline{\mathrm{MS}}$ scheme using the supertrace formula \cite{supertrace}, it reads
\begin{equation}
    V_{\mathrm{CW}}^{(1)}(\phi)=\sum_i\frac{(-1)^{2s_i}n_i}{64\pi^2}m_i^4(\phi)\left[\ln\left(\frac{m_i^2(\phi)}{Q^2}\right)-c_i\right],
\end{equation}
where $i$ runs over the particle content --- scalars, fermions, and gauge bosons --- of the model (we note that we work here in the Landau gauge, so that ghosts do not contribute to $V_\text{CW}^{(1)}$, nor its field derivatives), and $Q$ is the renormalisation scale. The constants $c_i$ are $3/2$ for scalars and fermions, and $5/6$ for gauge bosons (as we use here the $\overline{\text{MS}}$ scheme). For the rest of this work we choose to take $Q=v$, unless otherwise specified. 
Although one could in principle use the $\overline{\mathrm{MS}}$ scheme result, we choose instead to convert our expressions to the more convenient and physically-motivated 
OS-like scheme defined in Refs.~\cite{Basler:2018cwe,Basler:2020nrq,Basler:2024aaf}. To do so, we add a finite shift to our effective potential, corresponding to a finite conversion of the parameters entering the tree-level potential. This counterterm potential is defined as
\begin{equation}
V_{\mathrm{CT}}^{(1)}(\phi)=\sum_i\frac{\partial V^{(0)}}{\partial x_i}\delta^{\mathrm{CT}}x_i+\sum_j(\phi_j+v_j)\delta^{\mathrm{CT}}t_j\,,
\end{equation}
where $i$ runs over the parameters of the Lagrangian and $j$ over the scalar particles of the RxSM. 
The idea of this OS-like scheme is to maintain the location of the minimum of $V_\text{eff}$ at $T=0$ as well as the effective-potential masses of all the scalars (also at $T=0$) to their tree-level values --- hence the ``OS-like'' name of the scheme. In terms of derivatives of the potential, this means that we impose that the first and second derivatives of the zero-temperature effective potential should be equal to the tree-level ones, from which we obtain the following renormalisation conditions
\begin{gather}
    \partial_{\phi_i}V_{\mathrm{CT}}^{(1)}(\phi)|_{\langle\phi\rangle_{T=0}}=-\partial_{\phi_i}V_{\mathrm{CW}}^{(1)}(\phi)|_{\langle\phi\rangle_{T=0}},\label{eq:BSMPTscheme_1pt}\\
     \partial_{\phi_i}\partial_{\phi_j}V_{\mathrm{CT}}^{(1)}(\phi)|_{\langle\phi\rangle_{T=0}}=-\partial_{\phi_i}\partial_{\phi_j}V_{\mathrm{CW}}^{(1)}(\phi)|_{\langle\phi\rangle_{T=0}},\label{eq:BSMPTscheme_2pt}
\end{gather}
where $i$ and $j$ run over the scalar degrees of freedom. By applying these conditions to the case of the RxSM, one can see that there are not sufficiently many conditions to define a unique set of counterterms and that three additional degrees of freedom remain. Following the original notation of this scheme in Ref.~\cite{Basler:2018cwe}, we call these degrees of freedom $t_1, t_2, t_3$, and absorb them into the counterterms (although in the end, we set them to $0$). Finally, 
we obtain the following set of counterterms
{\allowdisplaybreaks
\begin{align}
    \delta^{\mathrm{CT}}\mu^2&= 0\,,\nonumber\\
    \delta^{\mathrm{CT}}M_S^2&= 0\,,\nonumber\\
    \delta^{\mathrm{CT}}\kappa_S&=\frac{1}{v_S^3}\left(\frac{3v^2}{2}H_{G^0G^0}-\frac{v^2}{2}H_{hh}+\frac{vv_s}{2}H_{hs}+v_S^2H_{ss}-3v_SN_s\right) \,,\nonumber\\
    \delta^{\mathrm{CT}}\kappa_{SH}&=\frac{H_{hh}-3H_{G^0G^0}}{v_S}+\frac{H_{hs}}{v} \,,\nonumber\\
    \delta^{\mathrm{CT}}\lambda&= \frac{H_{G^0G^0}-H_{hh}}{v^2}\,,\nonumber\\
    \delta^{\mathrm{CT}}\lambda_S&= \frac{1}{v_S^4}\left(\frac{-3v^2}{4}H_{G^0G^0}+\frac{v^2}{4}H_{hh}-\frac{v_S^2}{2}H_{ss}+v_SN_s\right)\,,\nonumber\\
    \delta^{\mathrm{CT}}\lambda_{SH}&=\frac{3H_{G^0G^0}-H_{hh}}{v_S^2}-\frac{2H_{hs}}{vv_S} \,,\nonumber\\
    \delta^{\mathrm{CT}}t_{G^0}&=-N_{G^0} \,,\nonumber\\
    \delta^{\mathrm{CT}}t_{G^+}&=-N_{G^+} \,,\nonumber\\
    \delta^{\mathrm{CT}}t_{G^-}&=-N_{G^-} \,,\nonumber\\
    \delta^{\mathrm{CT}}t_h&=H_{G^0G^0}v-N_h \,,\nonumber\\
    \delta^{\mathrm{CT}}t_s&=0 \,,
\end{align}}
where $N_{\phi_i}$ and $H_{\phi_i\phi_j}$ are the first and second derivatives of the CW potential evaluated at the minimum of the potential, i.e.
\begin{equation}
    N_{\phi_i}=\left.\frac{\partial V_{\mathrm{CW}}^{(1)}}{\partial\phi_i}\right|_{\langle\phi\rangle_{T=0}}\,,\qquad\text{and}\qquad H_{\phi_i\phi_j}=\left.\frac{\partial^2 V_{\mathrm{CW}}^{(1)}}{\partial\phi_i\partial\phi_j}\right|_{\langle\phi\rangle_{T=0}}\,.
\end{equation}
Turning next to the temperature-dependent corrections, the one-loop thermal potential \cite{efft1,efft2} is given by  
\begin{equation}
    V_{\mathrm{T}}^{(1)}(\phi,T)=\sum_i\frac{n_iT^4}{2\pi^2}J_{\pm}\left(\frac{m_i^2(\phi)}{T^2}\right),
\end{equation}
where $i$ runs again over the particle content of the model, and $J_{\pm}$ are the fermionic and bosonic thermal integrals, respectively, and are defined as:  
\begin{equation}
    J_{\pm}(x)=\mp \int_0^{\infty}dz\: z^2\log\left[1\pm\exp\left(-\sqrt{z^2+x}\right)\right]\,.
\end{equation}

Finally, we have to solve the IR-divergence problem that arises due to the massless bosonic zero Matsubara modes of finite-temperature field theory and the infrared pole of the propagator of these modes.  
To solve this problem, there are different ways of resumming relevant higher-order effects in order to dress the propagator of the zero-modes and avoid the pole divergence. 
The minimal set of higher-order diagrams that must be resummed has been shown to be those known as Daisy diagrams. 

There are different methods that one can follow to resum these diagrams. In this work, we have followed the Arnold-Espinosa resummation method~\cite{Arnold:1992rz} to disentangle the UV physics from the IR one. In this method, the mass of the zero Matsubara modes is shifted by the thermal mass and, therefore, substitute the propagator of the zero mode with a dressed propagator, avoiding the divergence of the pole. This leads to a contribution to the potential with the following structure,
\begin{equation}
    V_{\mathrm{daisy}}^{(1)}(\phi,T)=-\sum_i\frac{T}{12\pi}\mathrm{Tr}\left\{[m_i^2(\phi_i)+\Pi_i^2(T)]^{\frac{3}{2}}-[m_i^2(\phi_i)]^{\frac{3}{2}}\right\},
\end{equation}
where $i$ runs over the different bosonic degrees of freedom, and $\Pi_i^2(T)$ is the squared thermal mass of the boson $\phi_i$, generated by the thermal potential. The hard thermal masses for the scalars and the longitudinal components of the gauge fields read
\begin{align}
\Pi_{\Phi}(T)
&= 
\left(
\frac{3 g_{2}^{2}+g_{1}^{2}}{16}
+\frac{\lambda}{2}
+\frac{\lambda_{SH}}{12}
+\frac{y_{t}^{2}}{4}
\right)T^{2}\,,
\nn\\
\Pi_{S}(T)
&= 
\left(
\frac{\lambda_{SH}}{3}
+\lambda_{S}
\right)T^{2}\,,
\nn\\
\Pi_{W_L}(T)
&= 
\frac{11}{6}\, g_{2}^{2}\, T^{2}\,,
\nn\\
\Pi_{B_L}(T)
&= 
\frac{11}{6}\, g_{1}^{2}\, T^{2}\,.
\end{align}
Here $g_1$ and $g_2$ denote the $U(1)_Y$ and $SU(2)_L$ gauge couplings, respectively, and $y_t$ is the top Yukawa coupling of the detected Higgs boson.


\subsection{Electroweak Phase Transition dynamics}
Employing \cref{VT} 
we can compute the thermal evolution of the Higgs potential in order to study the dynamics of the EWPT. Two main ways in which the EWPT could have occurred can be distinguished. On the one hand, 
it could have proceeded as a second-order phase transition or a cross-over, meaning that the transition from the false vacuum to the EW one is smooth and there is no associated energy gap. On the other hand, the EWPT could have happened as a first-order phase transition.  
In the latter case, which we denote as FOEWPT, during the thermal evolution of the potential a barrier forms between the false and the EW minima. The only way to transition is through quantum tunnelling, leading to an energy gap in the transition. This event occurs nearly simultaneously at different points in the Universe as it expands, creating bubbles of the EW vacuum that nucleate and then expand in the background of the false vacuum. 
The temperature at which this transition occurs is called the nucleation temperature $T_n$, at which the tunnelling decay rate from the false to the EW minima per Hubble volume matches the Hubble rate. 

\subsubsection*{SFOEWPT characterisation}
As discussed in the introduction, we are interested in the realization of a FOEWPT. 
However, FOEWPTs by themselves are not sufficient to explain the origin of the BAU --- they must be a strong FOEWPT (SFOEWPT). We consider a FOEWPT to be a SFOEWPT if the suppression of the sphaleron transitions \cite{Klinkhamer:1984di} in the broken-phase vacuum is large enough to avoid the wash-out of the generated baryon asymmetry. This leads to the following condition
\begin{equation}
    \xi_n\equiv\frac{v_n}{T_n}\gtrsim 1,
\end{equation}
where $v_n\equiv v(T_n)$. Thus, the first step to determine whether we have an SFOEWPT is to compute $T_n$, which is the temperature at which the tunnelling decay rate from the false to the EW minima per Hubble volume matches the Hubble rate 
\begin{equation}
    \frac{\Gamma(T_n)}{H^4(T_n)}\overset{!}{=}1\,,\label{nucleation}
\end{equation}
where the tunnelling decay rate is given by~\cite{tunrate}
\begin{equation}
    \Gamma(T)=A(T)\,e^{-S_E(T)}\,.
    \label{GaT}
\end{equation}
In this expression, $A(T)$ is a temperature-dependent prefactor (discussed below)
and $S_E(T)$ is the Euclidian action (expressed in terms of the Euclidean time $\tau$),
\begin{equation}
    S_E(T)=\int d\tau d^3x\left[\frac{1}{2}(\partial_{\mu}\phi)(\partial^{\mu}\phi)+V(\phi,\tau)\right]\,.
\end{equation}
In finite-temperature field theory, the Euclidean time evolution replaces the $T$ evolution, and there are periodic boundary conditions for this coordinate, i.e.\ $\tau\in[0,\frac{1}{T}]$. Thus, 
by integrating over this coordinate and performing a 
change of variables to spatial spherical coordinates, with $\rho=\sqrt{\sum_{i\leq3}x_i^2}$, one can obtain a three-dimensional action that is known as the bounce action~\cite{Linde:1980tt,s4}
\begin{equation}
    S_3(T)=4\pi\int^{\infty}_0d\rho \,\rho^2\left[\frac{1}{2}\left(\frac{d\phi}{d\rho}\right)^2+V(\phi,T)\right]\,,
\end{equation}
in terms of which the tunnelling decay rate is expressed as
\begin{align}
    \Gamma(T)=A(T)\,e^{-S_3(T)/T}\,.
\end{align}
In this equation, the prefactor $A(T)$ can be computed, following Ref.~\cite{Linde:1980tt}, as
\begin{align}
    A(T)=T^4\left(\frac{S_3(T)}{2\pi T}\right)^{3/2}\,.
\end{align}
Imposing that, far from the true vacuum bubble, the false vacuum remains unaffected by the phase transition, and that the transition occurs at $\rho=0$ as boundary conditions to solve the equations of motion of the bounce action, one can find the bounce solution for the fields $\phi_B$, which describes a bubble of the true vacuum nucleating and expanding within the false vacuum. 

With the computed bounce solution one
can calculate the transition rate for different temperatures and thereby determine the nucleation temperature.  
The problem is that one has to trace the real and the false vacua for different temperatures because the potential is a 
thermodynamic quantity, and so are the minima obtained by solving the equations of motion. Since this cannot be done 
analytically, one resorts to use numerical methods to solve the bounce action each time. 
This adds a level of computational difficulty to our investigations, and to 
overcome this, we have used the public tool \texttt{BSMPTv3} \cite{Basler:2024aaf}. In comparison to earlier versions~\cite{Basler:2018cwe,Basler:2020nrq}, \texttt{v3}\footnote{More precisely, we have employed \texttt{v3.1.1}, which contains improved features in the GW computation, in particular in the case of strong supercooling.} additionally allows the computation of the background spectrum 
of stochastic GWs produced during the SFOEWPT and features improved numerical stability. 


\subsection{Stochastic GW background}
The nucleation of the bubbles during a SFOEWPT would have been a very violent event in the early Universe, giving rise to a stochastic background of primordial GWs that could be detectable nowadays. 
In this section we briefly review the computation of
the spectrum of GWs created during the SFOEWPT and verify whether it would be observable in a future GW observatory, such as the space-based GW interferometer LISA~\cite{Caprini:2019egz,LISACosmologyWorkingGroup:2022jok}. 
However, since GWs are a macroscopic phenomenon, the first step is to describe the phase transition dynamics at the thermodynamical level.

\subsubsection{Macroscopic characterisation of the phase transition}
There are four main macroscopic parameters that characterise the GW spectrum generated during a phase transition. The first of them is the temperature, $T_*$, at which the transition takes place. Two temperatures are used in the literature for the transition temperature: the first one is the nucleation temperature $T_n$, which was defined in the previous section. However, we use here instead the more appropriate percolation temperature $T_p$, which represents the temperature at which at least $29\%$ of the Universe has tunnelled to the true vacuum. At this false vacuum fraction, a macroscopic cluster of true vacuum bubbles has formed, driving the system irreversibly toward the true vacuum phase \cite{Broadbent:1957rm}. This choice is motivated by the nature of the phenomena sourcing of the GWs, which will be discussed below. Once the transition temperature has been defined, a second parameter, $\alpha$, measures the strength of the phase transition, 
\begin{equation}
    \alpha = \frac{30}{g_*\pi^2T_*^4}\left[V(\phi_f)-V(\phi_t)-\frac{T}{4}\left(\frac{\partial V(\phi_f)}{\partial T}-\frac{\partial V(\phi_t)}{\partial T}\right)\right]_{T=T_*},\label{eq:def_alpha}
\end{equation}
where $g_*$ is the effective number of relativistic degrees of freedom and is a function of $T$ and $\phi_t$ and $\phi_f$ are the true and false phases respectively (the false phase being the one where the transition begins and the true phase the one where it finishes). 
The third parameter that characterises the phase transition is the inverse duration of the phase transition in units of the Hubble rate $H_*$ at the moment of the transition,
\begin{equation}
    \frac{\beta}{H_*}=T_*\frac{d}{dT}\left(\frac{S_3(T)}{T}\right)\Bigg|_{T_*}\,.
\end{equation}
In this case, $H_*$ is the Hubble parameter in a radiation-dominated universe, i.e.
\begin{equation}
    H_*\equiv T_*^2\sqrt{\frac{g_*\pi^2}{90\widetilde{M}^2_{\mathrm{Pl}}}}\,,
\end{equation}
where $\widetilde{M}_{\mathrm{Pl}}$ is the reduced Planck mass.
    Finally, the last parameter to be computed is the wall velocity, \vw. There exist several approaches to compute the bubble dynamics \cite{Branchina:2024rva, DeCurtis:2024hvh, DeCurtis:2023hil, DeCurtis:2022hlx,Branchina:2025adj}, however all of them suffer from large theoretical uncertainties (for a general discussion of theoretical uncertainties in studies of EWPT dynamics, see e.g.\ Ref.~\cite{Biekotter:2025npc}) that can have a considerable impact on final results for the produced GW spectrum. 
In several studies of SFOEWPT in models with extended Higgs sectors~\cite{Ai:2023see,Ekstedt:2024fyq,Ellis:2022lft}, it has been shown that the value is typically within the range of $\vw \in [0.2,1]$. In this work, we will assume as a first approach the bubble wall velocity to be equal to $0.95$, which is a pessimistic choice for the observability of the GW spectrum. 
Later, in \cref{sec:vwdep}, we will study the dependence of the observability of the GW signal on the bubble wall velocity.

\subsubsection{GW spectrum}
Three main sources of GWs have been studied for a SFOEWPT~\cite{Caprini:2015zlo,Caprini:2019egz}. First, there are the collisions of the bubbles: this happens when bubbles of the true vacuum nucleate and expand into the false vacuum, until they collide with each other, thereby generating GWs. Additionally, the dynamics of the EWPT can create pressure waves in the plasma around the bubble walls, which act like sound waves that propagate, leading to a second source generating GWs. Finally, the sound waves induce magnetohydrodynamical turbulence in this plasma. The plasma is then far from equilibrium during the SFOEWPT and, similarly to how turbulence in fluids can generate waves, GWs are sourced by turbulence after sound-wave decay.\footnote{We note that we employ the default value of $\epsilon_\text{turb}=0.1$ from \texttt{BSMPT} for the calculation of the spectrum of GW sourced by turbulence.} The total spectrum of produced gravitational waves can be described as
\begin{equation}
    h^2\Omega_{\mathrm{GW}}(f)=h^2\Omega_{\mathrm{col}}(f)+h^2\Omega_{\mathrm{SW}}(f)+h^2\Omega_{\mathrm{turb}}(f)\,,
\end{equation}
where
\begin{equation}
    h^2\Omega_{\mathrm{GW}}(f)\equiv\frac{h^2}{\rho_c}\frac{\partial\rho_{\mathrm{GW}}}{\partial\log f}\,,
\end{equation}
$\rho_\text{GW}$ being the energy density of the stochastic background of GWs, $h$ the reduced Hubble constant, and $f$ denotes the frequency of the GWs. 
In order to understand which contributions are more relevant in the total spectrum, we need to know if we have runaway bubble walls. For this purpose, we have checked the Bödecker-Moore criterion \cite{Bodeker:2009qy} as well as the new criterion based on Ref.~\cite{Ai:2024shx}. We found that for all the points considered in our work we have non-runaway bubble walls, meaning that the bubbles expanding in the plasma can reach a relativistic terminal velocity. In this case, the energy stored in the scalar field is negligible, and therefore, the main sources of gravitational waves are related to the fluid dynamics, namely sound waves and plasma turbulence. On the other hand, for this type of bubble, collisions can be neglected in $\Omega_\text{GW}$. We can express the spectrum of each type of source in terms of its peak amplitude ($\Omega^\text{peak}$) and peak frequency ($f^\text{peak}$), which leads finally to the following expression for the GW spectrum:
\begin{align}
     h^2\Omega_{\mathrm{GW}}(f)\approx&\  h^2\Omega_{\mathrm{SW}}^{\mathrm{peak}}\left(\frac{4}{7}\right)^{-\frac{7}{2}}\left(\frac{f}{f_{\mathrm{SW}}^{\mathrm{peak}}}\right)^3\left[1+\frac{3}{4}\left(\frac{f}{f_{\mathrm{SW}}^{\mathrm{peak}}}\right)^2\right]^{-\frac{7}{2}}\nonumber\\
     &+h^2\Omega_{\mathrm{turb}}^{\mathrm{peak}}\left(\frac{(f/f_{\mathrm{turb}}^{\mathrm{peak}})^3}{(1+f/f_{\mathrm{turb}}^{\mathrm{peak}})^{11/3}(1+8\pi f/H_*)}\right).
\end{align}
Descriptions of the peak amplitudes and the spectral functions for the different sources can be found in Ref.~\cite{Basler:2024aaf}.

\subsubsection{Observability}
Once having derived the power spectrum of the GW background, we can compare it with the sensitivity of different planned GW observatories to test if this signal could be detected. If we check the characteristic frequency range of the potential signals coming from a SFOEWPT, they lie within the frequency range of the future GW observatory LISA. Therefore, in this work, we have focused on this observatory. 
To determine whether the signal is observable, we compare the power spectrum of the signal with the sensitivity of LISA. To do so, we define a signal-to-noise ratio (SNR) following Ref.~\cite{Caprini:2019egz}, 
\begin{equation}
\mathrm{SNR}=\sqrt{\mathcal{\tau}\int^{f_{\mathrm{max}}}_{f_{\mathrm{min}}}df\left[\frac{h^2\Omega_{\mathrm{GW}}(f)}{h^2\Omega_{\mathrm{sens}}(f)}\right]^2},
\end{equation}
where $\tau$ is the data-taking time of LISA and $h^2\Omega_{\mathrm{sens}}$ is the nominal sensitivity of LISA to stochastic sources like phase transitions. Throughout this work, we have made the conservative assumption of a data-taking period of three years.


\section{Phenomenological implications}
\label{sec:pheno}

In this section, we investigate the dynamics of the EWPT in the RxSM employing the theoretical framework described in the previous section. Numerical computations are performed with the public code \texttt{BSMPTv3.1.1}~\cite{Basler:2024aaf}. 

\subsection{Different thermal histories}

\begin{figure}[ht]
    \centering
    \includegraphics[width=0.32\linewidth]{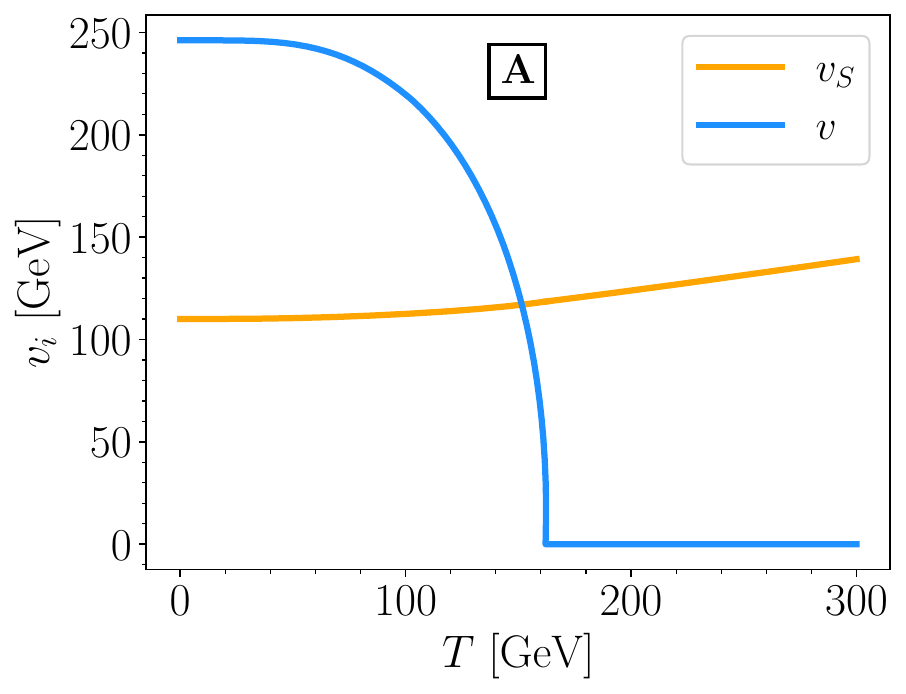}
    \includegraphics[width=0.32\linewidth]{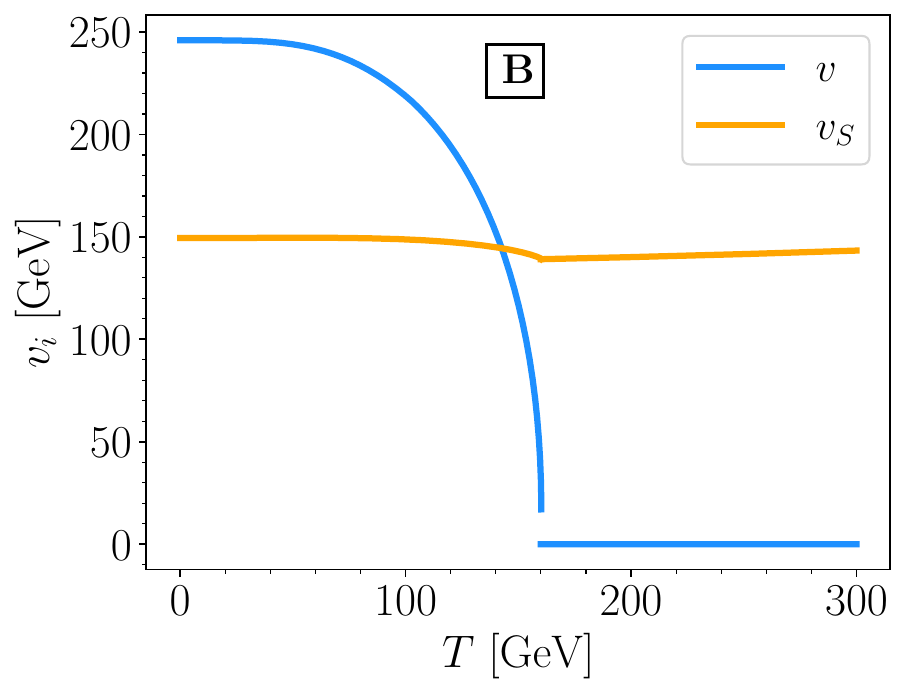}
    \includegraphics[width=0.32\linewidth]{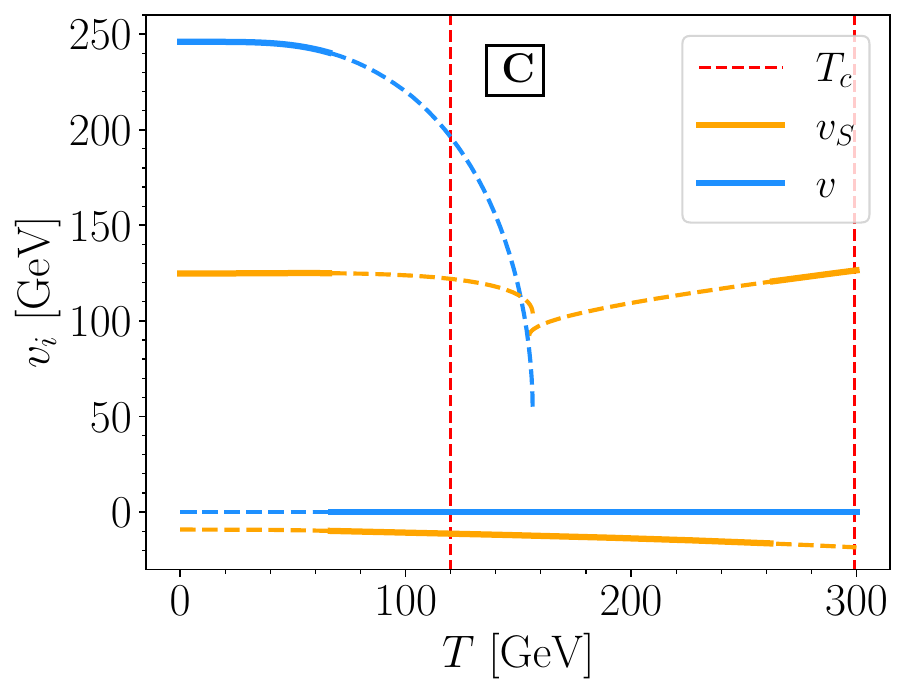}
    \includegraphics[width=0.32\linewidth]{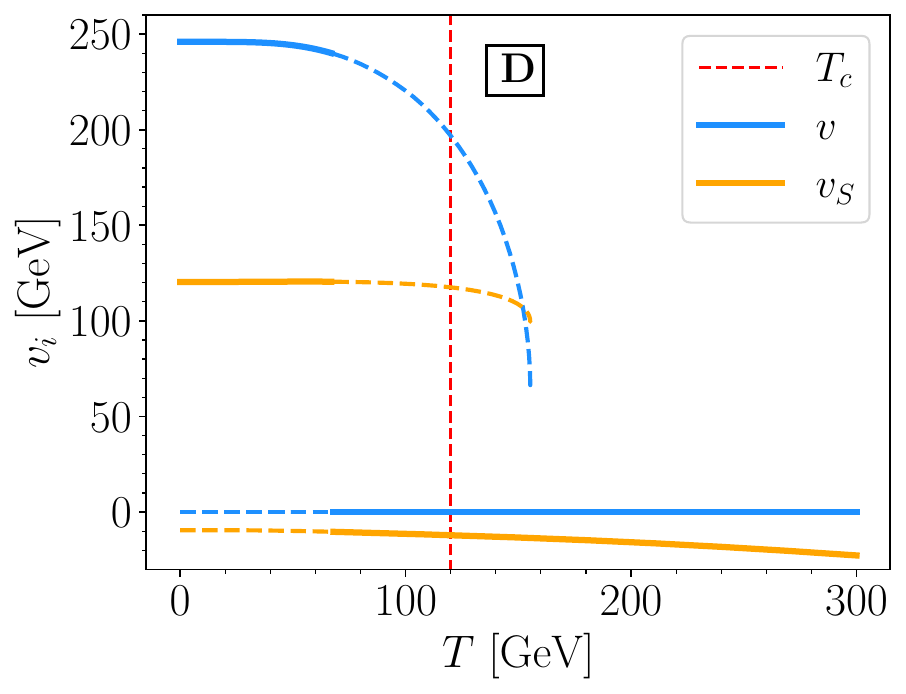}
    \includegraphics[width=0.32\linewidth]{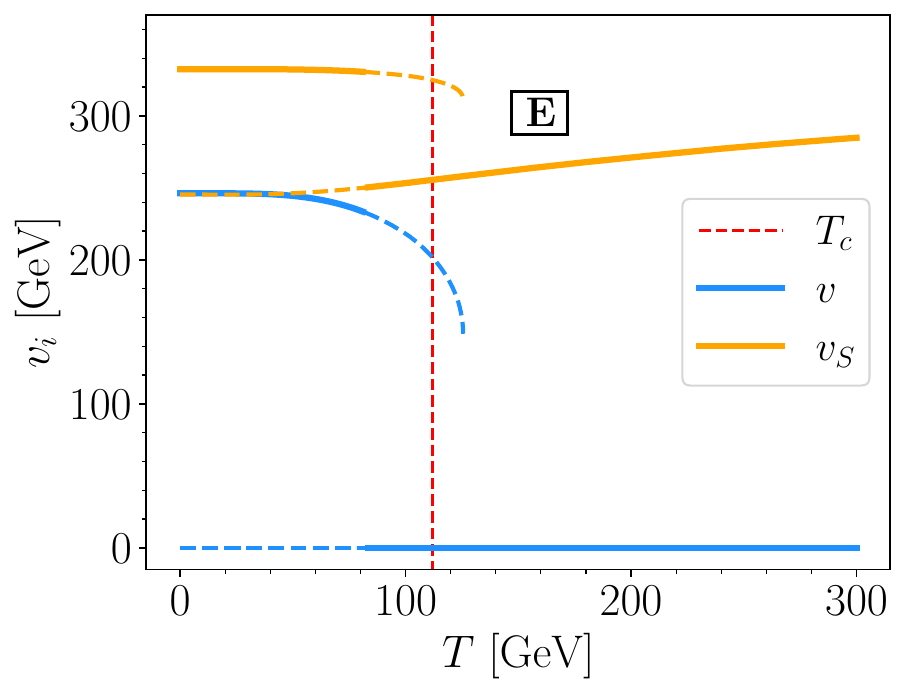}
    \includegraphics[width=0.32\linewidth]{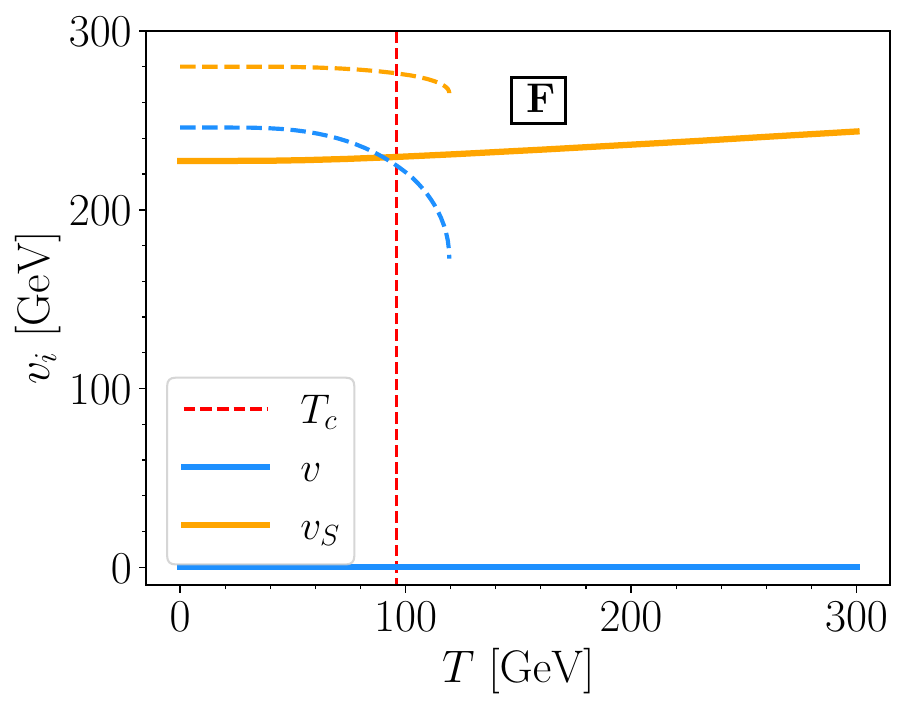}
    \caption{The tracing of the different minima of the potential with respect to the temperature for six different thermal histories. In blue we show the minima of the EW doublet and in orange the minima of the singlet field. The solid lines represent the path that the universe follows. The red dashed line indicates the critical temperature $T_c$.}
    \label{thermalhistories}
\end{figure}

We begin by investigating the different possible thermal histories in the RxSM (a similar analysis in the 2HDM can be found in Ref.~\cite{Biekotter:2022kgf}). By scanning of the parameter space of the RxSM, we find six different scenarios, which we label from A to F (similar to Ref.~\cite{Biekotter:2022kgf}), and which we illustrate in \cref{thermalhistories} for selected points listed in \cref{tab:thhist}.

\begin{table}[h]
\centering
\begin{tabular}{@{}cccccc@{}}
\toprule
 Id & $m_H$ [GeV] & $\cos\alpha$ & $\kappa_S$ [GeV] & $\kappa_{SH}$ [GeV] & $v_S$ [GeV] \\ \midrule
A & 342.8     & 0.9999         & $-1000$      & $-100$          & 110       \\
B & 288.5     & 0.9949         & $-900$       & $-54$           & 149       \\
C & 285.5     & 0.9891         & $-900$       & $-87$           & 125       \\
D & 290.7     & 0.9881         & $-900$       & $-93$           & 120       \\
E & 463.9     & 0.9955         & $-24$        & $-992$          & 332       \\
F & 600.1     & 0.9800         & $-300$       & $-1406$         & 280       \\ \bottomrule
\end{tabular}
\caption{RxSM benchmark points for the investigation of the different thermal histories of the Universe in \cref{thermalhistories}.}
\label{tab:thhist}
\end{table}

For each scenario, we trace in \cref{thermalhistories} the evolution of the EW and singlet minima as a function of the temperature in order to understand its specific dynamics. 

The blue and orange lines in \cref{thermalhistories} correspond respectively the tracing of the minima of the EW doublet and of the singlet. The parts of the line in solid style indicate the path that the Universe actually follows, i.e.\ at the temperature where 
a change from dashed to solid is found, a phase transition takes place.   
The first scenario (top left, \textbf{A}) is a \textit{second-order PT}. The second case (top middle, \textbf{B}) is a \textit{weak first-order PT}, where a first-order phase transition occurs along the EW direction while the transition along the singlet field direction is continuous. In case \textbf{B}, the first-order PT along the EW direction is so weak --- in the sense that the discontinuity in the evolution of the Higgs VEV is so small --- that this scenario cannot be phenomenologically distinguished from a second-order PT. The third scenario (top right, \textbf{C}) is a \textit{multi-step-first order PT}: specifically the PT occurs in two steps, starting with a first-order PT along the singlet direction at higher temperature, followed by another first-order phase transition where both VEVs experience a discontinuity. 
The latter transition corresponds to the EWPT. The fourth scenario (bottom left, \textbf{D}) is the \textit{low singlet VEV first-order PT}, where the EWPT occurs as a one-step transition with discontinuities along the EW and singlet directions concurrently. We name the fifth scenario (bottom middle, \textbf{E}) \textit{high singlet VEV first-order PT}. This scenario is similar to case \textbf{D}, but with the difference that the minimum of the singlet VEV at the moment of the transition is positive. We will see that this leads to weaker transitions than in case \textbf{D}. Finally, the last possible history that we find (bottom right, \textbf{F}) is \textit{vacuum trapping}~\cite{Biekotter:2022kgf}. In this scenario, the barrier between the symmetric and EW minima is so large that the tunnelling time is longer than the age of the Universe, meaning that one remains trapped in the symmetric phase. We note that we define here vacuum trapping in terms of the computed tunnelling rate, which is compared to the age of the Universe, rather than in terms of some threshold value of the bounce action.


\subsection{SFOEWPT}
\label{sec:fivedimscan}
As a first step to search for SFOEWPT scenarios in the RxSM, we perform a parameter scan in the five-dimensional parameter space of the model. 
We begin by checking theoretical and experimental constraints as outlined in \cref{sec:constraints}, 
and afterwards compute the EWPT dynamics for the allowed points. 
The scan ranges are as follows,
\begin{align}
    m_H&\in[260,1000]~\mathrm{GeV},\nonumber\\
    \cos\alpha&\in[0.95,1],\nonumber\\
    v_S&\in[30,300]~\mathrm{GeV},\nonumber\\
    \kappa_S&\in[-1000,1000]~\mathrm{GeV},\nonumber\\
    \kappa_{SH}&\in[-1000,0]~\mathrm{GeV}.
\end{align}
We choose as lower bound $m_H>260$~GeV in order to allow the decay channel $H
\to hh$ and enable a complementary analysis of di-Higgs production (including resonant contributions from $H$) for the scan points --- see \cref{sec:collider}. The requirement of $\kappa_{SH}$ to be negative comes from the theoretical constraint of boundedness-from-below of the tree-level potential.

\begin{figure}[ht]
    \centering
    \includegraphics[width=0.49\linewidth]{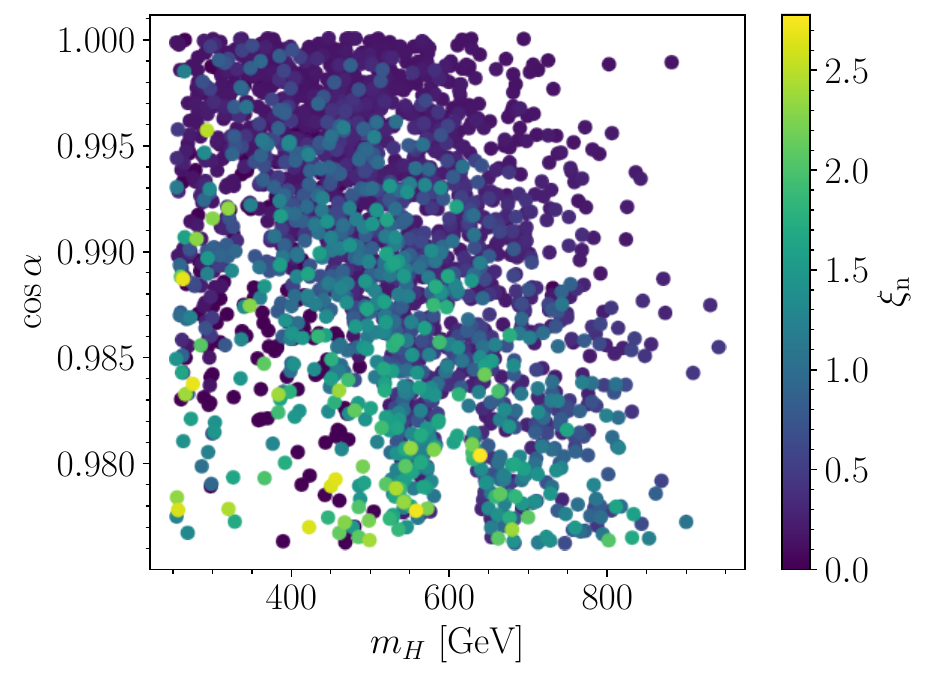}
    \includegraphics[width=0.49\linewidth]{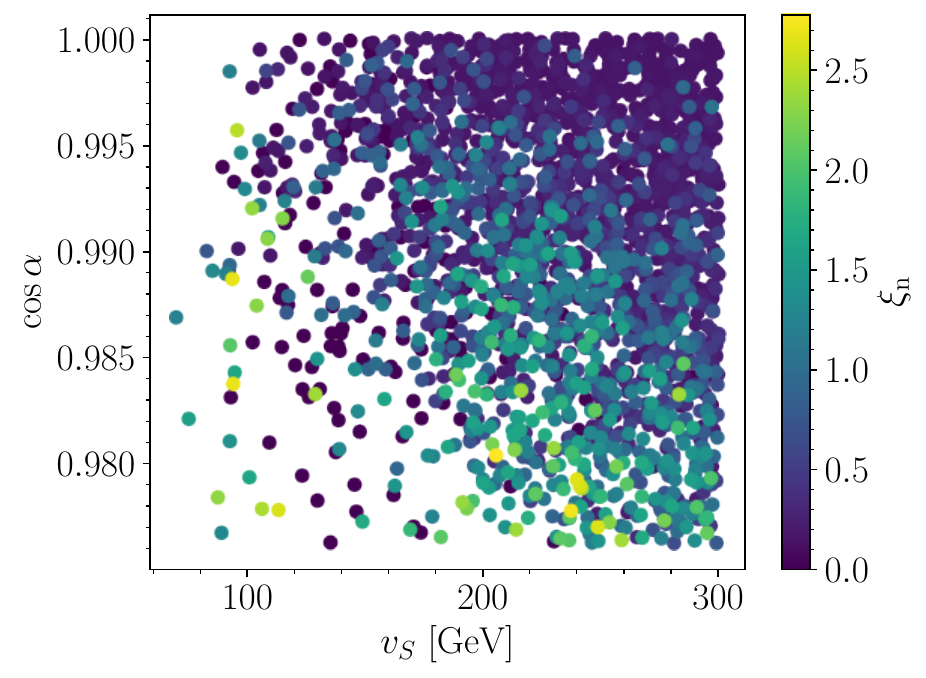}
    \includegraphics[width=0.49\linewidth]{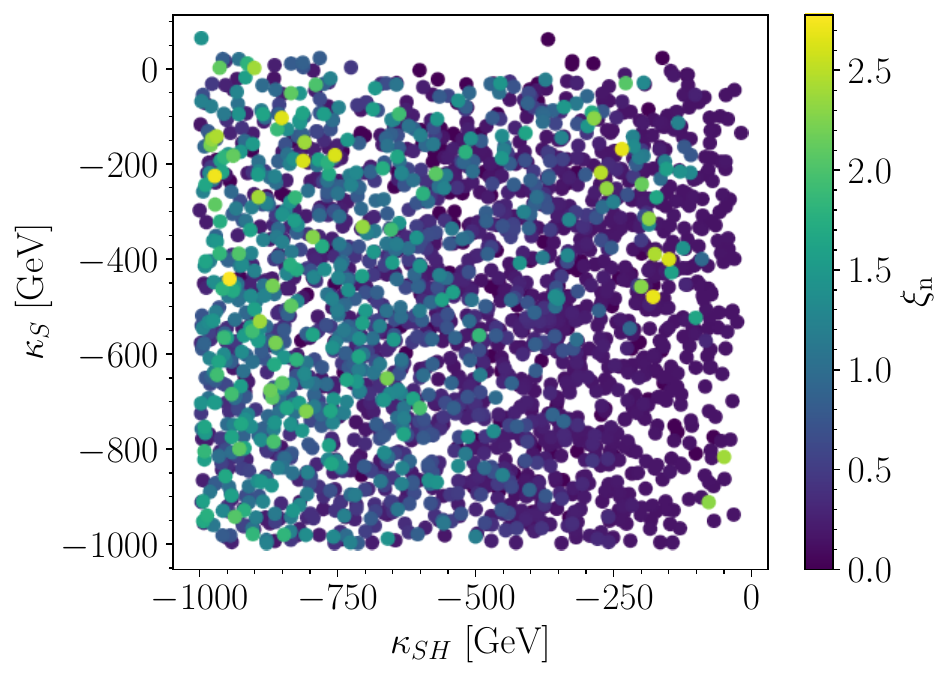}
    \includegraphics[width=0.49\linewidth]{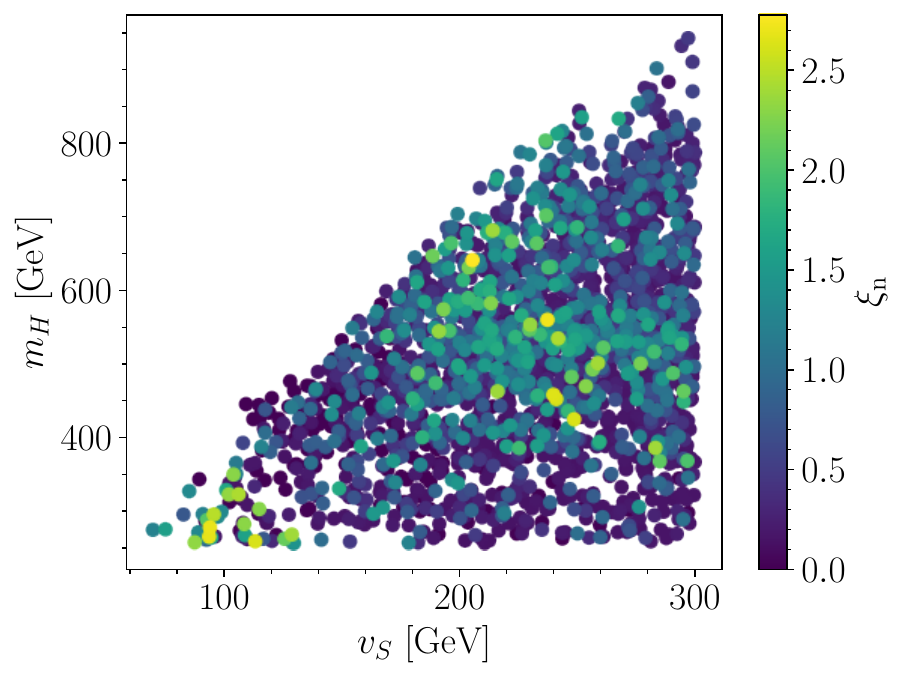}
    \caption{Results for $\xi_n\equiv v_n/T_n$ for the points from our RxSM parameter scan.  
    \textit{Top left}: $\{\cos \alpha,m_H\}$ plane; \textit{top right}: $\{\cos \alpha,v_S\}$ plane; \textit{bottom left}: $\{\kappa_S ,\kappa_{SH}\}$ plane; \textit{bottom right}: 
    $\{m_H,v_S\}$ plane.}
    \label{scan1}
\end{figure}

In \cref{scan1} we present as colour coding the ratio $\xi_n\equiv v_n/T_n$ 
in the planes $\{\cos\alpha,m_H\}$ (top left), $\{\cos \alpha,v_S\}$ (top right), $\{\kappa_S,\kappa_{SH}\}$ (bottom left) and $\{m_H,v_S\}$ (bottom right). Our first finding is to confirm the existence of points exhibiting a SFOEWPT, with $\xi_n>1$, in the scanned region. 
From the top row in \cref{scan1} we can see that the strongest phase transitions occur further from the alignment limit ($\cos\alpha=1$), while the weaker ones are closer to it. Another important observation is that we do not find points for large positive values of $\kappa_S$. Indeed, for points with large positive values of $\kappa_S$, the symmetric minimum of the NLO potential at $T=0$ is deeper than the EW minimum and therefore the EW vacuum
is not stable at NLO. We note that for the potential at NLO, we only check whether the EW minimum is deeper than the minimum of the origin of the potential. Further minima may appear also away from the origin of the potential. However, we do not expect the description of these minima to be sufficiently trustworthy, because the renormalisation scheme employed in \texttt{BSMPT} is defined in the vicinity of the electroweak vacuum. Thus, when moving far from the EW minimum, the renormalisation scheme can become less reliable. 

One can also see from \cref{scan1} that the strongest phase transitions (i.e.\ with $\xi_n>2$) arise in two distinct regions of the parameter space: a first region with $v_S\lesssim 150\gev$, $m_H\lesssim 400\gev$, $\kappa_{SH}\gtrsim -600\gev$ and $\kappa_S\gtrsim -300\gev$; and a second region for $v_S\gtrsim 150\gev$, $m_H\gtrsim 400\gev$ and $\kappa_{SH}\lesssim -500\gev$. The second region spans the entire range  $-1000\gev<\kappa_S\lesssim 0\gev$ that is populated by allowed points, although the largest values of $\xi_n$ are reached for points with $\kappa_S\gtrsim -600\gev$.

Investigating the two regions in more detail, one finds that the first region with low $v_S$ corresponds to case \textbf{D} in terms of the possible thermal histories of the Universe shown in \cref{thermalhistories}, while the second region with larger $v_S$ corresponds to case \textbf{E}.

\subsection{GW signal at LISA}
In the following, we focus on the two regions of parameter space found in the previous section that feature the strongest FOEWPT, and we investigate whether they can give rise to GW spectra observable at LISA. 
Investigating scenarios with five free parameters is however complicated, because of the interplay between effects from different parameters. 
Instead, we choose to define, for each region, sets of three phenomenological relations between the RxSM parameters that maximise $\xi_n$, and can then study two-dimensional benchmark planes.

\subsubsection{Benchmark plane 1}
\begin{figure}[ht!]
    \centering
    \includegraphics[width=0.75\linewidth]{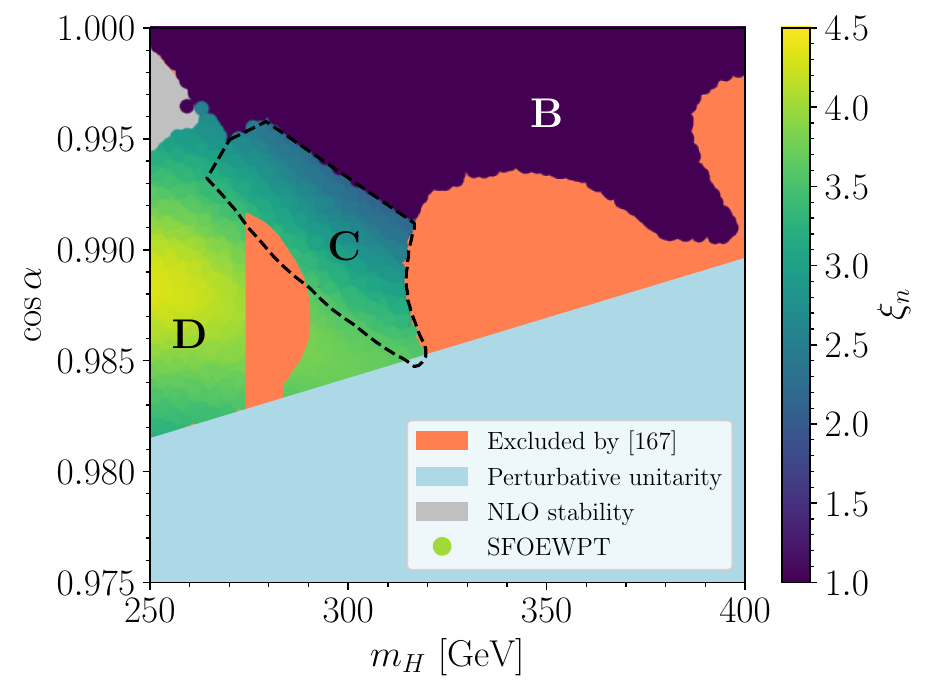}
    \caption{Parameter scan results in the RxSM for benchmark plane 1. The colour coding shows the ratio $\xi_n=v_n/T_n$. 
    The different thermal histories are labelled following \protect\cref{thermalhistories}.
    We indicate in light blue the region excluded by perturbative unitarity, in grey the region excluded by the stability of the EW vacuum at NLO and in light red the region excluded by direct searches for heavy Higgs bosons in Ref.~\cite{ATLAS:2020tlo}. We note that in region \textbf{B} $0<\xi_n<1$.}
    \label{plane1}
\end{figure}

We begin by considering 
the region with low values of $v_S$ and $m_H$. 
Using the results from the five-dimensional scan in the previous section, we find that the following conditions maximise $\xi_n$ in this region
\begin{align}
    \kappa_S&=-900~\mathrm{GeV}\,,\nonumber\\
    \kappa_{SH}&=(5662.9\cos\alpha-5688.4)~\mathrm{GeV}\,,\nonumber\\
    v_S&=(4239.5\cos\alpha-4067.6)~\mathrm{GeV}\,. \label{conditions1}
\end{align}
We are then left with only two free parameters, over which we scan with the ranges\footnote{We note that no point allowed by \texttt{HiggsSignals} was found below $\cos\alpha=0.975$, hence the lower bound quoted here.}
\begin{align}
    m_H&\in[260,1000]~\mathrm{GeV}\,,\nonumber\\
    \cos\alpha&\in[0.975,1]\,.
\end{align}
In \cref{plane1} we present the results of the scan in this benchmark plane. The colour coding shows the value of $\xi_n$
for the points allowed by the experimental and theoretical constraints. The region shaded in light blue is excluded by perturbative unitarity. It is important to note that the fact that perturbative unitarity excludes points farther away from the alignment limit is not a general feature of the RxSM, but is only an artefact of the considered plane. We indicate in grey the region where the tree-level potential is stable while at the one-loop level the EW vacuum is unstable (at $T=0$). 
Finally, we show in light red the parameter region that is excluded by the direct search of a heavy BSM Higgs boson decaying into a pair of $Z$ bosons in the $l^+l^-l^+l^-$ and $l^+l^-\nu\bar{\nu}$ channels by ATLAS \cite{ATLAS:2020tlo}.

In \cref{plane1} we can see that in the allowed parameter region three different thermal histories are possible. Following the labels in \cref{thermalhistories}, we name the corresponding regions \textbf{B}, \textbf{C} and \textbf{D}. First, in region \textbf{B} (dark purple) we find weak first-order phase transitions: these transitions are of first order because there is a small discontinuous transition in the EW doublet direction, and the nucleation condition of \cref{nucleation} is fulfilled. However, these points are not interesting for phenomenology because the phase transition is too weak. Next, regions \textbf{C} and \textbf{D} exhibit SFOEWPT --- the difference between the two regions is that region \textbf{C} features multi-step phase transitions, starting with a weak first-order phase transition followed by the EW phase transition occuring as a SFOEWPT,  while in region \textbf{D} the phase transitions occur in one step. We can observe that the behaviour of $\xi_n$ at the boundary between both regions is smooth. This is because in region \textbf{C} 
for the calculation of $\xi_n$ we consider 
the second step of the two-step phase transition, which is the step during which there is tunnelling in the EW doublet direction. Taken separately, this transition corresponds exactly to the single-step transition in region \textbf{D}. 
A similar continuous behaviour of $\xi_n$ is, however, not found at the boundary between regions \textbf{B} and \textbf{C}. The reason is that at some moment the two  branches for positive values of the singlet field in region \textbf{C} in \cref{thermalhistories} connect, and the corresponding phase remains deeper than the negative phase. As a consequence, there is then no tunnelling to the negative phase of the singlet field, meaning that the second step of the phase transition of region \textbf{C} never occurs, and thus there is no SFOEWPT.

\begin{figure}[h!]
    \centering
    \includegraphics[width=0.49\linewidth]{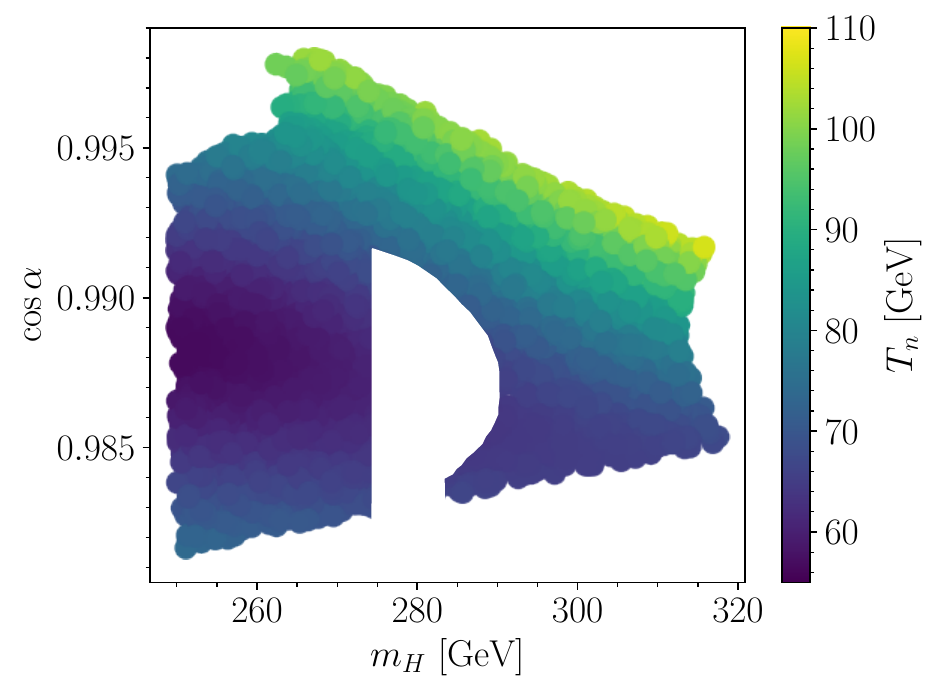}
    \includegraphics[width=0.49\linewidth]{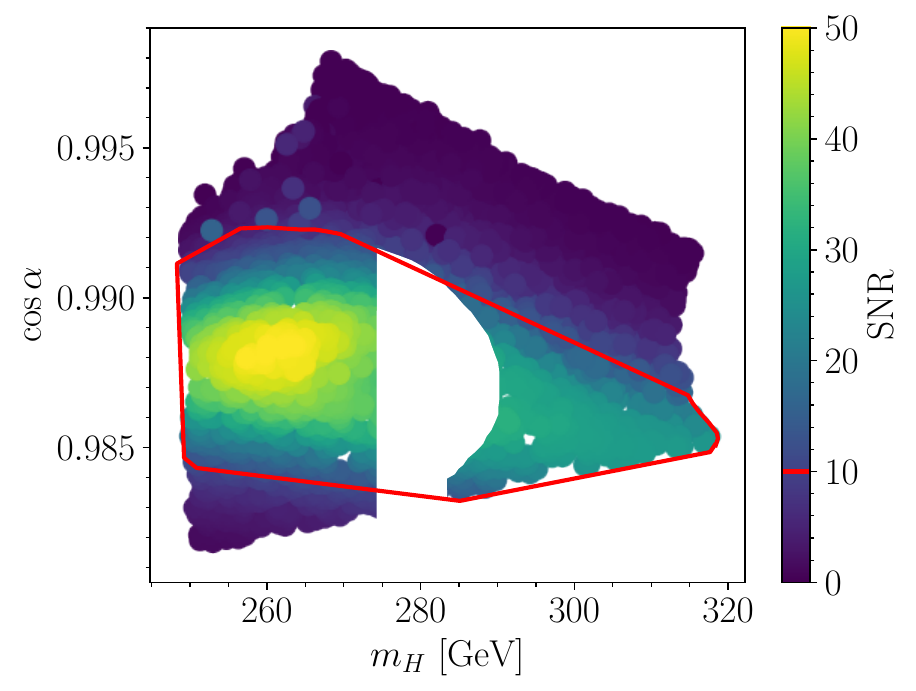}
    \caption{Parameter scan results in the RxSM for the points with a SFOEWPT in the benchmark plane 1. \textit{Left}: $T_n$ (in GeV); \textit{right}: SNR at LISA (assuming 3 years of observations) for $\vw=0.95$, where the region delimited in red represents the observable region (i.e.\ SNR$\;>10$). }
    \label{plane1variables}
\end{figure}

In \cref{plane1variables} we restrict our attention to the part of benchmark plane 1 where we find a SFOEWPT. In the left and right plots the colour coding represents, respectively, $T_n$ (in GeV) and the SNR at LISA for a bubble wall velocity of $\vw=0.95$ and 3 years observation time.%
\footnote{The shape of the shown areas in both plots is slightly different, because the calculation of the SNR fails for a few points.}
From the right plot, we find that there is a large region of the parameter space for which SNR $> 10$, i.e.\ where we find a stochastic GW background detectable at LISA. This result is obtained for $\vw=0.95$, which as will be discussed in \cref{sec:vwdep} is the most pessimistic assumption. Turning next to the left plot of \cref{plane1variables}, we can see that for the points that are observable at LISA $T_n$ is relatively low, $T_n\lesssim70$ GeV --- while the percolation temperature for these same points can be computed to be even significantly lower, $T_p\lesssim60\gev$. We can conclude from this that the high GW signals found in benchmark plane 1 are directly related to low values of $T_p$. We also checked the strength of the transition ($\alpha$, see \cref{eq:def_alpha}) in this region of the benchmark plane, and we discuss in Appendix A the assumption of working during a radiation dominated era and the possibility of supercooling.

\begin{figure}[ht!]
    \centering
    \includegraphics[width=.8\linewidth]{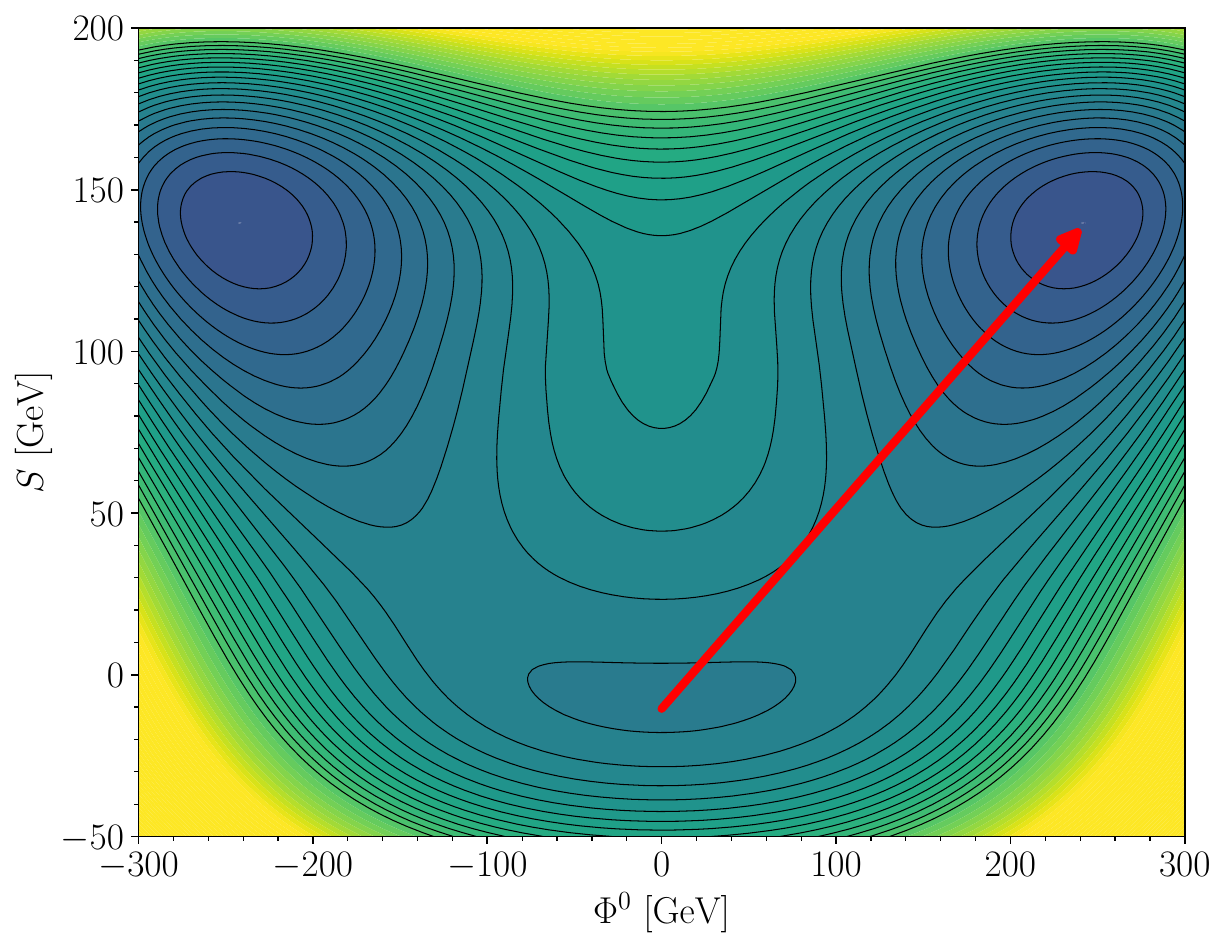}
    \caption{Temperature-dependent one-loop effective potential at the nucleation temperature, $V_\text{eff}(T_n)$, for a benchmark point featuring a SFOEWPT with a thermal history of type \textbf{D} (from \cref{thermalhistories}). The BP is defined by $m_H=262 \gev$, $c_{\alpha}=0.9915$, $v_S=139 \gev$, $\kappa_S=-833 \gev$, $\kappa_{SH}=-72 \gev$, and we obtain $T_n=68\gev$ as well as a SNR$\;>10$. $\Phi^0$ denotes the CP-even neutral component of the Higgs doublet.  }
    \label{pot1}
\end{figure}

An important characteristic of the SFOEWPT in this benchmark plane (in regions \textbf{C} and \textbf{D}) is that the transition along the singlet field direction starts from a negative value of $v_S$. To better visualise this, we display in \cref{pot1} the temperature-dependent effective potential at the nucleation temperature, $V_\text{eff}(T_n)$, for a specific benchmark point with an observable SNR$\;>10$ and $T_n=68\gev$. This point, while not being located directly in benchmark plane~1, features a SFOEWPT driven by the singlet field, with qualitatively the same behaviour as points in the benchmark plane. Because of the negative value of the singlet VEV at
the moment of the transition, there is a large difference between the initial and final values of $v_S$. Another important effect of the negative initial value of the singlet VEV is that it delays the phase transition, thereby lowering the nucleation temperature. This can be observed in the illustration of case \textbf{D} in \cref{thermalhistories}: one sees that at high temperatures the initial singlet phase is far from the final phase and gradually approaches as the temperature decreases. Starting from a larger negative $v_S$ therefore yields a lower $T_n$, and in turn a stronger phase transition. 

\subsubsection{Benchmark plane 2}
We now consider the second region, with high values of $v_S$ and $m_H$, and we define a two-dimensional benchmark plane accordingly. 
Based on the results from the five-dimensional scan in \cref{sec:fivedimscan}, we fix the following conditions
\begin{align}
    \cos\alpha&=0.98\,,\nonumber\\
    \kappa_S&=-300~\mathrm{GeV}\,,\nonumber\\
    v_S&=280~\mathrm{GeV}\,.\label{conditions2}
\end{align}
in order to maximise the strength of the SFOEWPT. For the remaining free RxSM parameters, we scan over the following ranges
\begin{align}
    m_H&\in[260,1000]~\mathrm{GeV}\,,\nonumber\\
    \kappa_{SH}&\in[-2000,-1000]~\mathrm{GeV}\,.
\end{align}

\begin{figure}[ht]
    \centering
    \includegraphics[width=0.75\linewidth]{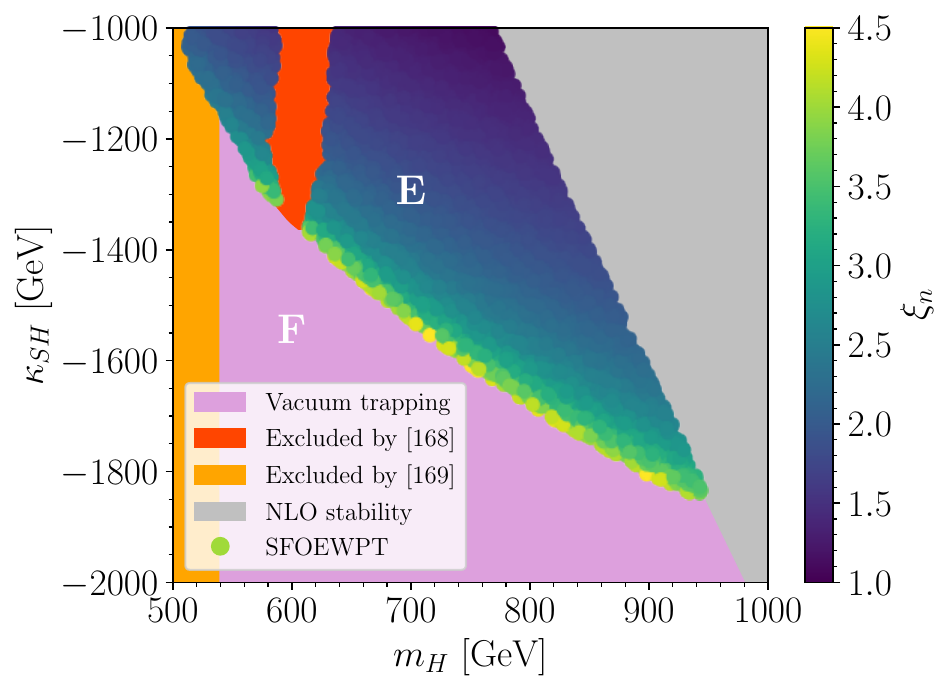}
    \caption{Parameter scan results in the RxSM for benchmark plane 2. The colour coding shows the ratio $\xi_n$. 
    The different thermal histories are labelled following \protect\cref{thermalhistories}. 
    The pink region indicates vacuum trapping, while the grey region is excluded by the stability of the potential at NLO. Regions ruled out by experimental searches are shown in red for the region excluded by direct searches of a heavy Higgs boson in Ref.~\cite{ATLAS:2018sbw}, and in orange for the region excluded by direct searches of two SM-like Higgs bosons in Ref.~\cite{CMS:2021yci}.}
    \label{plane2}
\end{figure}

Scan results for this second benchmark plane are shown in \cref{plane2}. The colour coding represents the ratio $\xi_n$ for the points allowed by experimental and theoretical constraints. The pink region indicates vacuum trapping, while the grey region corresponds to points for which the potential is stable at tree level but not at NLO (at $T=0$). The region highlighted in red is excluded by the combination of searches of heavy resonances decaying into bosonic and leptonic final states by ATLAS \cite{ATLAS:2018sbw}. Finally, the orange region is excluded by direct searches by CMS of a heavy BSM Higgs boson decaying into two lighter Higgs bosons in the $\tau\tau b\bar{b}$ channel \cite{CMS:2021yci}. 

In \cref{plane2} we can observe that in the allowed parameter space of benchmark plane 2 points with a phase transition only feature one type of thermal history, namely case \textbf{E}. Following the labelling in \cref{thermalhistories}, we name the corresponding part of the benchmark plane region \textbf{E}. With this thermal history, the EWPT is of first order, however, 
the initial value of $v_S$ at the time of the transition is positive. This leads to weaker phase transitions. From \cref{plane2} we also find that the range of $m_H$ for which a SFOEWPT occurs in benchmark plane 2, namely $m_H\in[550,950]$ GeV, is significantly larger than in the case of the benchmark plane 1, where it was  $m_H\in[250,320]$ GeV. Additionally, the strength of the phase transitions increases for lower values of $m_H$ and larger absolute values of $\kappa_{SH}$ up to a maximum value of $\xi_n\gtrsim 4$, where vacuum trapping occurs. This is the same behaviour as was observed for the 2HDM in Refs.~\cite{Biekotter:2022kgf,Bittar:2025lcr}. The explanation for this phenomenon is that the strength of the phase transition grows with the potential barrier between the minima, until the size of the barrier is such that tunnelling cannot happen, leading to the vacuum being trapped in the symmetric minimum.

\begin{figure}[ht!]
    \centering
    \includegraphics[width=0.49\linewidth]{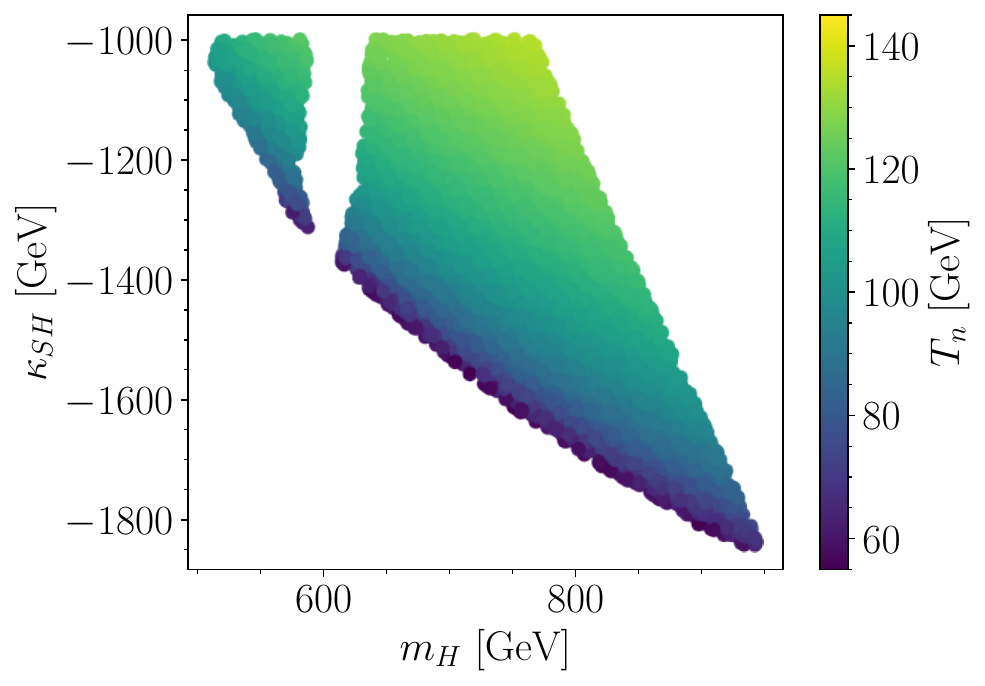}
    \includegraphics[width=0.49\linewidth]{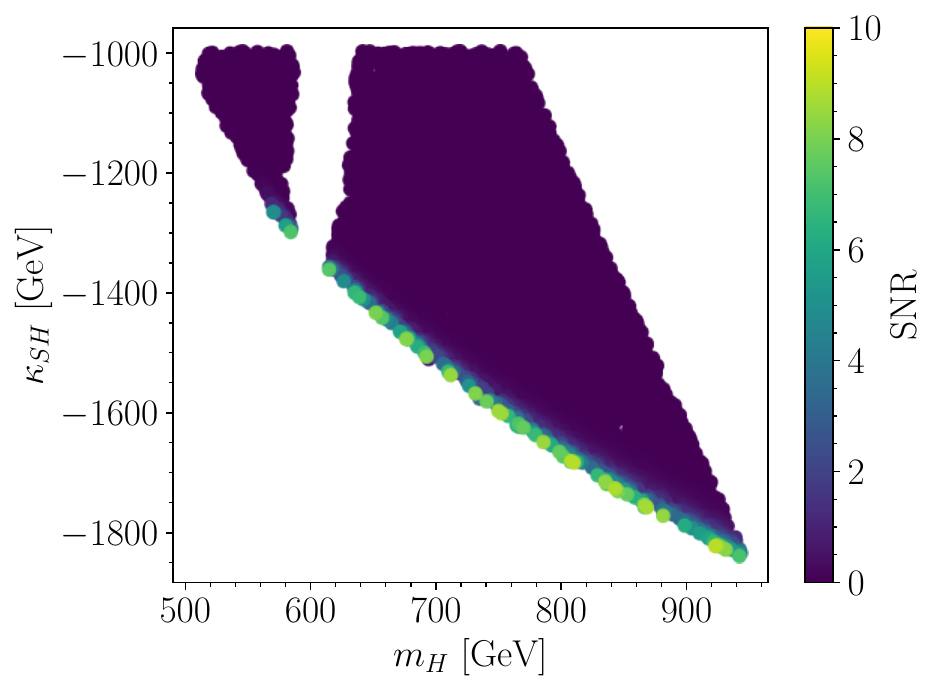}
    \caption{Parameter scan results for the points with a SFOEWPT in benchmark plane 2. \textit{Left}: $T_n$ (in GeV); \textit{right}: SNR at LISA for $\vw=0.95$ and three years of observation time. }
    \label{plane2variables}
\end{figure}

We concentrate in \cref{plane2variables} on the region of the benchmark plane 2 featuring a SFOEWPT. In the left plot the colour coding indicates $T_n$, while in the right plot it represents the SNR at LISA, calculated for a bubble wall velocity of $\vw=0.95$ and three years of observation time.  
At first, we observe from the right plot that for $\vw = 0.95$ no point reaches an SNR larger than 10, even though points at the boundary with the vacuum trapping region are close, with SNR\;$\approx 8$. The observability of these points depends on the uncertainty coming from $\vw$, which will be discussed in \cref{sec:vwdep}, as well as on the threshold value of the SNR that one defines as observable. From the left plot, we find that, like in the case of the benchmark plane 1, the strongest phase transitions occur for low values of $T_n\approx60 \gev$.

\begin{figure}
    \centering
    \includegraphics[width=.8\linewidth]{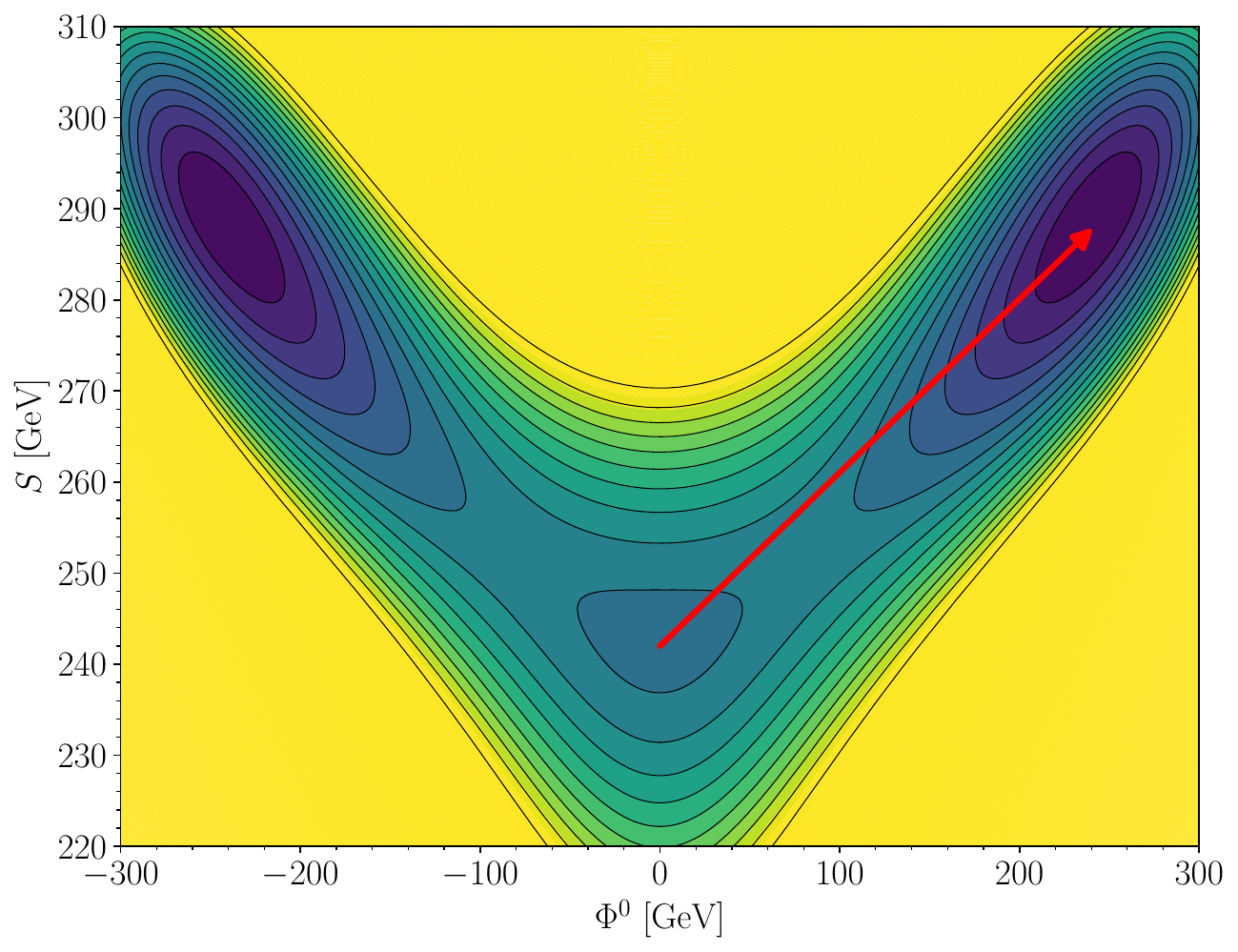}
    \caption{Temperature-dependent one-loop effective potential at the nucleation temperature, $V_\text{eff}(T_n)$, for a benchmark point featuring a SFOEWPT with a thermal history of type \textbf{E} (from \cref{thermalhistories}). The point is defined by $m_H=639 \gev$, $c_{\alpha}=0.9777$, $v_S=289 \gev$, $\kappa_S=-205 \gev$, $\kappa_{SH}=-1403 \gev$ and we find $T_n=64\gev$ and a SNR$\;=4.5$. Here $S$ and $\Phi^0$ denote the singlet field and the CP-even neutral component of the doublet, respectively. }
    \label{pot2}
\end{figure}

\Cref{pot2} provides an illustration of the temperature-dependent effective potential evaluated at the nucleation temperature $V_\text{eff}(T_n)$ for a  point with a SFOEWPT with a SNR\;$=4.5$ and $T_n=64\gev$. For this point, while not being located directly in benchmark plane~2, the SFOEWPT is driven by the doublet field, and has the same type of dynamics as what we observe for points in the benchmark plane. 
In this scenario, the singlet field direction does not play an important role at the time of the transition, and the difference between the initial and final phases of the singlet VEV is relatively small. On the other hand, the EW doublet direction is the most important one as the tunnelling distance can be seen to be much larger. We can therefore distinguish the situation in benchmark plane 2, where the EW doublet direction plays the most important role in the transition, with that of benchmark plane 1, where it is the singlet field direction. This finding also helps understand the behaviour, similar to that of the 2HDM~\cite{Biekotter:2022kgf}, observed for the transition between SFOEWPT and vacuum trapping in \cref{plane2}. Indeed, in the 2HDM, the EWPT is driven by the EW VEV, like in our benchmark plane 2. On the contrary, in benchmark plane 1 the singlet phase plays a more important role and we observe different dynamics of the EWPT compared to the 2HDM.

\subsubsection{\boldmath{$\vw$} dependence}
\label{sec:vwdep}

\begin{figure}[ht]
    \centering
    \includegraphics[width=0.6\linewidth]{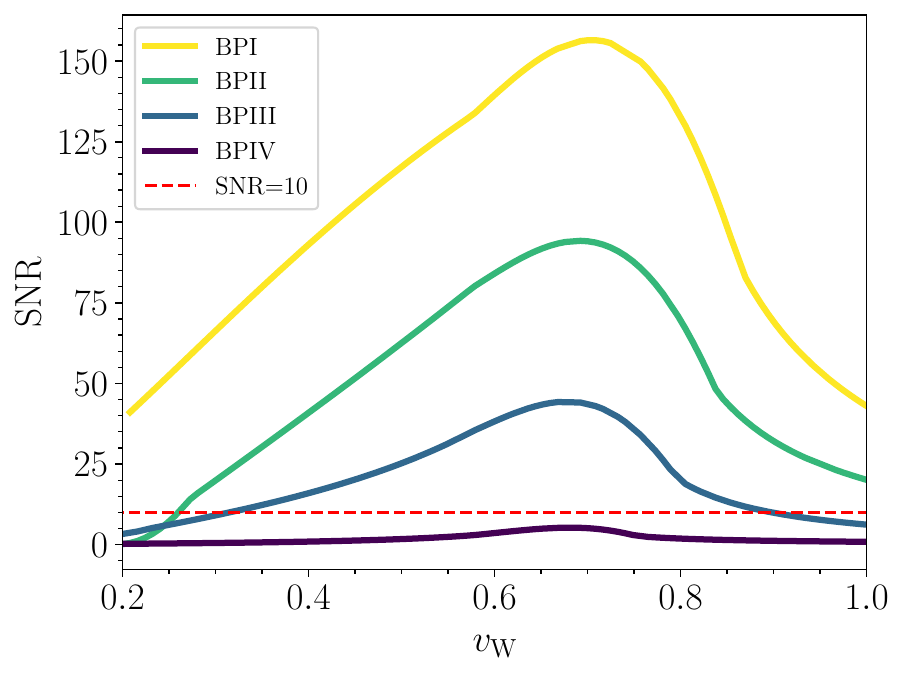}
    \caption{SNR at LISA after three years of data taking as a function of $\vw$ for the benchmark points BPI, BPII, BPIII and BPIV from \cref{vwpoints}. The dashed red line represents the threshold that we consider as observable at LISA, i.e.\ SNR $=10$.}
    \label{vw}
\end{figure}

The objective of this section is to estimate from a qualitative point of view the uncertainty in the computation of the stochastic GW background due to the bubble wall velocity. There have been several investigations of the best way to estimate the velocity of the wall \cite{Ai:2023see,Ekstedt:2024fyq}, however all of them are associated with large uncertainties, which are propagated to the predictions of GW spectra and in turn to the signal-to-noise ratios. 
This is illustrated in \cref{vw}, where we show 
the SNR at LISA (solid lines) after three years of data taking with respect to $\vw$ for the benchmark points BPI, BPII, BPIII and BPIV from \cref{vwpoints}, which are representative examples of different values of the SNR, 
while the red dashed line indicates the observability threshold, SNR\;$= 10$. As a first observation, the behaviour of the curves for the different BPs demonstrates that the assumption of using $\vw\sim1$ is in each case nearly the most pessimistic choice possible, whereas a maximum is reached in all four BPs for $\vw \sim 0.7$. This happens due to a larger efficiency factor $\kappa_{\mathrm{sw}}$, corresponding to a more efficient transfer of the released vacuum energy into bulk fluid motion. We find that any point that is observable at $\vw\sim1$ is also observable for $\vw\gtrsim 0.2$, with this lower value corresponding to phase transitions with a large degree of supercooling \cite{Guth:1979bh}. 
Next, for points that are close to be observable --- i.e.\  with $6\lesssim\mathrm{SNR}\lesssim10$ --- for $\vw=1$, we obtain an observable SNR for $0.3\lesssim \vw\lesssim 0.9$. If for $\vw=1$ $1\lesssim\mathrm{SNR}\lesssim6$, then the SNR could be observable for some range of values of $\vw$.
Finally, when we obtain a value of $\mathrm{SNR}\lesssim1$ for $\vw\sim1$, the signal is not observable for any value of $\vw$.

\begin{table}[ht]
\centering
\begin{tabular}{@{}lccccc@{}}
\toprule
Id  & $m_H$ [GeV] & $\cos\alpha$ & $v_S$ [GeV] & $\kappa_S$ [GeV] & $\kappa_{SH}$ [GeV] \\ \midrule
BPI & 259.3     & 0.988        & 121.5     & $-900$           & $-91$               \\
BPII & 289.2     & 0.987        & 114.4     & $-900$           & $-100$              \\
BPIII & 638.6     & 0.978        & 289.4     & $-204$           & $-1403$             \\
BPIV & 251.4     & 0.981        & 94.2      & $-900$           & $-127$              \\ \bottomrule
\end{tabular}
\caption{RxSM benchmark points for the investigation of the SNR dependence on $\vw$.}
\label{vwpoints}
\end{table}

\begin{figure}[ht]
    \centering
    \includegraphics[width=0.49\linewidth]{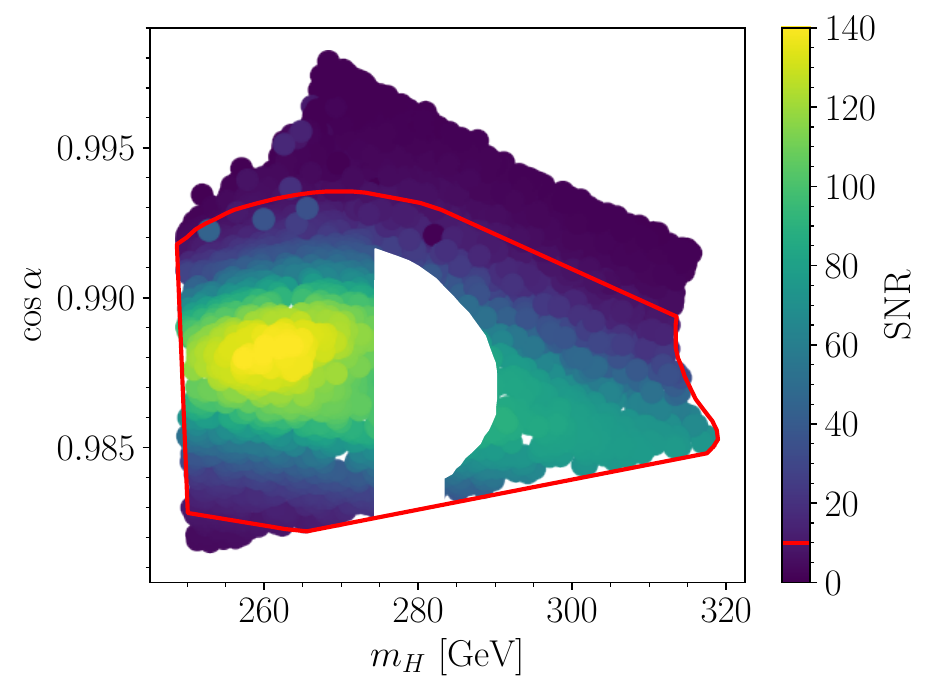}
    \includegraphics[width=0.49\linewidth]{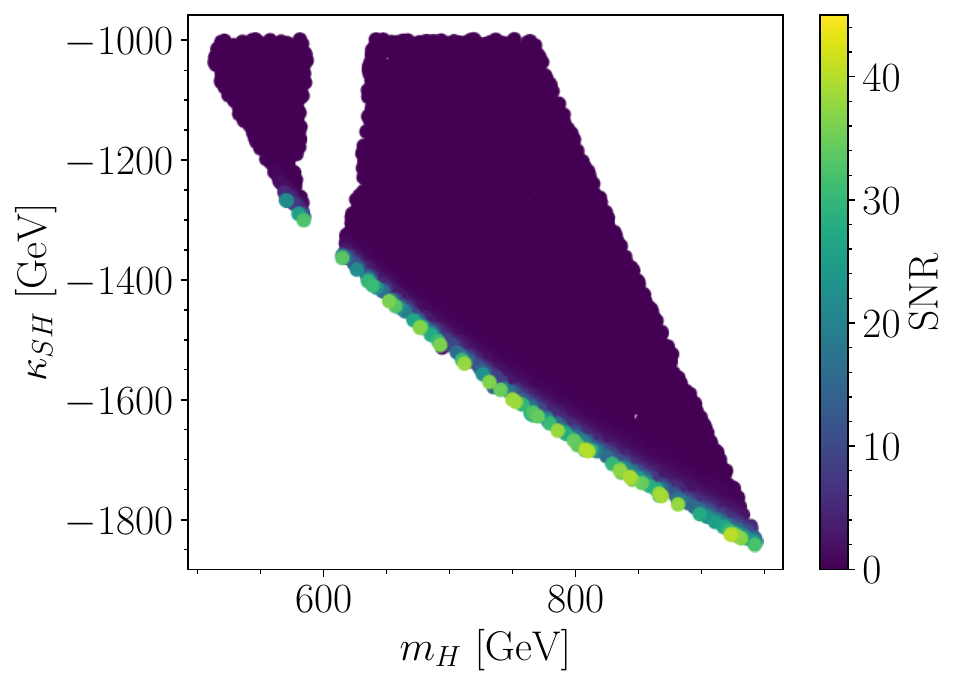}
    \caption{SNR at LISA for a bubble wall velocity of $\vw=0.6$. \textit{Left}: 
    benchmark plane~1, where the region delimited by the red line indicates the region with an observable SNR$\;>10$; 
    \textit{right}: benchmark plane~2.}
    \label{vwplanes}
\end{figure}

Lastly, in \cref{vwplanes} the colour coding indicates the SNR at LISA after 3 years of data taking and using now a bubble wall velocity of $\vw=0.6$, in benchmark planes~1 (left) and~2 (right). In benchmark plane 1, while the parameter region producing an observable signal has only grown moderately compared to the case of $\vw=0.95$ (c.f.\ \cref{plane1variables}), the maximal achievable SNR has increased significantly (by a factor of about 3). In benchmark plane~2, the region with the highest values of the SNR (which were however not observable for $\vw=0.95$) becomes observable for $\vw=0.6$, as expected from \cref{vw}. On the other hand, the region with a sizeable SNR remains confined to a small strip at the border to the parameter space where vacuum trapping occurs. 

\section{Collider analysis}
\label{sec:collider}

In this section, we investigate the possibility of probing scenarios with a SFOEWPT at future colliders such as the HL-LHC and future $e^+e^-$ colliders, as well as the complementarity with a GW analysis with data from 
the future GW observatory LISA. 
For consistency with the analysis of the EWPT dynamics and GW production in the previous sections, we also include one-loop corrections to the trilinear scalar couplings involved in di-Higgs production in the RxSM, following Ref.~\cite{Braathen:2025qxf}. To compute the loop corrections, we perform complete one-loop diagrammatic calculations of both $\lambda_{hhh}$ and $\lambda_{hhH}$, employing the public code \texttt{anyH3}~\cite{Bahl:2023eau,anyHH} using the full on-shell (OS) scheme defined in Ref.~\cite{Braathen:2025qxf}.

Here we have to comment on the different choices of renormalisation schemes for the calculation of the EWPT and the collider analyses. The scheme implemented in \texttt{BSMPT}~\cite{Basler:2018cwe,Basler:2020nrq,Basler:2024aaf}, often referred to as  ``OS-like'', serves to maintain at the one-loop level the same location of the EW minimum and curvature of the potential around it. It however retains a significant renormalisation scale dependence, unlike the full OS scheme of \citere{Braathen:2025qxf} employed in our collider analysis (this can be understood from the number of OS conditions that are actually imposed\footnote{To make the counting explicit: on the one hand, the scheme of \texttt{BSMPT} defines five ``OS-like'' conditions, two for one-point functions and three for two-point functions as can be seen from \cref{eq:BSMPTscheme_1pt,eq:BSMPTscheme_2pt}. As the RxSM scalar sector contains eight parameters to renormalise --- namely $t_\phi$, $t_S$, $m_h$, $m_H$, $\alpha$, $v$, $\kappa_S$, $\kappa_{SH}$ (while $v_S$ does not require a UV-divergent counterterm as it does not run~\cite{Sperling:2013eva,Braathen:2025qxf}) --- the scheme from \texttt{BSMPT} does not have enough conditions to renormalise all parameters OS. On the other hand, the full OS scheme of \citere{Braathen:2025qxf} provides eight OS conditions: two (three) from scalar one-point (two-point) functions, one for the standard OS renormalisation of the EW VEV, and two additional conditions are defined in terms of scalar three-point functions. All renormalisation scale dependences can therefore be compensated with this scheme.}). For the four representative benchmark points of \cref{bps} (discussed below), we have extracted the values of the trilinear scalar couplings computed with \texttt{BSMPT} and estimated the corresponding theoretical uncertainty by varying the renormalisation from $m_H/2$ to $2m_H$. We have found that, for the same points, the trilinear scalar couplings calculated using the full OS scheme of \citere{Braathen:2025qxf} always lie inside the $1\sigma$ uncertainty band from the \texttt{BSMPT} results. In other words, the difference between the two schemes is within the renormalisation scale dependence of the \texttt{BSMPT} predictions. This result justifies skipping the cumbersome step of performing a parameter conversion between the two schemes, when going from the EWPT study to the collider analysis. It should be noted that the dependence on the renormalisation scale $Q$ does of course affect the results for the dynamics of the EWPT. However, it was shown in Ref.~\cite{Lewicki:2024xan} that even after improving the calculation of the effective potential to reduce the $Q$ dependence, the phenomenological results in terms of GW productions remain, although potentially shifted in the parameter space of the model. Keeping this observation in mind, we emphasise that our work is meant to be a proof of concept for the opportunity of probing SFOEWPT scenarios using the complementarity of GW signals and collider searches.

\subsection{Loop corrections to the trilinear scalar couplings}
\begin{figure}[ht!]
    \centering
    \includegraphics[width=0.49\linewidth]{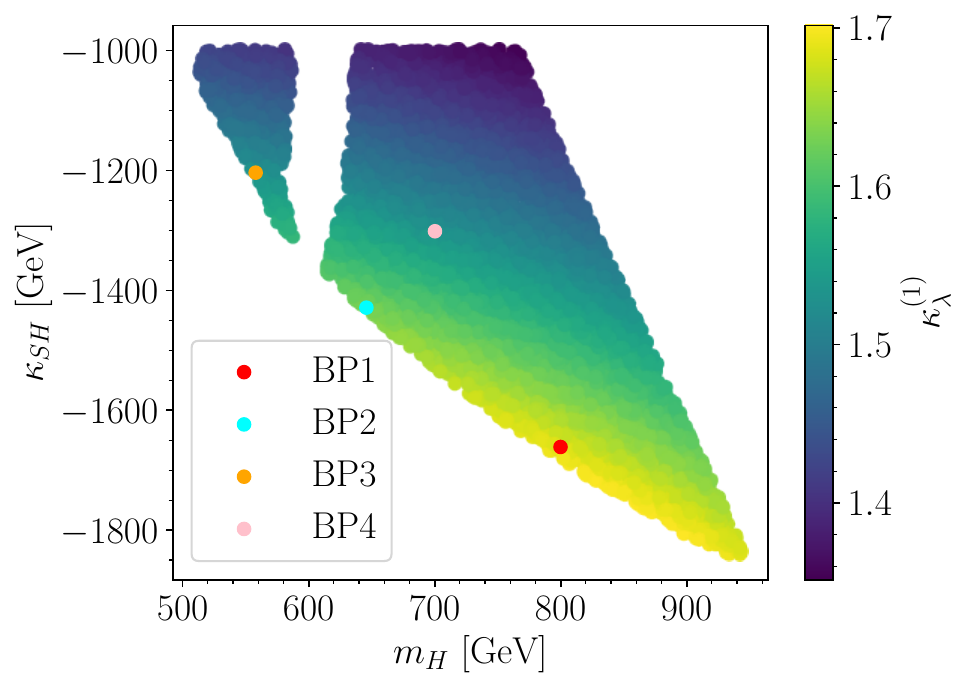}
    \includegraphics[width=0.49\linewidth]{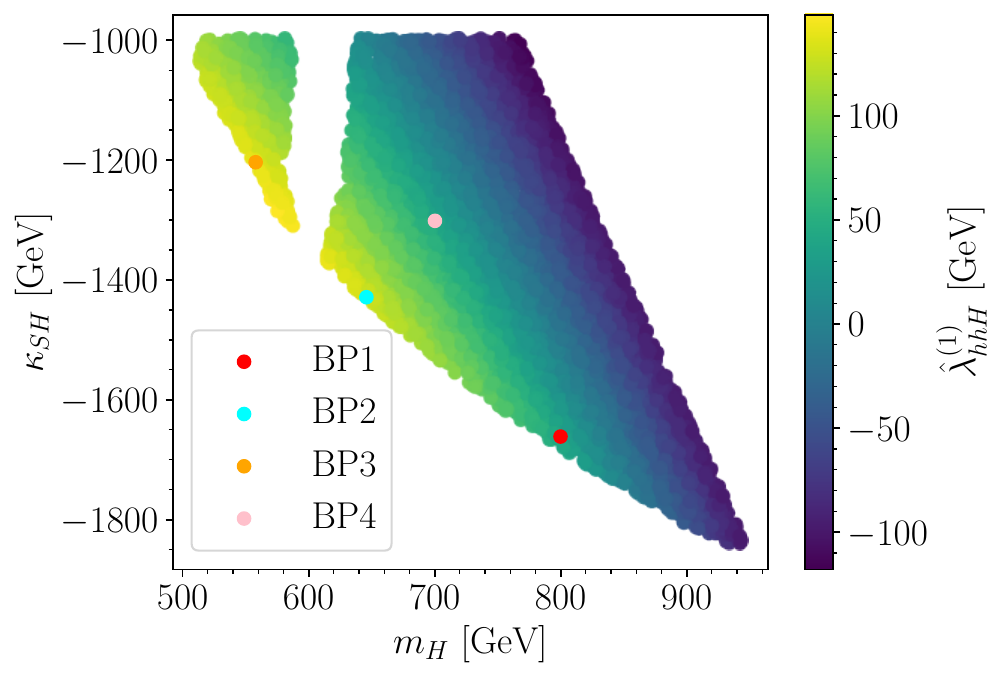}
    \caption{One-loop corrected trilinear scalar couplings in the benchmark plane~2. \textit{Left}: $\kappa_\lambda^{(1)}$; \textit{right}: $\hat\lambda_{hhH}^{(1)}$ (in GeV).}
    \label{tri1}
\end{figure}

\begin{figure}[ht]
    \centering
    \includegraphics[width=0.49\linewidth]{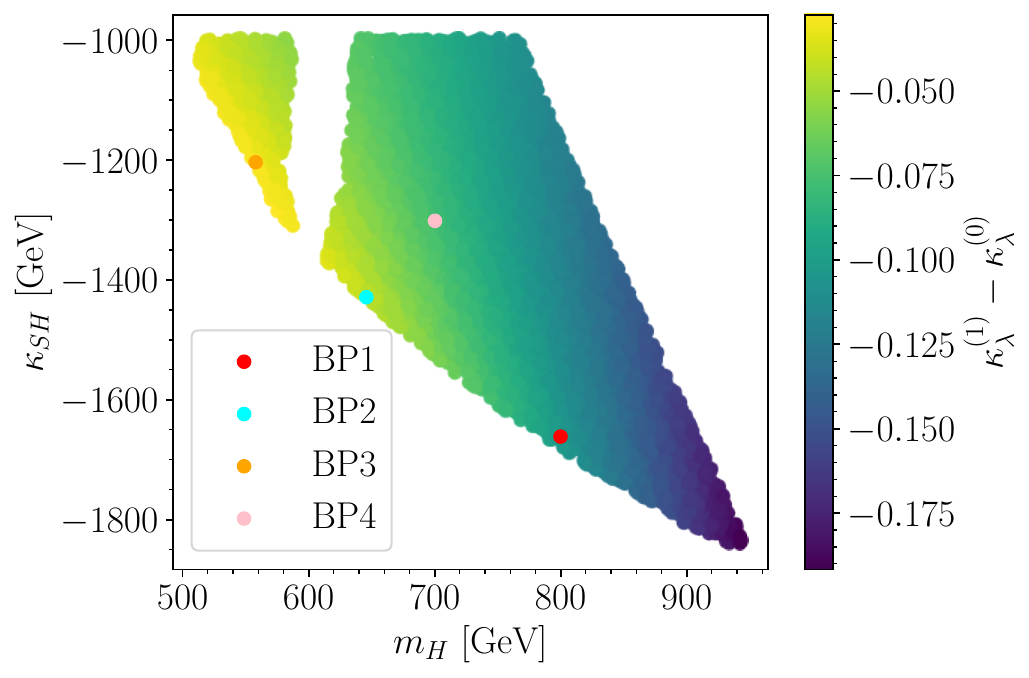}
    \includegraphics[width=0.49\linewidth]{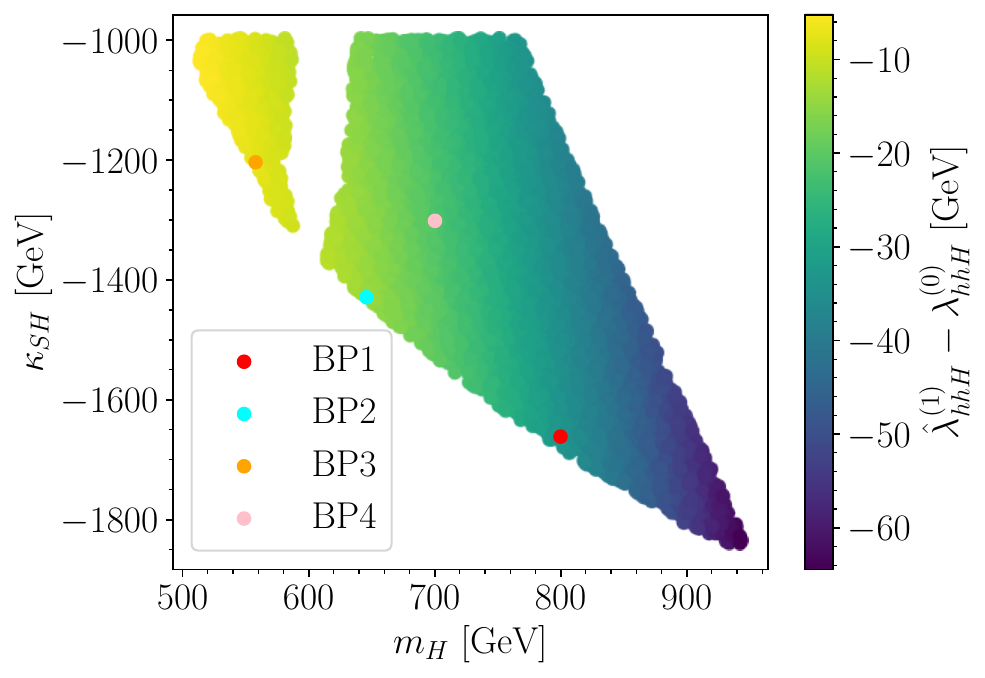}
    \caption{One-loop corrections to the trilinear scalar couplings in the benchmark plane~2. \textit{Left}: $\kala^{(1)}-\kala^{(0)}$; \textit{right}: $\rlahhH{1}-\lahhH^{(0)}$ (in GeV).}
    \label{tri2}
\end{figure}

We begin by investigating the size of the loop corrections to the trilinear scalar couplings $\lahhh$ and $\lahhH$, in the two benchmark planes defined in \cref{sec:pheno}. 
These corrections have automatically been taken into account in our analysis in the previous section by using the effective
 potential as defined in \cref{VT}.
For benchmark plane~1, we find that the trilinear scalar couplings $\lahhh$ and $\lahhH$ are very close to their SM values, i.e.\ $\kala=1$ and $\lahhH=0$, 
both at tree level and when including one-loop corrections. This can be understood from the fact that, in this plane, the SFOEWPT is driven by the singlet field direction, which is however largely secluded from the direction corresponding to the detected Higgs boson (due to the significant constraints on the amount of mixing allowed by experimental results on properties of the detected Higgs boson). Therefore, the di-Higgs production cross sections in this benchmark plane are essentially undistinguishable from the SM prediction, and we conclude that benchmark plane~1 is not suited for investigations via di-Higgs production at colliders.
On the other hand, for benchmark plane~2, we observe relevant deviations from the SM in the trilinear scalar couplings and the di-Higgs cross section, as illustrated in \cref{tri1,tri2}.

In \cref{tri1}, we show as colour coding the one-loop predictions for $\kappa_{\lambda}$ (left panel) and $\lambda_{hhH}$ (right panel) in benchmark plane~2. We also indicate with coloured points four benchmark points for which we perform in the next subsections the analysis of the differential cross-section distributions; the set of parameters defining these points are given in \cref{bps}. 
In the left panel of \cref{tri1}, we observe that, across the entire benchmark plane, $\kala^{(1)}$ deviates by $40-70\%$ from its SM value. This is in turn sufficient to induce noticeable changes in di-Higgs production. 
From the right panel of \cref{tri1}, we find that $\rlahhH{1}$ takes both positive and negative values, within the range $[-100, 150]~\mathrm{GeV}$. This has an impact on the structure of the interference between resonant and non-resonant contributions. Moreover, for the points with $\rlahhH{1}\simeq 0$, no resonance will be observable.

In order to understand whether the BSM deviations originate from tree-level effects or one-loop corrections, we present in the colour coding of \cref{tri2} the differences $\kala^{(1)}-\kala^{(0)}$ (left panel) and $\rlahhH{1}-\lahhH^{(0)}$ (right panel) in benchmark plane~2.
From the left panel of \cref{tri2}, we observe that the one-loop corrections to $\kappa_\lambda$ range between $-2\%$ and $-10\%$ of the full one-loop values, and that the corrections are negative across the entire benchmark plane. We also find that the largest correction corresponds to the largest value of $\kala$.
In the case of $\lambda_{hhH}$, the corrections are likewise negative throughout the whole benchmark plane, as can be seen in the right panel of \cref{tri2}. However, unlike for $\kala$, the corrections to $\rlahhH{1}$ can become significant in parts of the plane, with values ranging from $-6\%$ to $-60\%$. Finally, we note that for $\lahhH$, the correlation between large radiative corrections and large (negative) values of the coupling is even more important than for $\kala$. 

\begin{table}[ht]
\centering
\begin{tabular}{ccccccccccc}
\toprule
BP  & $m_H$ & $\cos\alpha$ & $v_S$ & $\kappa_S$ & $\kappa_{SH}$ & $\kappa_{\lambda}^{(0)}$ & $\kappa_{\lambda}^{(1)}$ & $\lambda_{hhH}^{(0)}$ & $\rlahhH{1}$& $\xi_n$ \\
&  [GeV] &  & [GeV] & [GeV] & [GeV] & & & [GeV] & [GeV] \\ \midrule
1 & 800     & 0.98         & 280      & $-300$           & $-1661$              & 1.8                      & 1.7                      & 72.1                      & $38.6$                   & 3.5\\
2 & 646     & 0.98          & 280      & $-300$           & $-1429$              & 1.7                      & 1.6                      & 143.9                       & $129.8$                  & 3.9 \\
3 & 558     & 0.98         &280    & $-300$           & $-1204$              & 1.6                      & 1.5                      & 145.0                     & 137.4                   & 2.7 \\
4 & 700     & 0.98         & 280      & $-300$           & $-1301$              & 1.6                      & 1.5                      & $61.5$                     & $40.0$                   & 1.9 \\ \bottomrule
\end{tabular}
\caption{Definitions of the RxSM benchmark points considered for the study of di-Higgs production, in terms of the five free BSM parameters of the model: $m_H,\ \cos\alpha,\ v_S,\ \kappa_S$ and $\kappa_{SH}$. Additionally, tree-level and one-loop predictions for $\kappa_\lambda$ and $\lahhH$ are included, as well as $\xi_n$.}
\label{bps}
\end{table}


\begin{figure}[ht]
    \centering
    \includegraphics[width=0.49\linewidth]{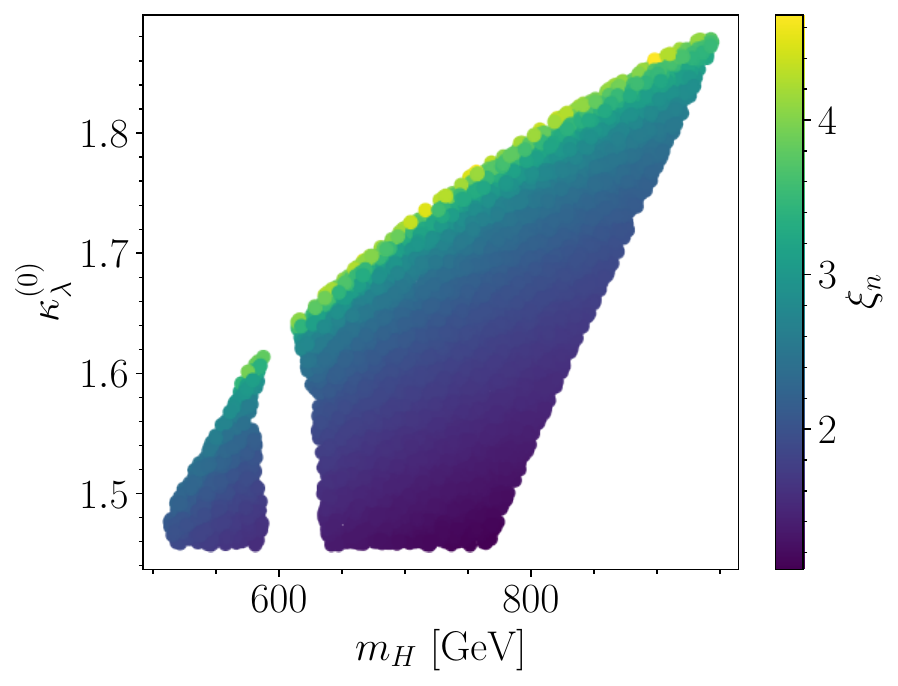}
    \includegraphics[width=0.49\linewidth]{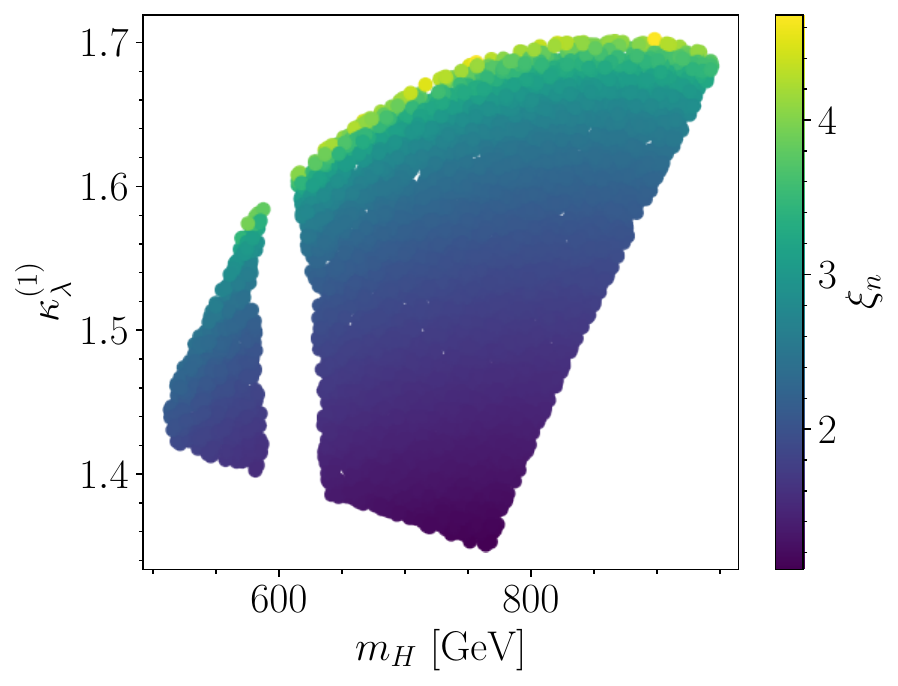}
    \caption{Results for $\xi_n$ in benchmark plane~2. \textit{Left}: projected in the plane $\{m_H,\kala^{(0)}\}$; \textit{right}: projected in the plane $\{m_H,\kala^{(1)}\}$. }
    \label{kmx}
\end{figure}

To investigate the correlation between a BSM deviation in $\kala$ and the strength of the phase transition, we show $\xi_n$ in the colour coding of \cref{kmx} for the scan points of benchmark plane~2, projected in the planes $\{m_H,\kala^{(0)}\}$ (left plot) and $\{m_H,\kala^{(1)}\}$ (right plot). For a fixed value of $m_H$, the strongest EWPT correspond to the largest allowed values of $\kala$, both at tree level and at one loop. Even larger values of $\kala$ would be associated with vacuum trapping. Lastly, we observe that for benchmark plane~2, a SFOEWPT is correlated with a BSM deviation of $35\%-70\%$ at one loop ($45\%-90\%$ at tree level), as is known to be the case for scenarios where the SFOEWPT is driven by the Higgs doublet (see e.g.\ Ref~\cite{Biekotter:2022kgf}). 

\subsection{Di-Higgs production at the HL-LHC}
In this section, we present the results obtained for di-Higgs production cross-sections and differential distributions at the HL-LHC using the same framework as in Ref.~\cite{Braathen:2025qxf}. For our theoretical predictions, we employ a modified version of \texttt{HPAIR}~\cite{Dawson:1998py,Nhung:2013lpa,Grober:2015cwa,Grober:2017gut,Abouabid:2021yvw,Arco:2022lai}, which was already used in Ref.~\cite{Arco:2025nii}. In this code, the three leading-order (LO) diagrams that contribute to the $gg \to hh$ process, as well as the interference between them, are taken into account. We also note that NLO QCD corrections are available for the total cross-section predictions in \texttt{HPAIR}, but not for the differential results. To simplify comparisons between results, we choose to work here only with results at LO in QCD (keeping in mind that total cross-sections are modified by a QCD $K$ factor of $\approx 2$). On the other hand, \texttt{HPAIR} takes numerical values for $\lahhh$ and $\lahhH$ as inputs. Using tree-level or one-loop values for these trilinear scalar couplings therefore allows us to obtain predictions at LO or leading NLO in terms of BSM effects.

\begin{figure}[ht!]
    \centering
    \includegraphics[width=0.49\linewidth]{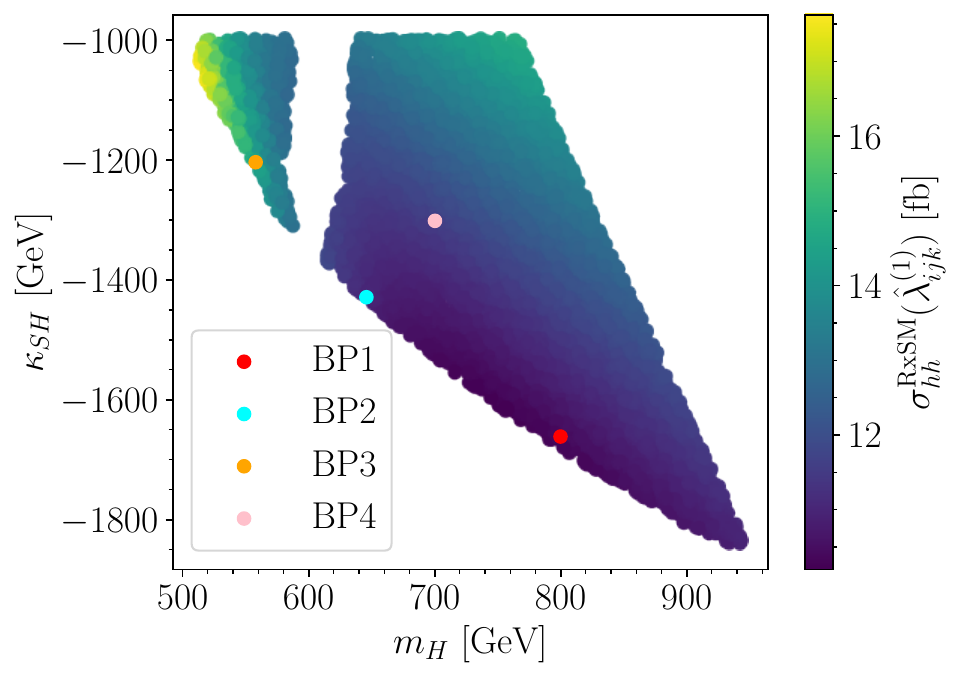}
    \includegraphics[width=0.49\linewidth]{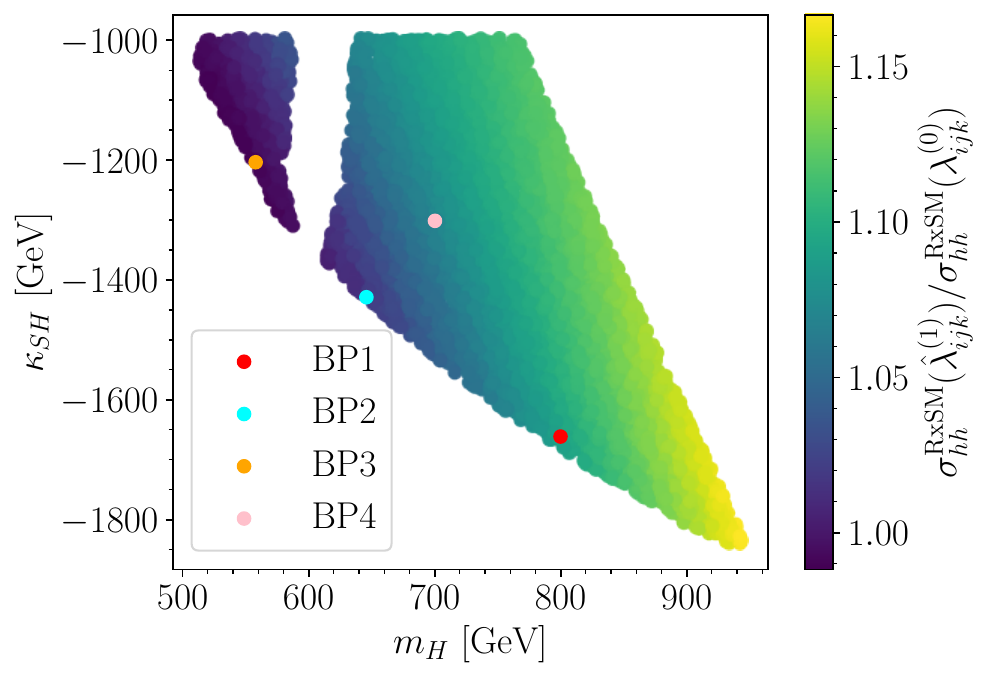}
    \caption{Results for the di-Higgs cross-section at the HL-LHC in the benchmark plane~2. \textit{Left}: $\sihh(\rlaijk{1})$ (in fb); \textit{right}: the ratio between the total di-Higgs production cross section using one-loop and tree-level trilinear Higgs couplings $\sihh(\rlaijk{1})/\sihh(\laijk^{(0)})$. The SM prediction is $\sigma^{\mathrm{SM}}_{hh} = 19.76$~fb, above the RxSM results in the entire benchmark plane. }
    \label{tcrosslhc}
\end{figure}

We present in \cref{tcrosslhc} our predictions for the total di-Higgs production cross-section in benchmark plane~2. 
In the left panel, the colour coding indicates the cross-section including one-loop corrections to the trilinear Higgs couplings, which we denote $\sihh(\rlaijk{1})$, and in the right panel it shows the ratio between the total cross-section including one-loop corrections and the same cross-section using the tree-level values --- i.e.\ $\sihh(\rlaijk{1})/\sihh(\laijk^{(0)})$. 
As can be seen from the left panel of \cref{tcrosslhc}, over the entire benchmark plane, the predicted total cross-section is lower than its SM value, $\sigma^{\mathrm{SM}}_{hh} = 19.76$~fb~\cite{Abouabid:2021yvw}. 
The behaviour of $\sigma_{hh}$ as a function of $\kala$ is well known from studies of the SM with a free $\lambda_{hhh}$ (see e.g.\ 
\citeres{Baglio:2012np,LHCHiggsCrossSectionWorkingGroup:2016ypw}), and can be explained by the destructive 
interference between the $s$-channel Higgs-exchange contribution and the box diagram.
The cross-section prediction in the RxSM is affected by the enhancement of $\lahhh$ w.r.t.\ its SM value, as well as by the contribution of the heavy Higgs-boson resonance. Overall, the values found for $\kala$ in benchmark plane~2 lead to an increase of the destructive interference of the SM-type contributions and thus to a decrease in the cross-section prediction. On the other hand, the heavy Higgs-boson resonance leads to an increase in the cross-section, which contributes for momenta in the $s$-channel around its mass. For all of benchmark plane 2, the first effect dominates, resulting in lower values of the total cross-section compared to the SM. With decreases in $\sihh(\rlaijk{1})$ of up to 50\%, a measurement of the di-Higgs cross-section could allow distinguishing the RxSM from the SM in parts of benchmark plane 2 (for larger values of $|\kappa_{SH}|$), however, the measurement itself would then be more challenging than in the SM. For the rest of the plane, the RxSM results are close to the SM, meaning that once experimental uncertainties are taken into account it would be difficult to distinguish the two models using the total cross-section alone. 

The right panel of \cref{tcrosslhc} shows that the largest contribution from the loop corrections to $\rlaijk{1}$ reaches up to $15\%$ of the value obtained with tree-level trilinear scalar couplings, and occurs in the region with the lowest total cross-section. On the other hand, we can see that in the region where the total cross-section is largest in benchmark plane~2, the loop corrections do not appear to have a significant impact on $\sihh$.

As discussed in previous works~\cite{Arco:2022lai,Heinemeyer:2024hxa,Arco:2025nii,Frank:2025zmj,Braathen:2025qxf}, the differential cross-section with respect to the di-Higgs invariant mass, \mhh, provides much more information for distinguishing BSM models like the RxSM from the SM than the total cross-section alone. For this reason, we have selected four benchmark points with different phenomenological features,
which are marked as coloured dots in \cref{tri1,tri2}. The benchmark points are labelled
BP1, BP2, BP3, and BP4, with the parameters summarised in \cref{bps}. 

\begin{figure}[ht!]
    \centering
    \includegraphics[width=1\linewidth]{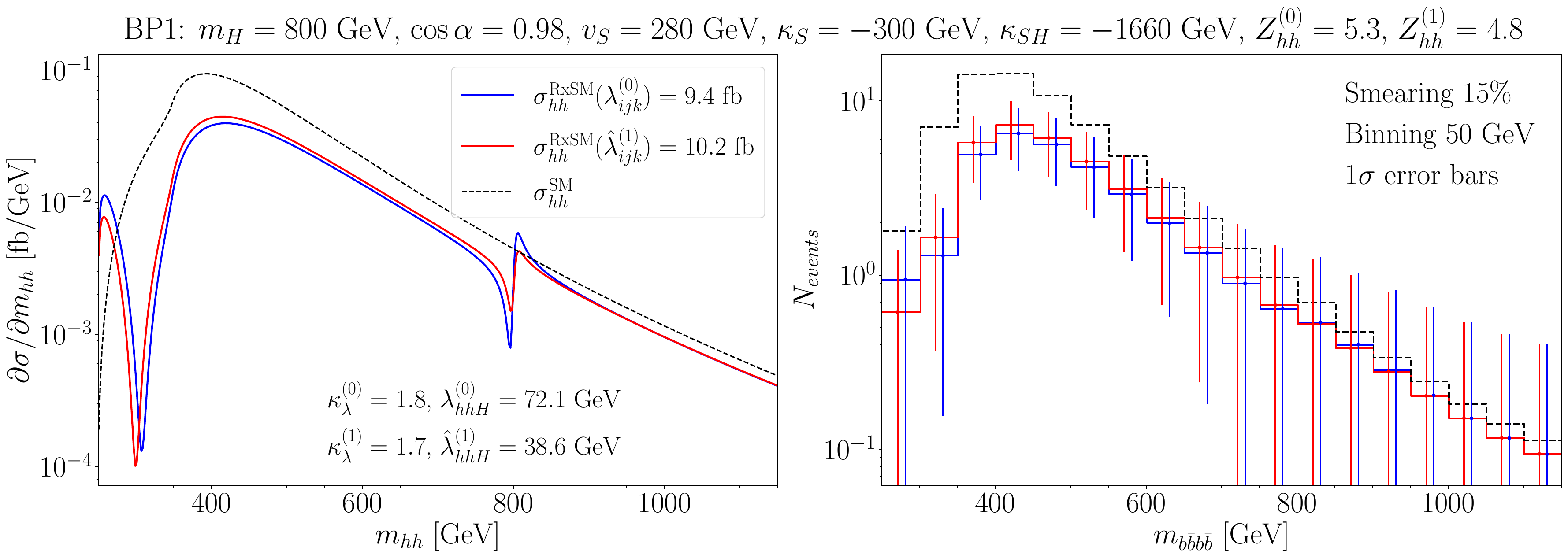}
    \includegraphics[width=1\linewidth]{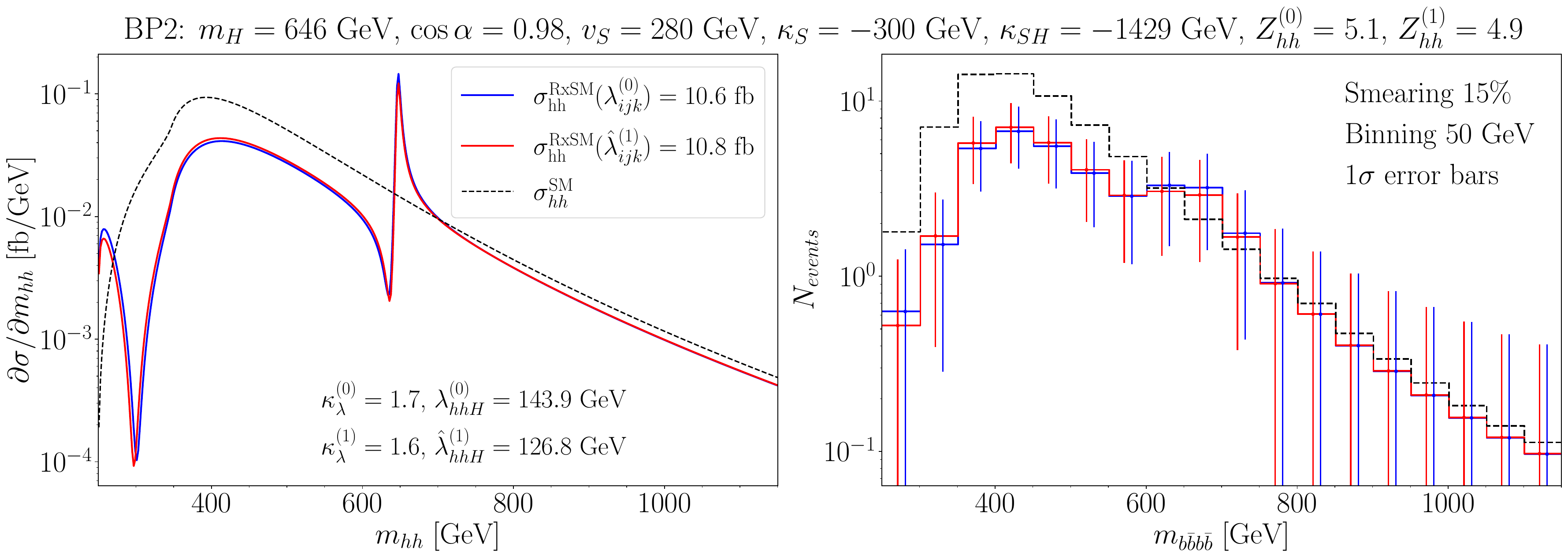}
    \caption{\textit{Left:} Differential di-Higgs production cross-section distributions w.r.t.\ \mhh. \textit{Right:} Distribution of the number of di-Higgs events, taking into account the $\br(h \to b\bar b)$, as well as smearing and binning effects, with respect to the invariant mass of the four reconstructed $b$ quarks, \mbbbb. 
    The error bars indicate statistical errors, assuming Poisson distributions for the number of events in each bins. 
    Blue curves show results using $\laijk^{(0)}$ in the computation of di-Higgs production, while red curves represent results using $\rlaijk{1}$. The black dashed line indicates the SM result. \textit{Top:} results for BP1; \textit{bottom:} results for BP2 from \cref{bps}.}
    \label{bp12}
\end{figure}

\begin{figure}[ht!]
    \centering
    \includegraphics[width=1\linewidth]{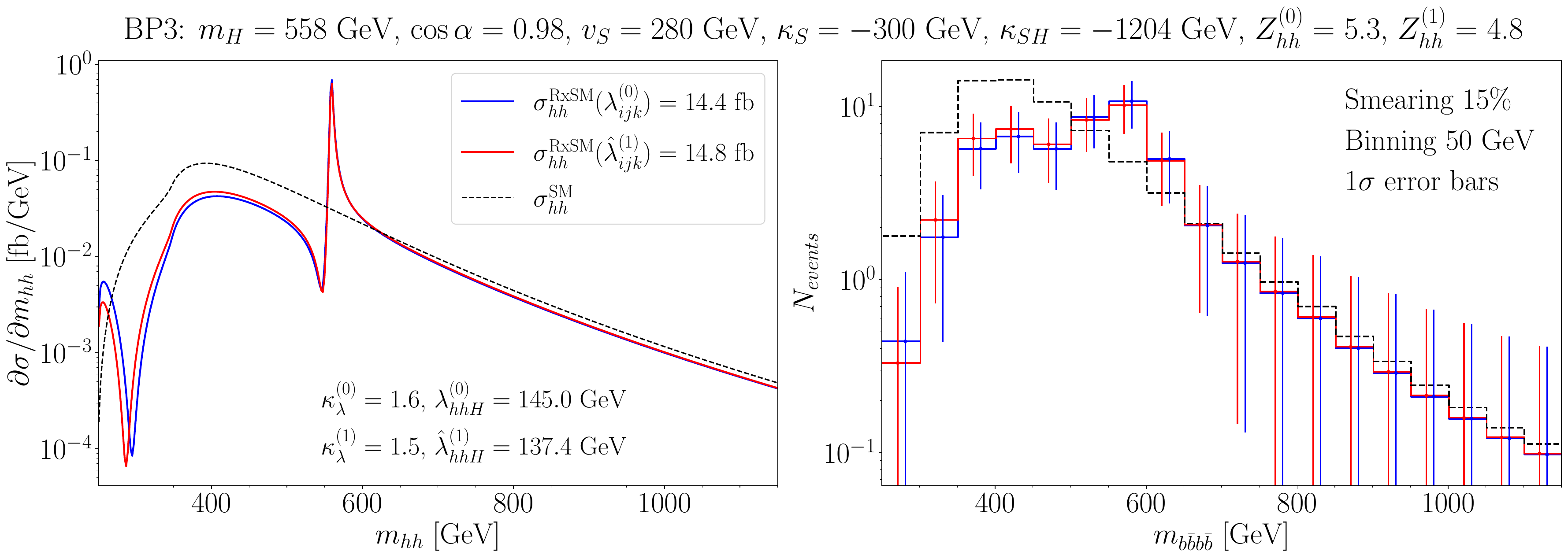}
    \includegraphics[width=1\linewidth]{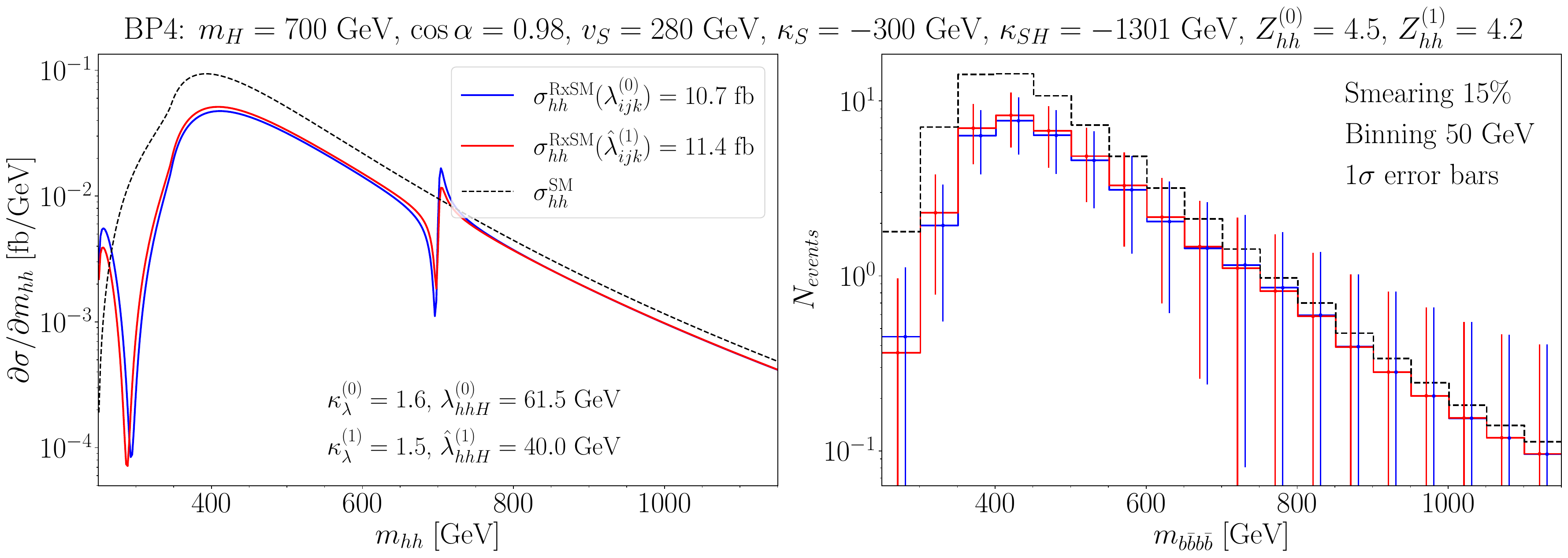}
    \caption{Plots and line styles as in \protect\cref{bp12}.
    \textit{Top:} results for BP3; \textit{bottom:} results for BP4 from \cref{bps}.}
    \label{bp34}
\end{figure}

For these four benchmark points we
compute the corresponding differential cross-section distributions, following the framework of \citere{Braathen:2025qxf}. 
The results for BP1 and BP2 are presented in \cref{bp12}, and for BP3 and BP4 in \cref{bp34}. The left plots display the theoretical differential cross-section distributions with respect to $m_{hh}$, while the right plots show the number of events, including experimental uncertainties. Here we assume an integrated luminosity of $\cLint = 6000 \ifb$ corresponding to the end of the HL-LHC. We also include the main decay channel, $h \to b\bar b$, into the calculation, with $\br(h \to b\bar{b}) = 0.58$, and the results are displayed as a function of the four-$b$ invariant mass, \mbbbb. For these plots we also include a smearing of $15\%$ to take into account the experimental errors, as well as a binning of $50 \gev$ to account for the detector resolution, see \citere{Braathen:2025qxf} for details. The error bars represent the $1\sigma$ statistical uncertainties on the signal (computed as Poisson-distribution errors). The dashed black line corresponds to the SM prediction, the blue line shows the RxSM result using $\laijk^{(0)}$, and the red line is the RxSM prediction including $\rlaijk{1}$.

In \cref{bp12}, we observe that for BP1 the theoretical RxSM distributions using $\laijk^{(0)}$ (blue curve) and using $\rlaijk{1}$ (red curve) have similar shapes.  
When taking into account the experimental uncertainties and error bands, we find in the right plot that the two curves are indistinguishable from each other, but can be distinguished from the SM distribution. Using the definition from Eqs.\ (47) and (48) of \citere{Braathen:2025qxf}, we obtain a significance to discriminate the RxSM distribution using $\rlaijk{1}$ from the SM curve of $\zohh = 4.8$. From this we conclude that using the differential distributions, the RxSM can to a good degree of confidence be distinguished from the SM in BP1. This significance arises from the modification of the interference at low values of $m_{hh} \lesssim 400~\mathrm{GeV}$, due to the BSM deviation in $\kala$. 
It is interesting to note that the resonant peak, visible in the theoretical curve for $m_{hh}\simeq 800\gev$, is completely washed away once experimental effects are taken into consideration.

In the case of BP2 (lower row of \cref{bp12}), the clear 
resonant peak at $m_{hh} \approx 650 \gev$ in the theoretical curves (left plot), 
is somewhat washed out after the inclusion of experimental uncertainties (right plot) but may remain distinguishable 
from the continuum, i.e.\ the SM distribution.  
The deviation of $\kappa_\lambda$ from the SM is smaller in BP2 than in BP1, which reduces the change in the 
\mhh\ distribution 
near the di-Higgs threshold and tends to lower the significance. On the other hand, the resonant peak also contributes more noticeably to the significance, resulting in a value of $\zohh = 4.9$. 
While not sufficient for a discovery, this could already be interpreted as an indication of new physics.

The result for BP3, shown in the upper row of \cref{bp34}, is quite similar to that of BP2. In this case, the heavy Higgs mass is smaller, $m_H = 558 \gev$, and therefore the resonant peak is shifted to lower $m_{hh}$. Due to the slope of the continuum contribution, we observe in the right plot that, after taking into account experimental uncertainties, the peak is enhanced compared to BP2. However, this is compensated the reduction of the significance due to the smaller value of $\kala$, yielding $\zohh = 4.8$. Additionally, the total cross-section is closer to the SM result for BP3 than BP2, which implies that by only considering the total cross-section this scenario might be more complicated to distinguish from the SM. However, the differential cross-section allows us to differentiate the two models.

Finally, the result for BP4 (lower row of \cref{bp34}) is similar to that of BP1. In this case, the coupling $\lambda_{hhH}$ is negative, which causes the interference around the resonant peak to have a dip–peak structure rather than a peak–dip structure. This suppresses the peak, making it indistinguishable from the continuum once experimental uncertainties are included. Moreover, $\kappa_\lambda$ is closer to its SM value, which reduces the modification of interference at low values of $m_{hh} \sim 400~\mathrm{GeV}$. In total, these two effects lead to a smaller significance, $\zohh = 4.2$. This result is the lowest among the four considered benchmark points, however, it would still be sufficient to hint at the presence of new physics. A summary of the HL-LHC di-Higgs production cross-sections and the significances can be found in \cref{bpscross_lhc_ee}. 

In summary, while it may be difficult to probe scenarios in the RxSM with a SFOEWPT using the total cross-section $\sihh$, having access to the differential \mhh~distributions would allow to investigate some of these scenarios and to distinguish them from the SM.


\begin{table}[htb!]
\centering
\begin{tabular}{@{}ccccccc@{}}
\toprule
BP  &$\sigma^{\mathrm{RxSM}}_{hh}(\lambda_{ijk}^{(0)})$ [fb] & $\sigma^{\mathrm{RxSM}}_{hh}(\rlaijk{1})$ [fb] &$\zthh$ &$\zohh$ \\
 \midrule
1 & 9.4                      & 10.2                      & 5.3                      & 4.8 \\
2 & 10.6                      & 10.8                      & 5.1                      & 4.9  \\
3 & 14.4                      & 14.8                      & 5.3                      & 4.8 \\
4 & 10.7                      & 11.4                     & 4.5                     & 4.2  \\
\bottomrule
\end{tabular}
\caption{Summary of the predictions for di-Higgs production cross-sections and significances at the HL-LHC for the benchmark points in \protect\cref{bps}.}
\label{bpscross_lhc_ee}
\end{table}


\subsection{Di-Higgs production at $e^+e^-$ colliders}

In this section, we complement our HL-LHC results with an analysis of di-Higgs production at future high-energy $e^+e^-$ colliders. We consider the double Higgs-strahlung channel, $e^+e^- \to Zhh$, which is the dominant production mode of two SM-like Higgs bosons up to centre-of-mass energies slightly above 1~TeV. In addition, we also consider the $WW$-fusion channel, $e^+e^- \to \nu \bar{\nu} hh$, which includes the diagrams of the channel $Zhh$ considering that $Z$ decaying to $\nu \bar{\nu}$, and also Vector Boson Fusion (VBF)-like diagrams, with $W$~bosons mediating the interaction. 
To compute the cross-section, which we denote $\siZhh$ and $\sivvhh$, we use the public code \texttt{Madgraph5\_aMC v3.5.9}~\cite{Alwall:2014hca} (which we will from now on refer to as \texttt{MadGraph}). The \texttt{UFO} model file for the RxSM required as input by \texttt{Madgraph} was generated using the \texttt{Mathematica} package \texttt{SARAH-4.15}~\cite{Staub:2008uz,Staub:2009bi,Staub:2010jh,Staub:2012pb,Staub:2013tta}. 

\begin{figure}[ht]
    \centering
    \includegraphics[width=0.49\linewidth]{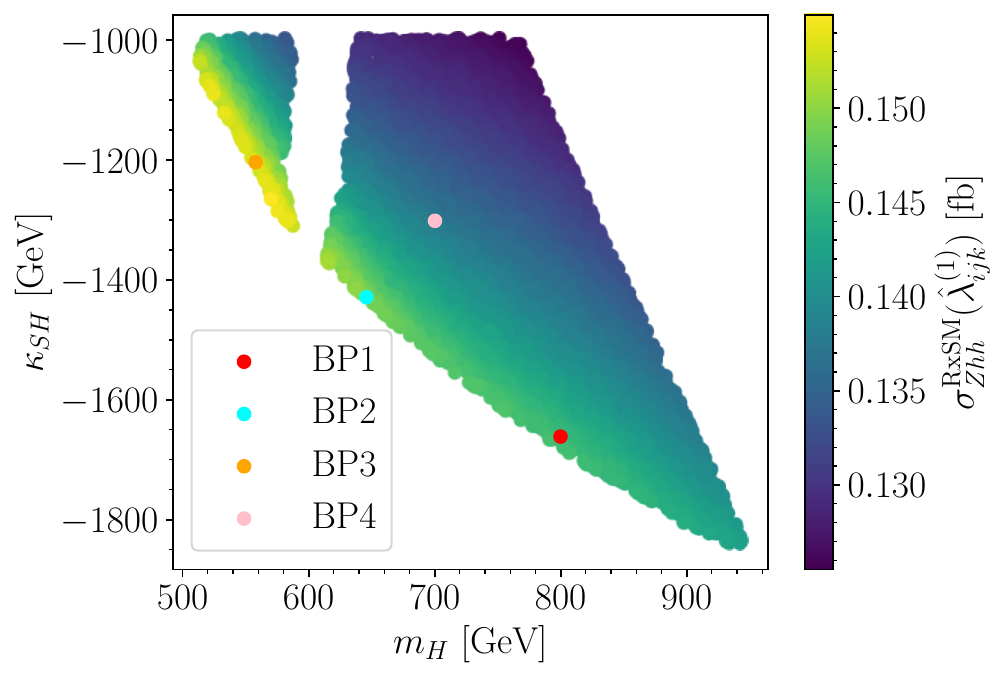}
    \includegraphics[width=0.49\linewidth]{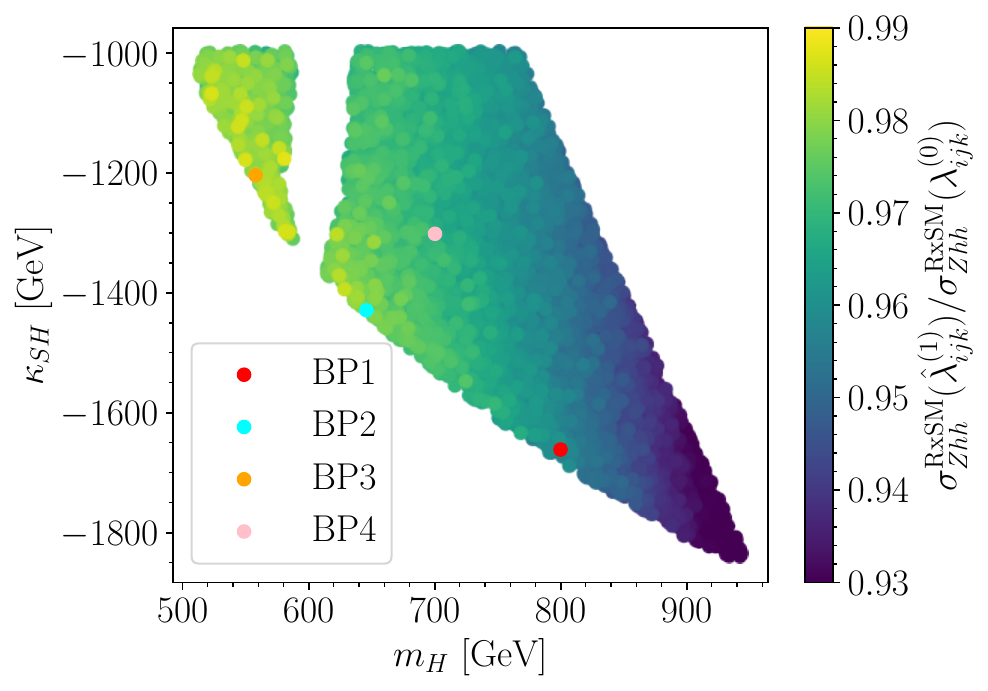}
    \caption{ Results of the di-Higgs cross-section computation for a 1 TeV $e^+e^-$ collider in the benchmark plane~2. \textit{Left}: $\siZhh(\rlaijk{1})$ (in fb); \textit{right}: $\siZhh(\rlaijk{1})/\siZhh(\laijk^{(0)})$.
    We note that in the left plot the value of $\sigma_{Zhh}^\text{SM} = 0.121~\mathrm{fb}$ is below the lower bound in the colour bar.}
    \label{totalee}
\end{figure}

We compute the cross-section for a future $e^+e^-$ collider operating at a centre-of-mass energy of $\sqrt{s} = 1~\mathrm{TeV}$, taking as an example the ILC1000~\cite{ILC:2013jhg,Moortgat-Pick:2015lbx,Bambade:2019fyw}. For other centre-of-mass energies such as $\sqrt{s}=550 \gev$ for the LCF550~\cite{LinearCollider:2025lya}, we would not expect to see interesting effects since in the benchmark plane studied here, $m_H>500 \gev$, and effects from $\lahhH$ in the $Zhh$ channel are typically visible for $m_H+m_Z>\sqrt{s}$, so we would not be able to access the resonant effects.
Concerning the $\nu\bar\nu hh$ channel, due to the energy taken by $\nu\bar\nu$ and the experimental cuts on missing energy also no relevant effects from the resonant $H$-channel diagram are expected. On the other hand, effects of $\kala \neq 1$ would be accessible at LCF550.

In \cref{totalee} we show the benchmark plane~2 where the colour coding indicates our predictions for $\siZhh$ using $\rlaijk{1}$ (left panel), and the ratio between the values of $\siZhh$ computed using $\rlaijk{1}$ and $\laijk^{(0)}$ (right panel), i.e.\ $\siZhh(\rlaijk{1})/\siZhh(\laijk^{(0)})$. 
In the left panel, we observe that the results lie in the range $\sigma^{\mathrm{RxSM}}_{Zhh}(\hat{\lambda}_{ijk}^{(1)}) \approx [0.125, 0.155]~\mathrm{fb}$. Considering that the SM prediction for the same process is $\sigma^{\mathrm{SM}}_{Zhh} = 0.121~\mathrm{fb}$%
\footnote{Including a $-80\%$ ($+30\%$) polarisation for electrons (positrons), these numbers go up by a factor of $1.476$, see below.} 
the RxSM result in the benchmark plane~2 is therefore $3\%-24\%$ larger than the SM value. This enhancement is due to the deviation of $\kappa_\lambda$ from its SM value. Unlike the HL-LHC case, where $\sigma_{hh}$ exhibits a minimum around $\kala \approx 2.5$, at an $e^+e^-$ collider in the process \eeZhh\ one finds a constructive interference between the SM-type diagrams, and $\sigma_{Zhh}$ increases monotonically with $\kala$. This feature makes an $e^+e^-$ collider a more promising setting to use the total di-Higgs cross section as a probe of SFOEWPT scenarios. Turning to the right panel of \cref{totalee}, we find that including one-loop corrections to the trilinear scalar couplings in the cross-section computation leads to deviations of up to $\sim7\%$ relative to the result obtained with tree-level couplings. 

\begin{figure}[ht]
    \centering
    \includegraphics[width=0.49\linewidth]{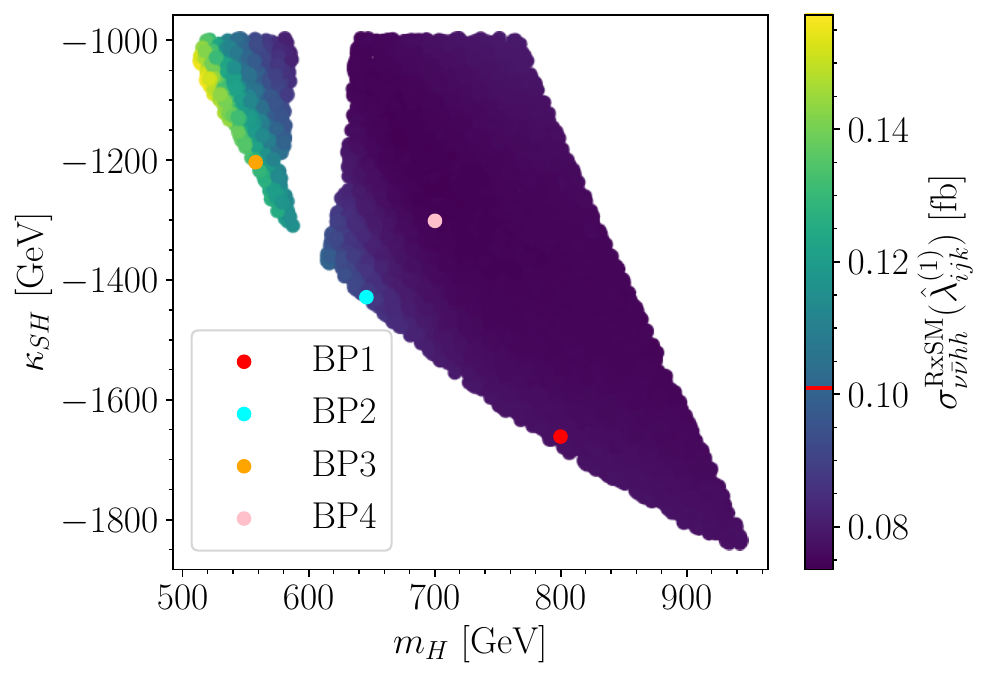}
    \includegraphics[width=0.49\linewidth]{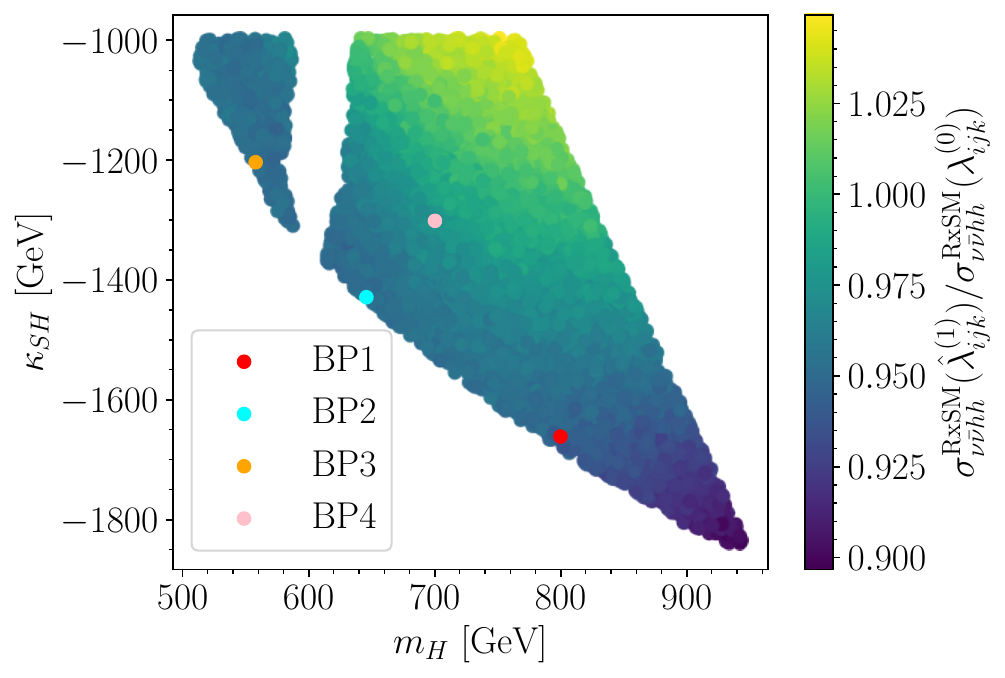}
    \caption{Results of the di-Higgs cross-section computation for a 1 TeV $e^+e^-$ collider in the benchmark plane~2. \textit{Left}: $\sivvhh(\rlaijk{1})$ (in fb); \textit{right}: $\sivvhh(\rlaijk{1})/\sivvhh(\laijk^{(0)})$.
    The red line in the colour bar of the left plot indicates the SM cross-section prediction. }
    \label{totalee_vvhh}
\end{figure}

In \cref{totalee_vvhh}, analogously to \cref{totalee}, we show the prediction in the benchmark plane~2 for $\sivvhh$ using $\rlaijk{1}$ (left panel), and the ratio between the value of $\sivvhh$ using $\rlaijk{1}$ and $\laijk^{(0)}$ (right panel), i.e.\ $\sivvhh(\rlaijk{1})/\sivvhh(\laijk^{(0)})$. 
The values of $\sigma^{\mathrm{RxSM}}_{\nu \bar{\nu}hh}(\hat{\lambda}_{ijk}^{(1)})$ range between $\approx [0.075,0.155]~\mathrm{fb}$, with a SM value of $\sivvhhSM = 0.101 \fb$. The $\kala$ values in our benchmark plane~2 range around $\sim 1.5$, see \cref{tri1}, where (unlike the $Zhh$ cross-section) the destructive interference of the SM-like diagrams has a maximum, i.e.\ the corresponding cross-section is smallest. On the other hand, the resonant $H$-channel diagram yields a positive contribution, which is larger for smaller values of $m_H$. These two effects partially cancel each other, leading to the cross-section range as found in the left plot of \cref{totalee_vvhh}. Approximately for $m_H \lesssim 600 \gev$ we find a $\sivvhh$ larger than in the SM. 
This makes this channel promising regarding the sensitivity to resonances, and hence to BSM trilinear scalar couplings, of heavy Higgs bosons in this mass range. In the right panel of \cref{totalee_vvhh}, we see that this ratio between the one-loop and tree-level cross-section predictions is in the range $\approx [0.90,1.04]$. In parts
of the plane (for larger $|\kappa_{SH}|$) we observe a decrease in the one-loop contributions of up to $10\%$. Nevertheless, unlike in \cref{totalee}, there are points in the parameter space where we find higher cross-section values at one loop than at tree level, mainly for smaller $|\kappa_{SH}|$ and larger $m_H$, with an increase of almost $5\%$.

Similarly to the HL-LHC case, differential $m_{hh}$ cross-section distributions at $e^+e^-$ colliders provide valuable information to explore SFOEWPT scenarios. We compute these distributions following the procedure described in \citere{Braathen:2025qxf}. As discussed there, we consider the scenario where the Higgs bosons decay as $hh \to b\bar{b}b\bar{b}$. 
We furthermore take into account the possible polarisations of the electron and positron beams, as described below.
Following \citere{Arco:2025pgx} for the $Zhh$ channel we include experimental cuts as,
\begin{equation}
    E_b>20 \gev, \quad |\eta_b|<2.5, \quad |\eta_Z|<2.5, \quad y_{bb}>0.001,  
\end{equation}
where $E_b$ is the energy of the $b$-tagged jets, $\eta_b$ and $\eta_Z$ are the pseudo-rapidity of the $b$-tagged jets and the $Z$~boson, respectively, and $y_{bb}$ is a variable used to perform the jet clustering procedure by the Durham algorithm \cite{Catani:1991hj}, where $y_{bb}$ gives us a notion of distance between the $b$-tagged jets, defined as $y_{ij}=2\text{min}(E_i^2,E_j^2)(1-\cos{\theta_{ij}})/s$, where $\theta_{ij}$ is the angle between the momenta of the particles $i$ and $j$. For the $\nu\bar{\nu}hh$ channel, we also consider similar experimental cuts based on \citeres{Arco:2025pgx,Abramowicz:2016zbo},
\begin{equation}
    E_b>20 \gev, \quad |\eta_b|<2.5, \quad E^T_{\mathrm{miss}}>20 \gev, \quad y_{bb}>0.001,  
\end{equation}
where we include a cut to the transverse missing energy, $E^T_{\mathrm{miss}}>20 \gev$~\cite{Abramowicz:2016zbo}, due to the presence of the neutrinos. 
The corresponding number of events is calculated as in~\citere{Braathen:2025qxf},
\begin{equation}
    N= \cLint \times \sigma\times\; \text{BR}(h \to b\bar{b})^2\times \mathcal{A}\times \epsilon_b,
\end{equation}
where $\sigma$ is the polarised cross-section for the respective channel.
Additionally, we define the acceptance as $\mathcal{A}=N^{\text{with cuts}}/N^{\text{w/o cuts}}$, which is calculated with \texttt{Madgraph}. We also consider $\epsilon_b=0.85$~\cite{Durig:2016jrs,Tian:2013qmi}, denoting the $b$-tagging efficiency for all the $b$-tagged jets in the final state. 
$\cLint$ denotes the integrated luminosity, which we define below. 
Concerning the size of the polarised cross-section, we assume two possible polarisations, $80\%$ for the electrons and $30\%$ for the positrons, with opposite signs, denoted by $(\pm80\%, \mp30\%)$. In the case of $\siZhh$ this can be taken into account by a simple scaling (see the discussion in \citere{Arco:2025pgx}), as
$\sigma(-80\%, +30\%) \simeq 1.476\,\sigma_{\mathrm{unpol}}$ and $\sigma(+80\%, -30\%) \simeq 1.004\,\sigma_{\mathrm{unpol}}$, with $\sigma_{\mathrm{unpol}}$ the cross-section without polarisation. In the case of $\sivvhh$ such a simple scaling is not possible, and we 
calculate the cross-section for each polarisation with \texttt{Madgraph}. For this channel, we also found that $(-80\%, +30\%)$ is the most favourable polarisation and its contribution to the $m_{hh}$ distribution is clearly predominant compared to the other polarisation. 
We additionally found that for each channel the shape of the $m_{hh}$ distributions for each polarisation is virtually the same, with the corresponding rescaling. Therefore, in what follows, for the $\mhh$ distributions we show the polarised cross-section for the most favourable polarisation $(-80\%, +30\%)$, while for the number of events we will consider the sum of both polarisations.
Here we have assumed an integrated luminosity of $\cLint = 3200 \ifb$ for each polarisation~\cite{Bambade:2019fyw}. Regarding the binning, for the $Zhh$ channel we set the size of the bins such that for the statistical significance calculations (see below), we meet the condition $N>2$ for most of the bins for each polarisation. For the $\nu\bar{\nu}hh$ channel, due to the considerably larger contribution of the $(-80\%, +30\%)$ polarisation, we prioritise applying this condition to this polarisation. 
In addition, we apply a smearing of $5\%$~\cite{Arco:2025pgx} to the cross-section distributions. 

The results for the $Zhh$ channel are shown in \cref{dist_mhh_e+e-_Zh1h1_Zbbbb_C_BP1BP2} (BP1 and BP2) and \cref{dist_mhh_e+e-_Zh1h1_Zbbbb_C_BP3BP4} 
(BP3 and BP4). In these plots, the red (green) curve corresponds to the RxSM result using $\rlaijk{1}$ ($\laijk^{(0)}$), the blue (orange) curve represents the contribution, taken alone, of the diagram with $H$ in the $s$-channel using $\hat{\lambda}_{hhH}^{(1)}$ ($\lahhH^{(0)}$), and finally the yellow curve indicates the SM prediction. 
Here we stress again that the cross-sections, as given in the legends and the left vertical axes, correspond to the case of 
$(-80\%, +30\%)$ polarisation, whereas the number of events shown on the right vertical axes are obtained as the sum of the two polarisations.

\begin{figure}[ht!]
    \centering  
    \includegraphics[width=0.66\linewidth]{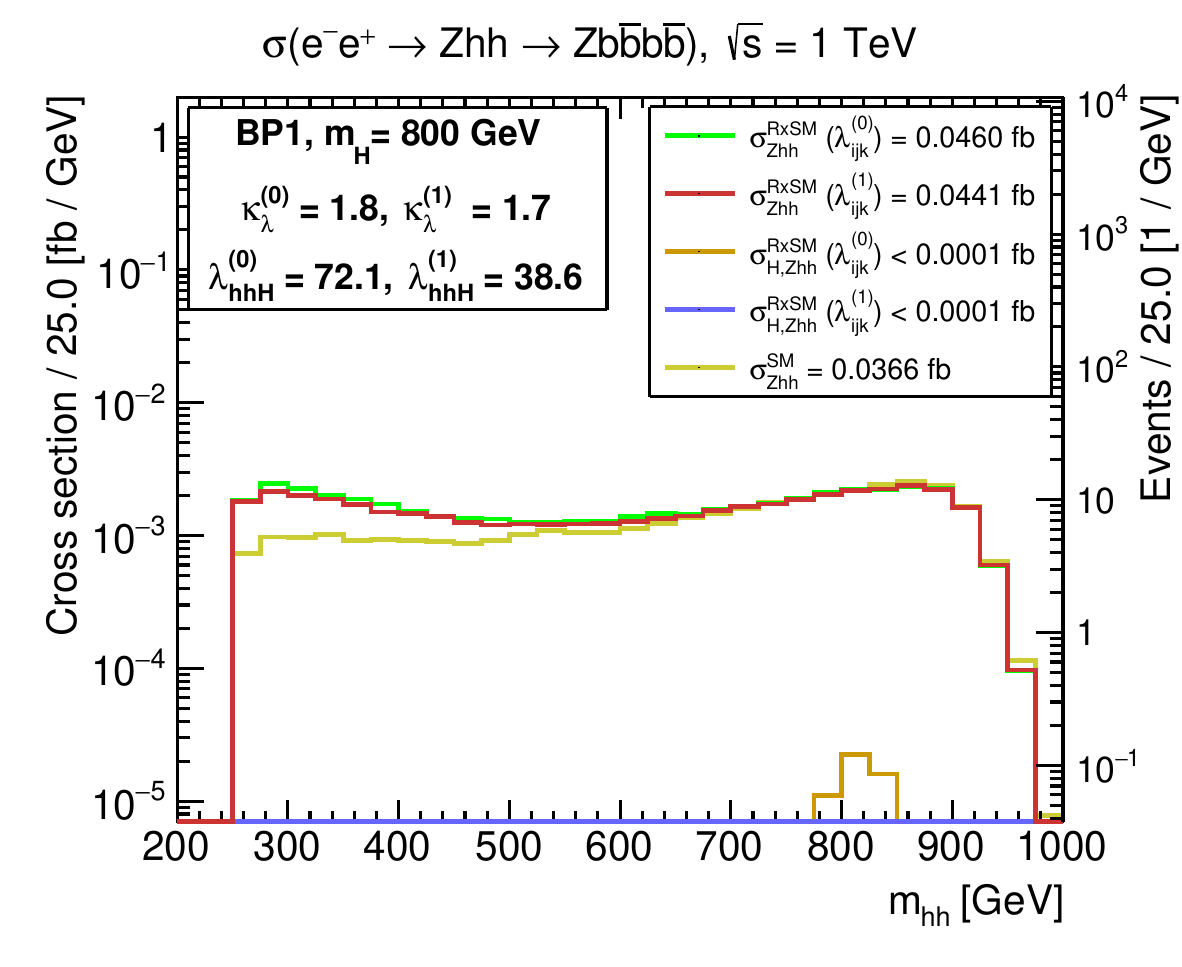}
    \includegraphics[width=0.66\linewidth]{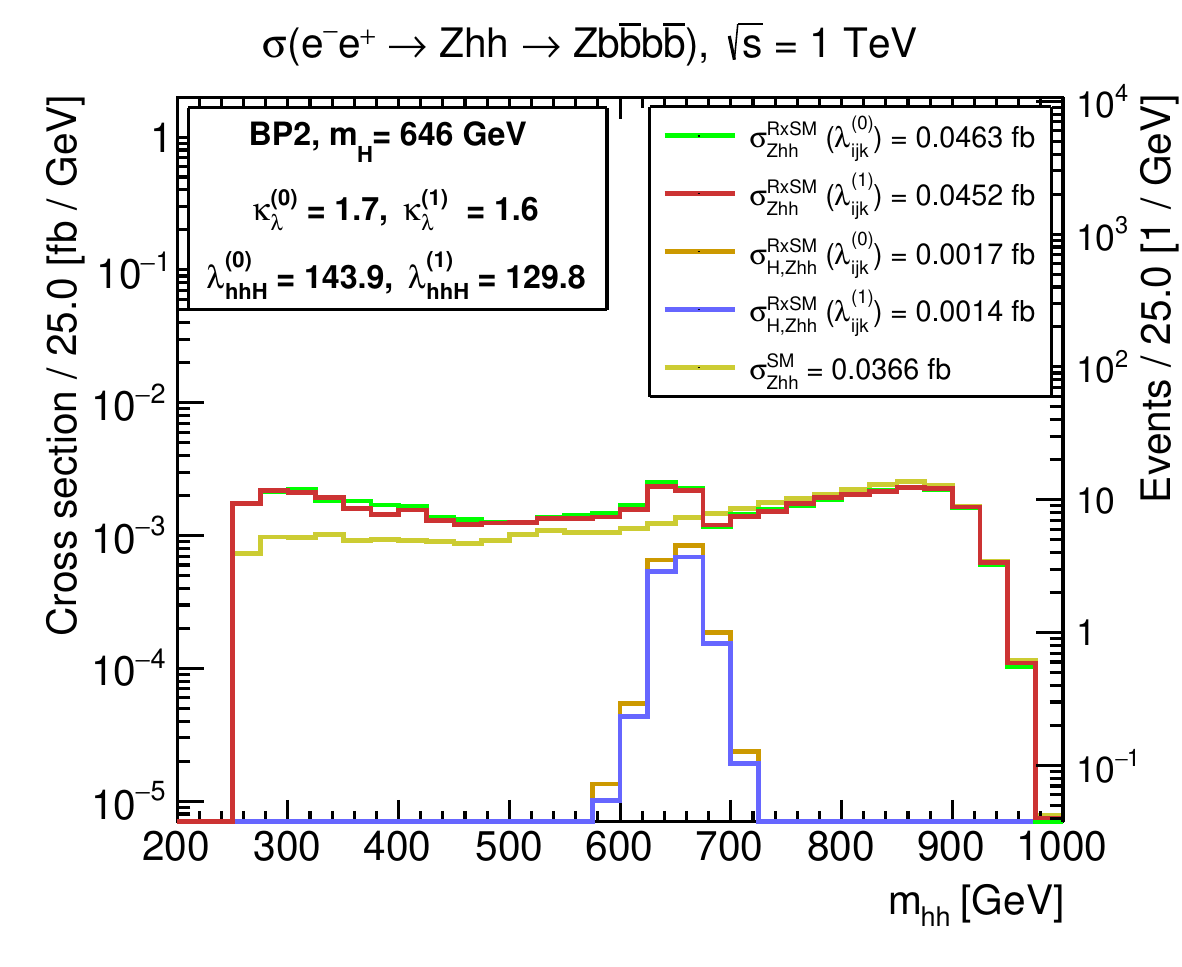}
    \caption{Differential polarised di-Higgs production cross-section distributions (for the polarisation $(-80\%,+30\%)$) after applying cuts and smearing (see text) as a function of the di-Higgs invariant mass $m_{hh}$ for the process $e^+e^- \to Zhh \to Zb\bar{b}b\bar{b}$ at the ILC1000. \textit{Top}: Results for BP1. \textit{Bottom}: Results for BP2. We plot the RxSM result using tree-level (one-loop) trilinear scalar couplings in green (red), 
    the contribution of the diagram with $H$ in the $s$-channel using tree-level (one-loop) trilinear scalar couplings in orange (purple),
    and the SM result in yellow. On the right axis we show the sum of the number of events for both polarisations considering $\cLint = 3200 \ifb$.
    Values of $\lahhH$ are shown in GeV. }
    \label{dist_mhh_e+e-_Zh1h1_Zbbbb_C_BP1BP2}
\end{figure}

\begin{figure}[ht!]
    \centering
    \includegraphics[width=0.66\linewidth]{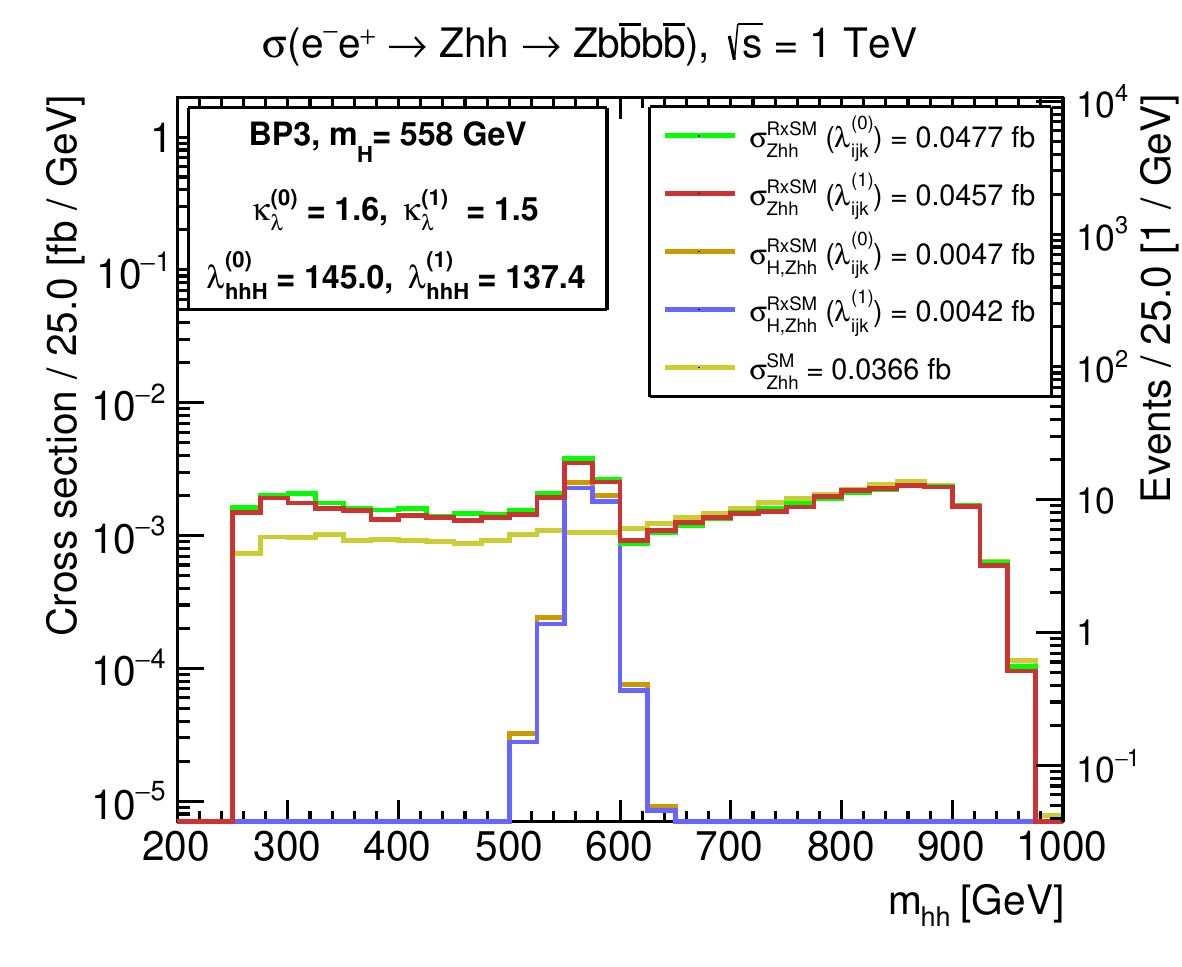}
    \includegraphics[width=0.66\linewidth]{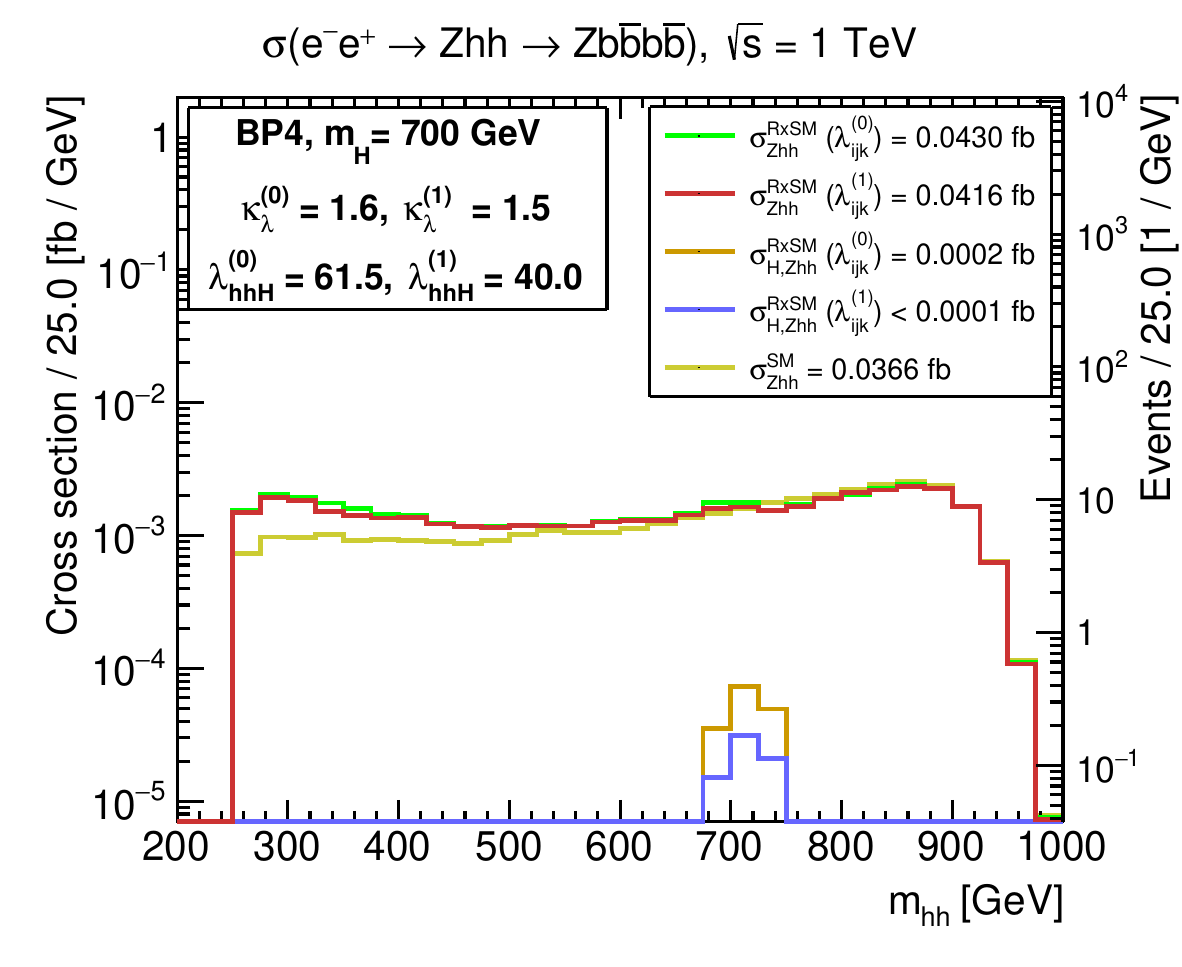}
    \caption{Plots and line styles as in \protect\cref{dist_mhh_e+e-_Zh1h1_Zbbbb_C_BP1BP2}. \textit{Top}: Results for BP3. \textit{Bottom}: Results for BP4. }
    \label{dist_mhh_e+e-_Zh1h1_Zbbbb_C_BP3BP4}
\end{figure}

In \cref{dist_mhh_e+e-_Zh1h1_Zbbbb_C_BP1BP2,dist_mhh_e+e-_Zh1h1_Zbbbb_C_BP3BP4}, we find larger differential cross-sections in the RxSM than in the SM at low $m_{hh}$, a significant non-resonant effect. One can furthermore observe a decrease when including trilinear couplings computed at one loop with respect to tree level, as expected from the corresponding small decrease of $\kala$. 
With regard to each BP, we find the highest cross-sections, and also the most pronounced resonances, in BP2 and BP3, which are the benchmark points with lower $m_H$ and higher $\lahhH$, an effect that is more pronounced in BP3. In these two BPs, we can distinguish a peak-dip structure, caused by the interference of the resonant $H$-channel diagram with the non-resonant diagrams in this process. This is illustrated by the orange and blue curves, showing the pure heavy Higgs-boson resonance cross-section. On the other hand, for BP1 and BP4 in the full calculation the resonances cannot be resolved, due to the small contribution of the resonance diagrams and the smearing.

In the next step, we calculate a statistical significance for distinguishing the RxSM $m_{hh}$ distribution from the SM prediction following \citere{Braathen:2025qxf}, defined as $Z=\sqrt{(Z_{-+})^2 + (Z_{+-})^2}$, with $Z_{\pm\mp}$  the statistical significance calculated for each polarisation $(\pm80\%,\mp30\%)$. The results are summarised in \cref{tab: dist_mhh_e+e-_Zh1h1_Zbbbb_C}, where $Z^{(0)}_{Zhh}$ 
denotes the tree-level and $Z^{(1)}_{Zhh}$ the one-loop result. 
In general we find
large significances which in the one-loop case lie in the range $ [3.9,6.9]$.  
One can also see a decrease in the significances from tree level to one loop, in agreement with the total cross-sections. In addition, we find the highest significances for BP3, with $7.8$ and $6.9$ at tree level and one loop, respectively. Thus, analysing the differential cross-section yields a promising method to study this type of BSM models.

Furthermore, we calculate the statistical significance of the heavy Higgs-boson resonance in the $m_{hh}$ distributions. This gives an indication of the sensitivity to $\lahhH$. For this calculation we follow the same procedure as before, but now calculating the significance of the full RxSM $\mhh$ distribution with respect to the RxSM prediction omitting the heavy Higgs-boson contribution. 
The results are given in the two right-most columns of \cref{tab: dist_mhh_e+e-_Zh1h1_Zbbbb_C} as $Z^{(0),\text{r}}_{Zhh}$ at tree level and $Z^{(1),\text{r}}_{Zhh}$ at one loop. 
As expected from our description of the differential cross-sections, BP1 and BP4 yield only marginal values for this significance. 
BP2 reaches values slightly below 3, 
and BP3 yields significances close to~6. The overall differences between the tree-level and one-loop results are small. These results indicate that at least for some parts of the RxSM parameter space favoured by a SFOEWPT a high-energy 
$e^+e^-$ collider may have access to the BSM trilinear scalar coupling $\lahhH$.  
Apart from significances, we also show in \cref{tab: dist_mhh_e+e-_Zh1h1_Zbbbb_C} our approximation for the detector efficiency, 
$\mathcal{A}_{Zhh}\times\epsilon_b$. We found that for the considered points at tree level and one loop the difference between the values of $\mathcal{A}_{Zhh}\times\epsilon_b$ for our benchmark points is below the numerical uncertainty of $\sim 0.5\%$ with values around $\sim 60\%$. This is related to the fact that the relative differences between the values of $\siZhh$ for the BPs, also before applying cuts, are small.

\begin{table}[htb!]
\centering
\begin{tabular}{@{}ccccccccc@{}}
\hline
BP  & $\sigma^{\text{RxSM}}_{Zhh} (\lambda^{(0)}_{ijk}) $ & $\sigma^{\text{RxSM}}_{Zhh} (\hat{\lambda}^{(1)}_{ijk})  $ & $Z^{(0)}_{Zhh}$ & $Z^{(1)}_{Zhh}$ & $\mathcal{A}^{(0)}_{Zhh}\times\epsilon_b$ & $\mathcal{A}^{(1)}_{Zhh}\times\epsilon_b$ & $Z^{(0),\text{r}}_{Zhh}$ & $Z^{(1),\text{r}}_{Zhh}$ \\
& [\text{fb}] & [\text{fb}] & & & & & & \\
\hline
1 & 0.0460 & 0.0441 & 5.8 & 4.8 & $ 60.5\%$ & $60.6\%$ & 0.7 & 0.7 \\  
2 & 0.0463 & 0.0452 & 6.2 & 5.7 & $60.6\%$ & $60.6\%$  & 2.9 &  2.8 \\
3 & 0.0477 & 0.0457 & 7.8 & 6.9 & $60.9\%$ & $60.9\%$ & 5.8 & 5.5 \\
4 & 0.0430 & 0.0416 & 4.4 & 3.9 & $60.6\%$ & $60.6\%$   & 1.1 & 0.8 \\
\hline
\end{tabular}
\caption{Summary of the predictions for $e^+e^-\to Zhh$ cross-sections, significances, and acceptance at the ILC1000 for the benchmark points in \protect\cref{bps}.  
From left to right: $(-80\%,+30\%)$ polarised cross-sections $\sigma (e^+e^- \to Zhh \to Zb\bar{b}b\bar{b})$ in the RxSM at tree level and one loop for ILC1000 after applying cuts; statistical significances at tree level and one loop to distinguish the RxSM distribution from the SM; values of acceptances times $\epsilon_b$ at tree level and one loop; statistical significances of the resonances, i.e.\ the RxSM compared to the RxSM process without the resonant heavy Higgs-boson diagram. For the SM, we have $\mathcal{A}\times\epsilon_b=61.4\%$.}
\label{tab: dist_mhh_e+e-_Zh1h1_Zbbbb_C}
\end{table}

As a final step in our analysis, we compute the differential $m_{hh}$ distributions for the $\nu\bar{\nu}hh$ channel, which are shown in \cref{dist_mhh_e+e-_vvh1h1_vvbbbb_C_BP1BP2} (BP1 and BP2) and \cref{dist_mhh_e+e-_vvh1h1_vvbbbb_C_BP3BP4} (BP3 and BP4). These plots follow the same colour coding and structure as \cref{dist_mhh_e+e-_Zh1h1_Zbbbb_C_BP1BP2,dist_mhh_e+e-_Zh1h1_Zbbbb_C_BP3BP4} for the $Zhh$ channel.
The corresponding significances and efficiencies are summarised in \cref{tab: e+e-_vvh1h1_vvbbbb_C}. For these efficiencies, we have performed the calculation for both polarisations. We have found that the contributions of the polarisation $(-80\%,+30\%)$ are significantly larger that for $(+80\%,-30\%)$, while the acceptances values are virtually the same, with variations of less than $1\%$. Therefore, we have used (and listed in \cref{tab: e+e-_vvh1h1_vvbbbb_C}) the results for the polarisation $(-80\%,+30\%)$.

\begin{figure}[ht!]
    \centering 
    \includegraphics[width=0.66\linewidth]{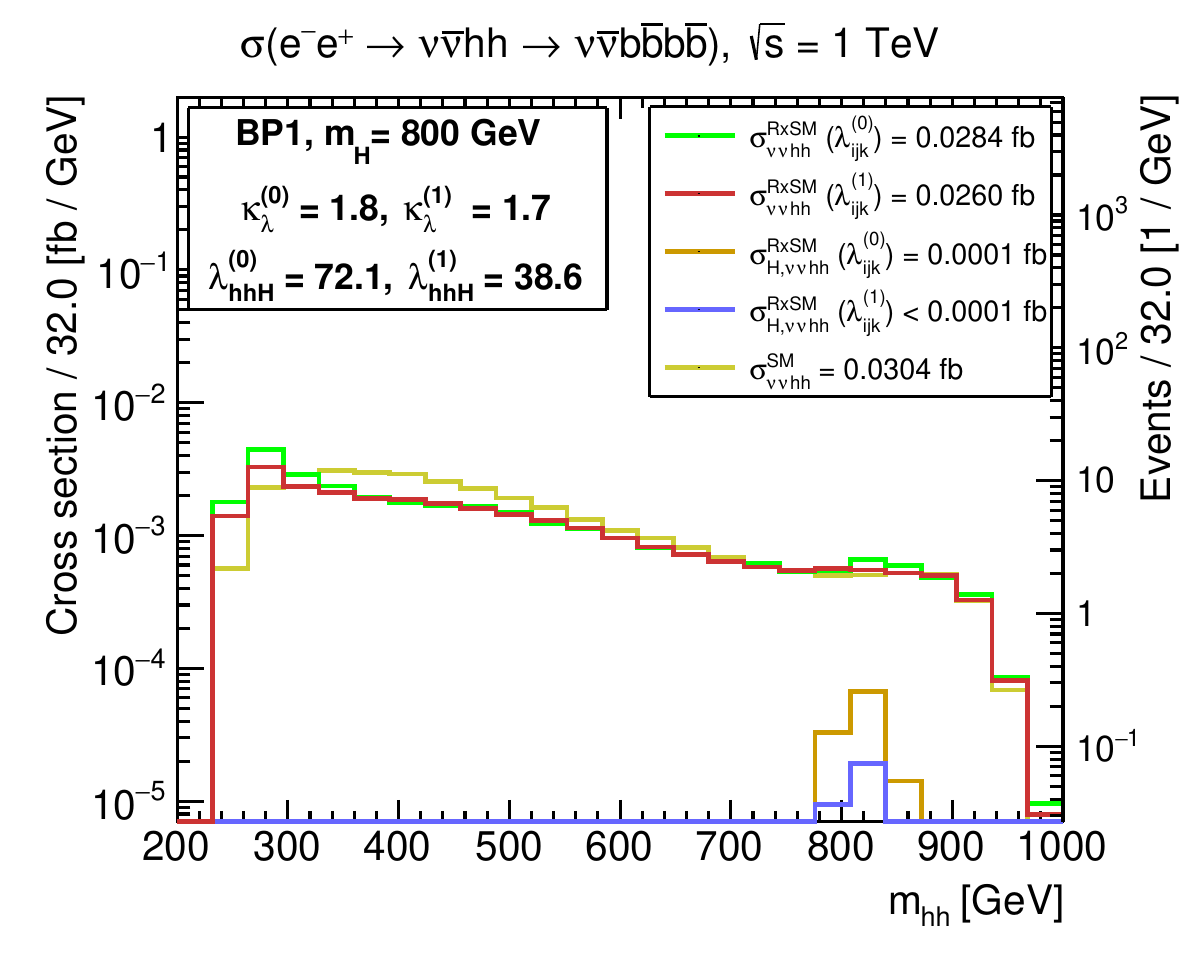}
    \includegraphics[width=0.66\linewidth]{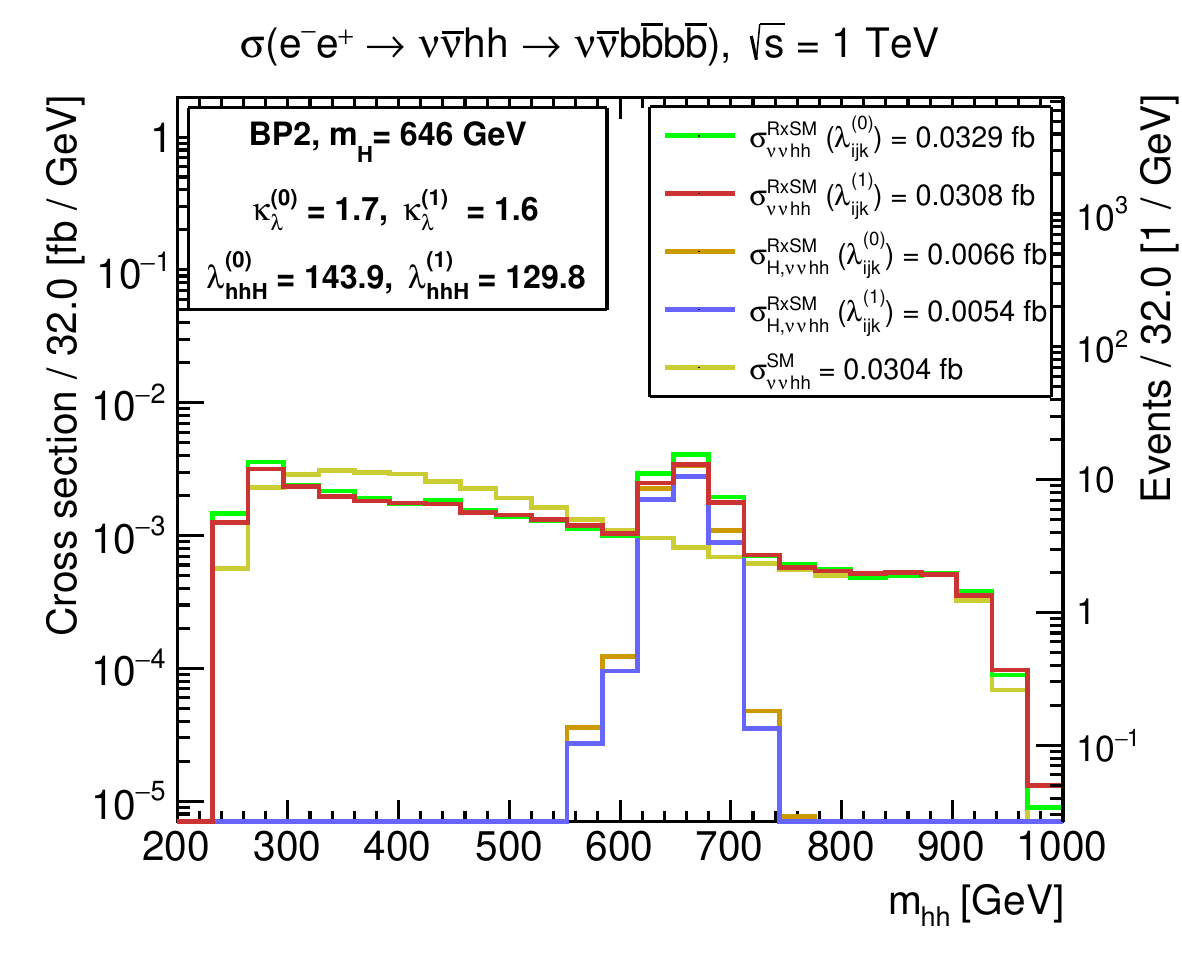}
    \caption{Plots and line styles as in \protect\cref{dist_mhh_e+e-_Zh1h1_Zbbbb_C_BP1BP2}, but for the process $e^+e^-\to \nu\bar{\nu}hh \to \nu\bar{\nu}b\bar{b}b\bar{b}$.
    \textit{Top}: Results for BP1. \textit{Bottom}: Results for BP2.}
    \label{dist_mhh_e+e-_vvh1h1_vvbbbb_C_BP1BP2}
\end{figure}

\begin{figure}[ht!]
    \centering
    \includegraphics[width=0.66\linewidth]{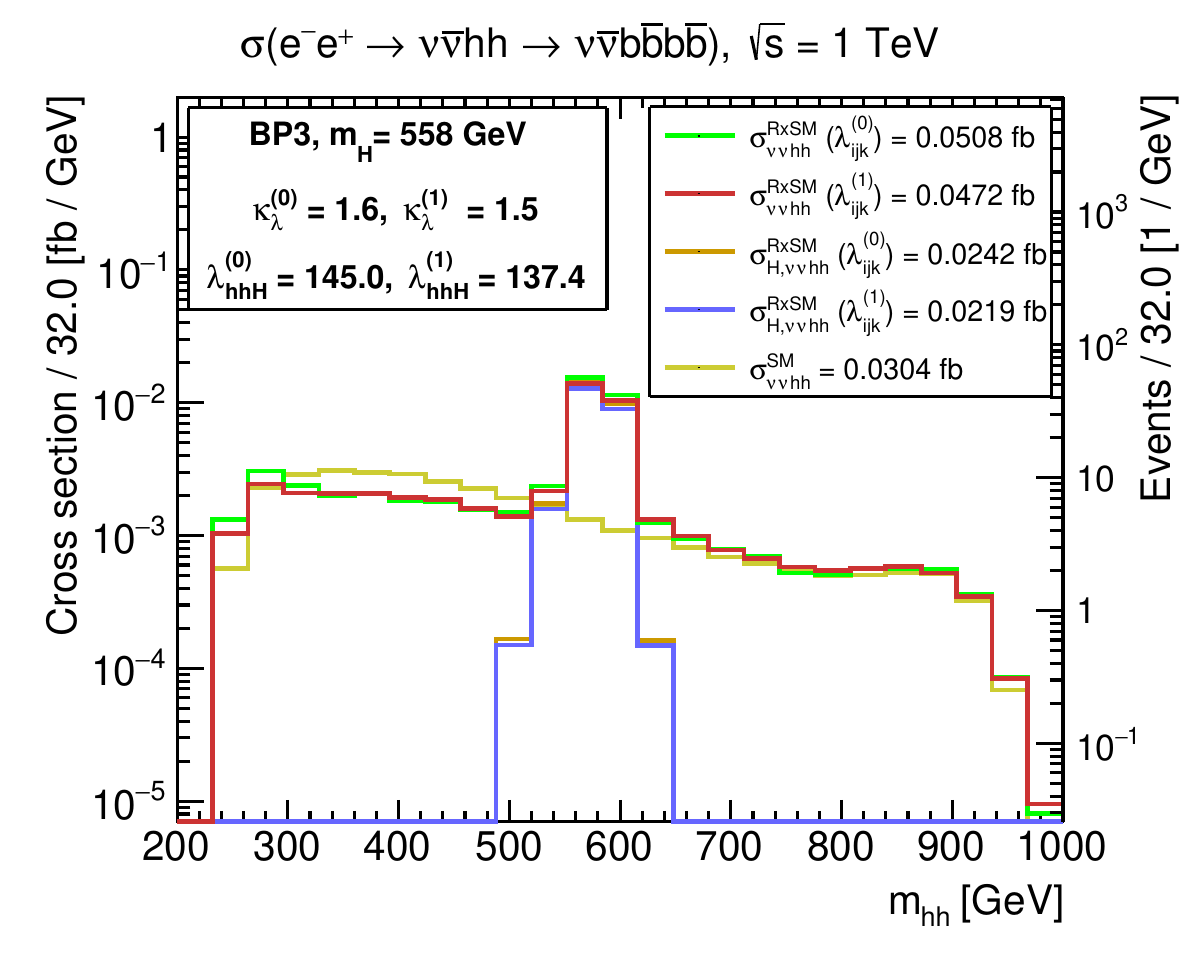}
    \includegraphics[width=0.66\linewidth]{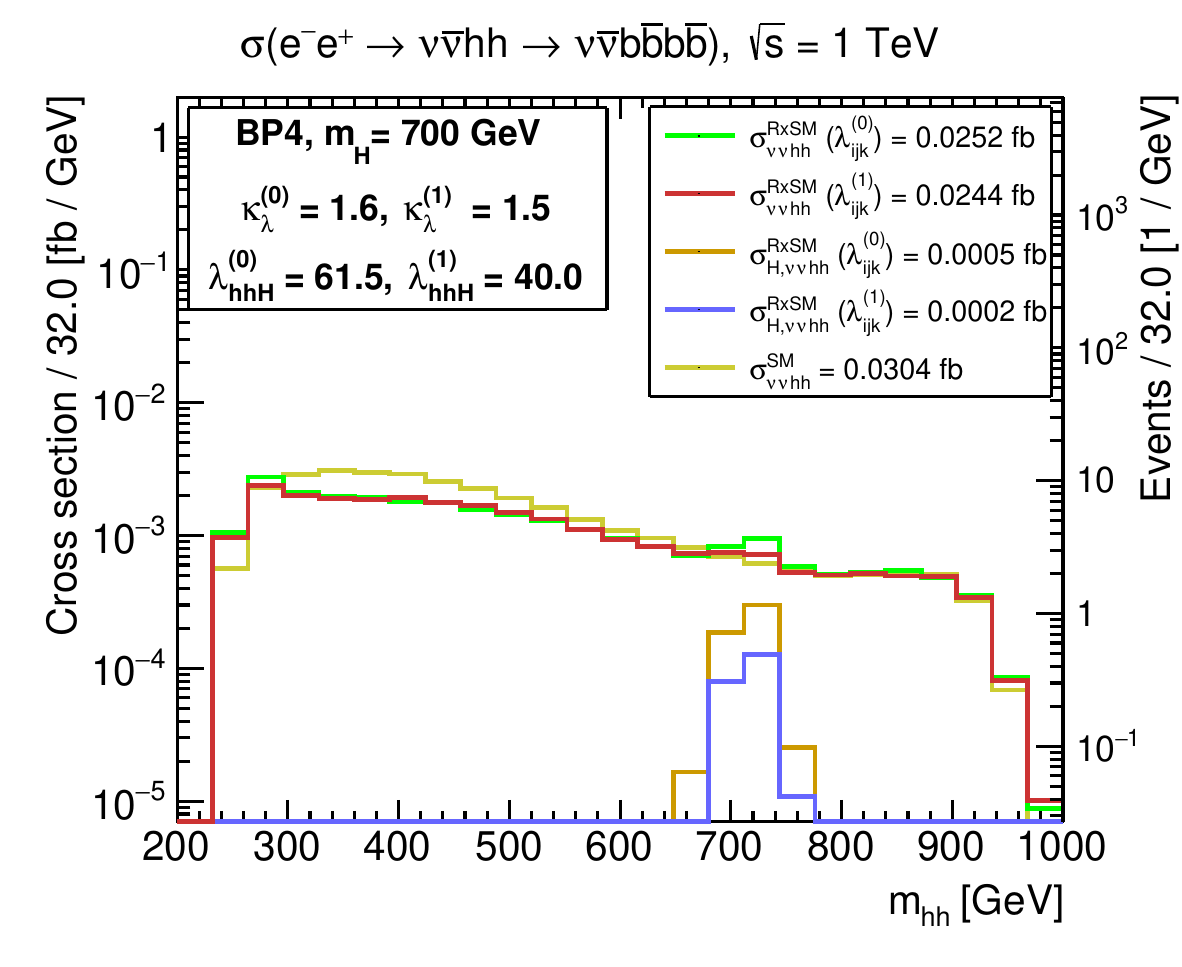}
    \caption{Plots and line styles as in \protect\cref{dist_mhh_e+e-_Zh1h1_Zbbbb_C_BP3BP4}, but for the process $e^+e^-\to \nu\bar{\nu}hh \to \nu\bar{\nu}b\bar{b}b\bar{b}$.
    \textit{Top}: Results for BP3. \textit{Bottom}: Results for BP4.}
    \label{dist_mhh_e+e-_vvh1h1_vvbbbb_C_BP3BP4}
\end{figure}

In \cref{dist_mhh_e+e-_vvh1h1_vvbbbb_C_BP1BP2,dist_mhh_e+e-_vvh1h1_vvbbbb_C_BP3BP4}, unlike 
\cref{dist_mhh_e+e-_Zh1h1_Zbbbb_C_BP1BP2,dist_mhh_e+e-_Zh1h1_Zbbbb_C_BP3BP4} for the $Zhh$ channel, we find a reduction of the differential cross-sections with respect to the SM prediction in most of the $m_{hh}$ range, observing again significant non-resonant effects in the differential cross-section. Following the analysis of the total cross-section in benchmark plane~2, we obtain lower values of the total cross-sections in the RxSM distributions than for the SM for BP1 and BP4. Meanwhile, for BP2, the cross-section is very close to the SM, and a noticeable increase is only found for BP3. The latter receives the largest enhancement from the heavy Higgs-boson resonance. The sizes and effects of the pure resonant heavy Higgs-exchange contribution follow the same pattern as in the $Zhh$ case, i.e.\ they yield visible results only for BP2 and BP3.

Finally, we compute the statistical significances to distinguish the RxSM differential distributions for this $\nu \bar{\nu} hh$ channel from that in the SM, summarised in \cref{tab: e+e-_vvh1h1_vvbbbb_C}, following the same procedure used for the results in \cref{tab: dist_mhh_e+e-_Zh1h1_Zbbbb_C} for the $Zhh$ channel~\cite{Braathen:2025qxf}.
We denote them as $Z^{(0)}_{\nu \bar{\nu} hh}$ and $Z^{(1)}_{\nu \bar{\nu} hh}$ at tree level and one loop. 
The overall results for the significances for distinguishing the RxSM from the SM are similar to the $Zhh$ case. However, BP2 and BP3 yield substantially higher values, due to the fact that the differences in the $\mhh$ distributions are spread over a larger range in $\mhh$. 
Concerning the significances for the heavy Higgs-boson resonance, defined as $Z^{(0),\text{r}}_{\nu \bar{\nu} hh}$ and $Z^{(1),\text{r}}_{\nu \bar{\nu} hh}$, there also the results are qualitatively in agreement with the ones of the $Zhh$ channel. Furthermore, BP2 and BP3 yield substantially higher values in the $\nu\bar{\nu}hh$ channel as compared to the $Zhh$ channel. This indicates that the $\nu\bar\nu hh$ channel may be better suited to extract information about $\lahhH$ at the ILC1000.

\begin{table}[htb!]
\centering\begin{tabular}{@{}ccccccccc@{}}
\hline
BP  & $\sigma^{\text{RxSM}}_{\nu \bar{\nu} hh} (\lambda^{(0)}_{ijk})\ $ & $\sigma^{\text{RxSM}}_{\nu \bar{\nu} hh} (\hat{\lambda}^{(1)}_{ijk})$ & $Z^{(0)}_{\nu \bar{\nu} hh}$ & $Z^{(1)}_{\nu \bar{\nu} hh}$ & $\mathcal{A}^{(0)}_{\nu \bar{\nu} hh}\times\epsilon_b$ & $\mathcal{A}^{(1)}_{\nu \bar{\nu} hh}\times\epsilon_b$& $Z^{(0),\text{r}}_{\nu \bar{\nu} hh}$ & $Z^{(1),\text{r}}_{\nu \bar{\nu} hh}$\\
& [\text{fb}] & [\text{fb}] & & & & & & \\
\hline
1 & 0.0284 & 0.0260 & 3.4 & 2.8 & $ 49.8\%$ & $48.3\%$ & 0.4 & 0.3\\ 
2 & 0.0329 & 0.0308 & 6.5 & 5.7 & $49.3\%$ & $48.9\%$ & 6.0 &   5.2 \\
3 & 0.0508 & 0.0472 & 16.4 & 15.0 & $51.0\%$ & $50.5\%$  & 16.4 &  15.0  \\
4 & 0.0252 & 0.0244 & 2.9 & 2.8 & $47.7\%$ & $47.3\%$  & 0.6 &  0.3 \\
\hline
\end{tabular}
\caption{Summary of the predictions for the benchmark points in \cref{bps} as in \protect\cref{tab: dist_mhh_e+e-_Zh1h1_Zbbbb_C}, 
but for the $\nu\bar{\nu}hh$ channel. For the SM, we have $\mathcal{A}\times\epsilon_b=45.3\%$.}
\label{tab: e+e-_vvh1h1_vvbbbb_C}
\end{table}


\section{Conclusions}
\label{sec:conclusions}

In this paper, we have explored the dynamics of the EWPT in the RxSM. Specifically, we analysed the possibility of probing scenarios with a SFOEWPT using the complementarity of GW signals and di-Higgs production at colliders, building on the previous studies in Refs.~\cite{Arco:2025nii,Braathen:2025qxf}. A novel aspect of our work is that we included, for consistency, one-loop radiative corrections both in the study of the thermal evolution of the vacuum and in the collider analyses. For the former, this was done using a new implementation of the RxSM into the public code \texttt{BSMPTv3.1.1}, including also higher-order corrections to the thermal potential. For the latter, we used the public code \texttt{anyH3}, employing the full on-shell renormalisation scheme devised in \citere{Braathen:2025qxf}, to compute one-loop corrected trilinear scalar couplings. In turn, these served as inputs for the computation of the di-Higgs production cross-sections for $gg\to hh$ at the HL-LHC, using \texttt{HPAIR}, and to calculate $\eeZhh$ and $\eenunuhh$ at the ILC1000, using \texttt{Madgraph5\_aMC}.  

By performing an extensive scan of the parameter space of the RxSM, we scrutinised the different possible thermal histories in the early Universe of this model. We found that first-order EWPTs are possible and can occur either as single- or multi-step transitions, with potentially various level of strengths. The most favourable scenarios in terms of realising a SFOEWPT are case \textbf{C} with a two-step transition, and cases \textbf{D} and \textbf{E} for a one-step transition (using here the same labelling as in \cref{thermalhistories}). Moreover, parts of the RxSM parameter space can be excluded due to vacuum trapping (case \textbf{F}). 

Inspired by this general scan, we devised two benchmark planes, representative of the specific regions where we found SFOEWPTs, and we investigated in detail their phenomenology. A first such plane features light BSM Higgs-boson masses and low values of $v_S$, and exhibits SFOEWPTs for an extended range of parameters --- both as one-step (case \textbf{D}) and two-step (case \textbf{C}) transitions. The second plane, with larger $v_S$ and heavy BSM scalar masses, features one-step SFOEWPTs (as well as vacuum trapping). For both benchmark planes we calculated predictions for the spectra of GWs produced during the SFOEWPT and the associated SNR at LISA, using \texttt{BSMPT}. We moreover investigated the dependence of the SNR on the bubble wall velocity (which we did not compute in our study), and found that for most scenarios considering $\vw=0.95$ is a sufficiently conservative choice, as long as the true $\vw\gtrsim 0.3$. 
In benchmark plane~1, the EWPT is largely driven by the singlet field direction, which allows for strong GW signals in significant parts of the plane. On the other hand, in benchmark plane~2, it is the doublet field direction that plays the most important role in the EWPT, and observable SNR are only found in a narrow band of the plane.  

Next, we considered whether searches for di-Higgs production processes at high-energy colliders, the HL-LHC or a 1~TeV $e^+e^-$ collider (ILC1000), can provide additional information to probe these scenarios. In the case of the first benchmark plane, with 
a singlet-driven EWPT, the BSM effects in trilinear scalar couplings are minute, and essentially no new information would be obtained from di-Higgs production searches. In contrast to this, for the second benchmark plane, with a doublet-driven EWPT, there is a clear correlation between the strength of the SFOEWPT and the BSM deviation in $\kala$ --- with $1.35\lesssim \kala^{(1)}\lesssim 1.7$ for points with a SFOEWPT. These values of $\kala$ result in a further suppression of the $gg\to hh$ and $\eenunuhh$ cross-sections compared to the SM, while the $\eeZhh$ cross-section is increased. Resonant $H$~contributions can yield an increase of the three cross-sections; for the case of $\eenunuhh$ this can suffice to exceed the SM prediction, but not for $gg\to hh$. In other words, a high-energy $e^+e^-$ collider could probe these SFOEWPT scenarios via the total di-Higgs production cross-section, but this would be significantly more difficult at the HL-LHC.

Moreover, additional information can be obtained from the study of differential distributions. For this reason, we devised four concrete benchmark points, selected from benchmark plane~2, in order to perform a detailed analysis of differential $\mhh$ distributions. While the relative level of significance of the different di-Higgs processes varied between the BPs, we found that all these scenarios could be distinguished from the SM, even though the corresponding total cross-sections were in many cases below the SM predictions. 
We also illustrated the possibility of discriminating the resonant $H$~contributions from the continuum, provided that the BSM scalar mass is not too heavy ($m_H\lesssim 650\gev$) --- and here the $\eenunuhh$ process is clearly the most promising. However, here it should be kept in mind that our study only takes into account parts of the experimental effects and uncertainties; a full experimental study is needed to analyse the sensitivity to the heavy Higgs-boson resonance.

To conclude, in this work we have demonstrated the crucial importance of using complementary sources of experimental data in order to probe SFOEWPT scenarios in the RxSM. Singlet-driven EWPT benchmarks typically feature comparatively stronger signals of GWs, sourced during the phase transition in the early Universe, but close to no signs of BSM effects in di-Higgs production. Meanwhile, the situation in scenarios with a doublet-driven SFOEWPT is more similar to that in models like the 2HDM (see e.g.\ Ref.~\cite{Biekotter:2022kgf}): i.e.\ with a significant correlation between $\xi_n$ and $\kala$, BSM effects in di-Higgs production that would likely be observable at colliders --- HL-LHC and/or an $e^+e^-$ machine --- at least in differential $\mhh$ distributions, but detectable GW signals only in a limited fraction of the parameter space. Our results clearly indicate that no single experimental direction can cover the entire parameter space of the RxSM that would give rise to a SFOEWPT --- and in particular singlet-driven SFOEWPTs would be extremely difficult to constrain only with collider experiments. This is somewhat in contrast to the results presented in Refs.~\cite{CMS:2025hfp,BriefingBook2025}. 

Finally, it would in principle be interesting to consider further probes of SFOEWPTs in the RxSM and to determine the additional parts of the model parameter space where these could be relevant. These could for example include other cosmological relics of FOEWPTs (primordial black holes or magnetic fields), or different collider processes (e.g.\ di-Higgs production including one or two BSM scalar(s) in the final state). We leave this for future work. 


\section*{Acknowledgements}
\sloppy{We thank Lisa Biermann, Christoph Borschensky, and Margarete M\"{u}hlleitner for communications and invaluable help with implementing the 
RxSM in \texttt{BSMPT}. We thank Andrea Parra Arnay for collaboration in the early stages of this work. 
J.B.\ and A.V.S.\ acknowledge support by the Deutsche Forschungsgemeinschaft (DFG, German Research Foundation) under Germany's Excellence Strategy --- EXC 2121 ``Quantum Universe'' --- 390833306. This work has been partially funded by the Deutsche Forschungsgemeinschaft (DFG, German Research Foundation) --- 491245950. J.B. and A.V.S are supported by the DFG Emmy Noether Grant No.\ BR 6995/1-1. The work of S.H.\ has received financial support from the
grant PID2019-110058GB-C21 funded by
MCIN/AEI/10.13039/501100011033 and by ``ERDF A way of making Europe'', 
and in part by by the grant IFT Centro de Excelencia Severo Ochoa CEX2020-001007-S
funded by MCIN/AEI/10.13039/501100011033. 
S.H.\ also acknowledges support from Grant PID2022-142545NB-C21 funded by
MCIN/AEI/10.13039/501100011033/ FEDER, UE. 
C.P.B.\ acknowledges support from DESY (Hamburg, Germany) and the DESY Summer Student Programme.
}

\appendix


\section*{Appendix}
\section{Check of the condition \boldmath{$\alpha<1$}}

In this appendix we analyse the condition of having $\alpha<1$, which relates to the assumption of working during the radiation domination 
era. For this purpose, we display in \cref{ap1} as colour coding the values of $\alpha$ (left) and $\beta/H$ (right)
in the benchmark plane~1, 
where we find the strongest EWPT. 

\begin{figure}[h!]
    \centering
    \includegraphics[width=0.49\linewidth]{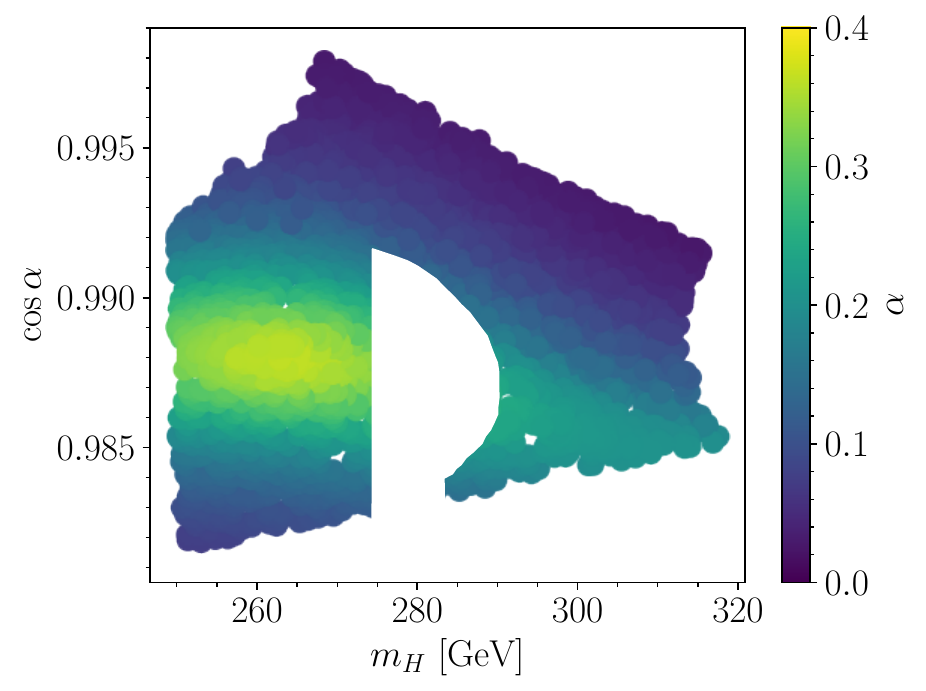}
    \includegraphics[width=0.49\linewidth]{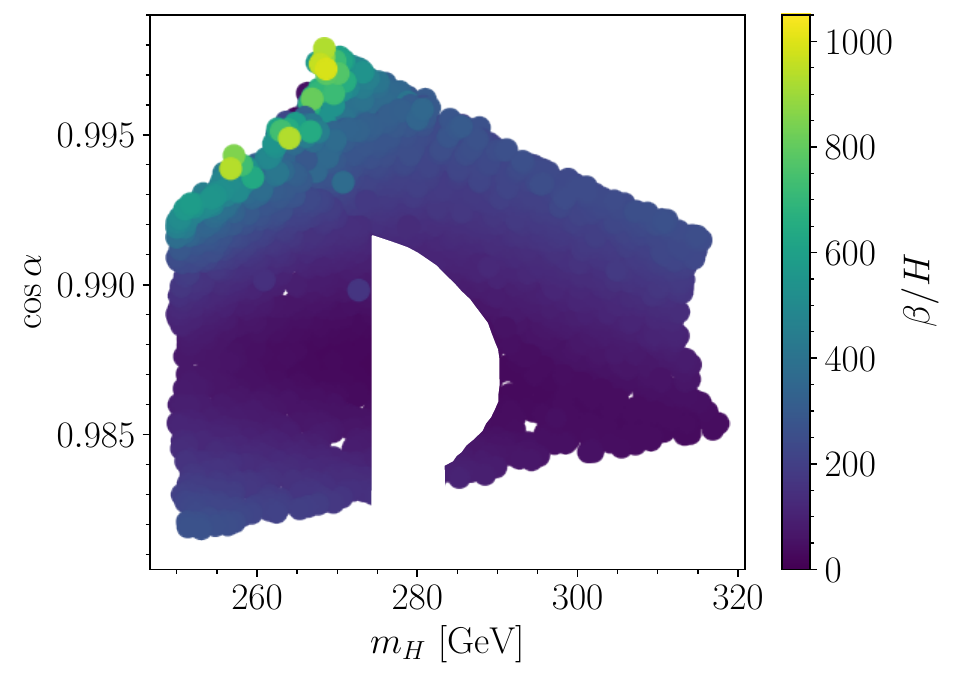}
    \caption{Results for the phase transition dynamics for the RxSM scan points in benchmark plane 1 with a SFOEWPT. \textit{Left}: Strength of the phase transition $\alpha$; \textit{right}: Inverse duration of the phase transition $\beta/H$.}
    \label{ap1}
\end{figure}
From the left and right panels of \cref{ap1}, respectively, we observe that $\alpha \lesssim 0.4$ and $\beta/H \lesssim 10^3$ for all viable points. 
Therefore, the condition $\alpha < 1$ is always satisfied, thus validating the assumption of being in the radiation domination area. Moreover, the fact that $\alpha \ll 1$ indicates that we are not in a supercooled scenario. 

%
%
%
%
%


\begin{thebibliography}{999}


\bibitem{Aad:2012tfa}
  G.~Aad {\it et al.} [ATLAS Collaboration],
  Phys.\ Lett.\ B {\bf 716} (2012) 1
  [arXiv:1207.7214 [hep-ex]].

\bibitem{Chatrchyan:2012xdj}
  S.~Chatrchyan {\it et al.} [CMS Collaboration],
  Phys.\ Lett.\ B {\bf 716} (2012) 30
  [arXiv:1207.7235 [hep-ex]].

\bibitem{Khachatryan:2016vau}
  G.~Aad {\it et al.} [ATLAS and CMS Collaborations],
  JHEP {\bf 1608} (2016) 045
  [arXiv:1606.02266 [hep-ex]].

\bibitem{Englert:1964et}
F.~Englert and R.~Brout,
Phys. Rev. Lett. \textbf{13} (1964), 321-323.

\bibitem{Higgs:1964pj}
P.~W.~Higgs,
Phys. Rev. Lett. \textbf{13} (1964), 508-509.

\bibitem{Guralnik:1964eu}
G.~S.~Guralnik, C.~R.~Hagen and T.~W.~B.~Kibble,
Phys. Rev. Lett. \textbf{13} (1964), 585-587.

\bibitem{Glashow:1961tr}
S.~L.~Glashow,
Nucl. Phys. \textbf{22} (1961), 579-588.

\bibitem{Weinberg:1967tq}
S.~Weinberg,
Phys. Rev. Lett. \textbf{19} (1967), 1264-1266.

\bibitem{Salam:1968rm}
A.~Salam,
Conf. Proc. C \textbf{680519} (1968), 367-377.

\bibitem{Kuzmin:1985mm}
V.~A.~Kuzmin, V.~A.~Rubakov and M.~E.~Shaposhnikov,
Phys. Lett. B \textbf{155} (1985), 36.

\bibitem{Cohen:1993nk}
A.~G.~Cohen, D.~B.~Kaplan and A.~E.~Nelson,
Ann. Rev. Nucl. Part. Sci. \textbf{43} (1993), 27-70
[arXiv:hep-ph/9302210 [hep-ph]].

\bibitem{Sakharov:1967dj}
A.~D.~Sakharov,
Pisma Zh. Eksp. Teor. Fiz. \textbf{5} (1967), 32-35.

\bibitem{Espinosa:1993bs}
J.~R.~Espinosa and M.~Quiros,
Phys. Lett. B \textbf{305} (1993), 98-105
[arXiv:hep-ph/9301285 [hep-ph]].

\bibitem{Ham:2004cf}
S.~W.~Ham, Y.~S.~Jeong and S.~K.~Oh,
J. Phys. G \textbf{31} (2005) no.8, 857-871
[arXiv:hep-ph/0411352 [hep-ph]].

\bibitem{Profumo:2007wc}
S.~Profumo, M.~J.~Ramsey-Musolf and G.~Shaughnessy,
JHEP \textbf{08} (2007), 010
[arXiv:0705.2425 [hep-ph]].

\bibitem{Espinosa:2007qk}
J.~R.~Espinosa and M.~Quiros,
Phys. Rev. D \textbf{76} (2007), 076004
[arXiv:hep-ph/0701145 [hep-ph]].

\bibitem{Barger:2008jx}
V.~Barger, P.~Langacker, M.~McCaskey, M.~Ramsey-Musolf and G.~Shaughnessy,
Phys. Rev. D \textbf{79} (2009), 015018
[arXiv:0811.0393 [hep-ph]].

\bibitem{Espinosa:2011ax}
J.~R.~Espinosa, T.~Konstandin and F.~Riva,
Nucl. Phys. B \textbf{854} (2012), 592-630
[arXiv:1107.5441 [hep-ph]].

\bibitem{Morrissey:2012db}
D.~E.~Morrissey and M.~J.~Ramsey-Musolf,
New J. Phys. \textbf{14} (2012), 125003
[arXiv:1206.2942 [hep-ph]].

\bibitem{Curtin:2014jma}
D.~Curtin, P.~Meade and C.~T.~Yu,
JHEP \textbf{11} (2014), 127
[arXiv:1409.0005 [hep-ph]].

\bibitem{Kurup:2017dzf}
G.~Kurup and M.~Perelstein,
Phys. Rev. D \textbf{96} (2017) no.1, 015036
[arXiv:1704.03381 [hep-ph]].

\bibitem{Ramsey-Musolf:2019lsf}
M.~J.~Ramsey-Musolf,
JHEP \textbf{09} (2020), 179
[arXiv:1912.07189 [hep-ph]].

\bibitem{Kajantie:1996mn}
K.~Kajantie, M.~Laine, K.~Rummukainen and M.~E.~Shaposhnikov,
Phys. Rev. Lett. \textbf{77} (1996), 2887-2890
[arXiv:hep-ph/9605288 [hep-ph]].

\bibitem{LHCHiggsCrossSectionWorkingGroup:2016ypw}
D.~de Florian \textit{et al.} [LHC Higgs Cross Section Working Group],
CERN Yellow Rep. Monogr. \textbf{2} (2017), 1-869
[arXiv:1610.07922 [hep-ph]].

\bibitem{Barklow:2017awn}
T.~Barklow, K.~Fujii, S.~Jung, M.~E.~Peskin and J.~Tian,
Phys. Rev. D \textbf{97} (2018) no.5, 053004
[arXiv:1708.09079 [hep-ph]].

\bibitem{LinearColliderVision:2025hlt}
D.~Atti{\'e} \textit{et al.} [Linear Collider Vision],
[arXiv:2503.19983 [hep-ex]].

\bibitem{Altmann:2025feg}
J.~Altmann, P.~Skands, A.~Desai, W.~Mitaroff, S.~Pl{\"a}tzer, D.~Dobur, K.~Skovpen, M.~Drewes, G.~Durieux and Y.~Georis, \textit{et al.}
2025,
ISBN 978-92-9083-700-8, 978-92-9083-701-5
[arXiv:2506.15390 [hep-ex]].

\bibitem{Grojean:2004xa}
C.~Grojean, G.~Servant and J.~D.~Wells,
Phys. Rev. D \textbf{71} (2005), 036001
[arXiv:hep-ph/0407019 [hep-ph]].

\bibitem{Kanemura:2004ch}
S.~Kanemura, Y.~Okada and E.~Senaha,
Phys. Lett. B \textbf{606} (2005), 361-366
[arXiv:hep-ph/0411354 [hep-ph]].

  \bibitem{Kakizaki:2015wua}
  M.~Kakizaki, S.~Kanemura, T.~Matsui,
  Phys.Rev.D \textbf{92} (2015) 11, 115007,
  [arXiv:1509.08394 [hep-ph]].

\bibitem{Hashino:2016rvx}
  K.~Hashino, M.~Kakizaki, S.~Kanemura, T.~Matsui,
  Phys.Rev.D \textbf{94} (2016) 1, 015005,
  [arXiv:1604.02069 [hep-ph]].

  \bibitem{Hashino:2016xoj}
  K.~Hashino, M.~Kakizaki, S.~Kanemura, P.~Ko, T.~Matsui,
  Phys.Lett.B \textbf{766} (2017) 49-54,
  [arXiv:1609.00297 [hep-ph]].

\bibitem{Basler:2017uxn}
P.~Basler, M.~M\"uhlleitner and J.~Wittbrodt,
JHEP \textbf{03} (2018), 061
[arXiv:1711.04097 [hep-ph]].

\bibitem{Biekotter:2022kgf}
T.~Biek\"otter, S.~Heinemeyer, J.~M.~No, M.~O.~Olea-Romacho and G.~Weiglein,
JCAP \textbf{03} (2023), 031
[arXiv:2208.14466 [hep-ph]].

\bibitem{Bittar:2025lcr}
P.~Bittar, S.~Roy and C.~E.~M.~Wagner,
[arXiv:2504.02024 [hep-ph]].

\bibitem{Gunion:2002zf}
J.~F.~Gunion and H.~E.~Haber,
Phys. Rev. D \textbf{67} (2003), 075019
[arXiv:hep-ph/0207010 [hep-ph]].

\bibitem{Kanemura:2002vm}
S.~Kanemura, S.~Kiyoura, Y.~Okada, E.~Senaha and C.~P.~Yuan,
Phys. Lett. B \textbf{558} (2003), 157-164
[arXiv:hep-ph/0211308 [hep-ph]].

\bibitem{Kanemura:2004mg}
S.~Kanemura, Y.~Okada, E.~Senaha and C.~P.~Yuan,
Phys. Rev. D \textbf{70} (2004), 115002
[arXiv:hep-ph/0408364 [hep-ph]].

\bibitem{Aoki:2012jj}
M.~Aoki, S.~Kanemura, M.~Kikuchi and K.~Yagyu,
Phys. Rev. D \textbf{87} (2013) no.1, 015012
[arXiv:1211.6029 [hep-ph]].

\bibitem{Kanemura:2015fra}
S.~Kanemura, M.~Kikuchi and K.~Yagyu,
Nucl. Phys. B \textbf{907} (2016), 286-322
[arXiv:1511.06211 [hep-ph]].

\bibitem{Kanemura:2015mxa}
S.~Kanemura, M.~Kikuchi and K.~Yagyu,
Nucl. Phys. B \textbf{896} (2015), 80-137
[arXiv:1502.07716 [hep-ph]].

\bibitem{Arhrib:2015hoa}
A.~Arhrib, R.~Benbrik, J.~El Falaki and A.~Jueid,
JHEP \textbf{12} (2015), 007
[arXiv:1507.03630 [hep-ph]].

\bibitem{Kanemura:2016sos}
S.~Kanemura, M.~Kikuchi and K.~Sakurai,
Phys. Rev. D \textbf{94} (2016) no.11, 115011
[arXiv:1605.08520 [hep-ph]].

\bibitem{Kanemura:2016lkz}
S.~Kanemura, M.~Kikuchi and K.~Yagyu,
Nucl. Phys. B \textbf{917} (2017), 154-177
[arXiv:1608.01582 [hep-ph]].

\bibitem{He:2016sqr}
S.~P.~He and S.~h.~Zhu,
Phys. Lett. B \textbf{764} (2017), 31-37
[erratum: Phys. Lett. B \textbf{797} (2019), 134782]
[arXiv:1607.04497 [hep-ph]].

\bibitem{Kanemura:2017wtm}
S.~Kanemura, M.~Kikuchi, K.~Sakurai and K.~Yagyu,
Phys. Rev. D \textbf{96} (2017) no.3, 035014
[arXiv:1705.05399 [hep-ph]].

\bibitem{Kanemura:2017gbi}
S.~Kanemura, M.~Kikuchi, K.~Sakurai and K.~Yagyu,
Comput. Phys. Commun. \textbf{233} (2018), 134-144
[arXiv:1710.04603 [hep-ph]].

\bibitem{Chiang:2018xpl}
C.~W.~Chiang, A.~L.~Kuo and K.~Yagyu,
Phys. Rev. D \textbf{98} (2018) no.1, 013008
[arXiv:1804.02633 [hep-ph]].

\bibitem{Basler:2018cwe}
P.~Basler and M.~M\"uhlleitner,
Comput. Phys. Commun. \textbf{237} (2019), 62-85
[arXiv:1803.02846 [hep-ph]].

\bibitem{Senaha:2018xek}
E.~Senaha,
Phys. Rev. D \textbf{100} (2019) no.5, 055034
[arXiv:1811.00336 [hep-ph]].

\bibitem{Braathen:2019pxr}
J.~Braathen and S.~Kanemura,
Phys. Lett. B \textbf{796} (2019), 38-46
[arXiv:1903.05417 [hep-ph]].

\bibitem{Kanemura:2019slf}
S.~Kanemura, M.~Kikuchi, K.~Mawatari, K.~Sakurai and K.~Yagyu,
Comput. Phys. Commun. \textbf{257} (2020), 107512
[arXiv:1910.12769 [hep-ph]].

\bibitem{Braathen:2019zoh}
J.~Braathen and S.~Kanemura,
Eur. Phys. J. C \textbf{80} (2020) no.3, 227
[arXiv:1911.11507 [hep-ph]].

\bibitem{Braathen:2020vwo}
J.~Braathen, S.~Kanemura and M.~Shimoda,
JHEP \textbf{03} (2021), 297
[arXiv:2011.07580 [hep-ph]].

\bibitem{Basler:2020nrq}
P.~Basler, M.~M\"uhlleitner and J.~M\"uller,
Comput. Phys. Commun. \textbf{269} (2021), 108124
[arXiv:2007.01725 [hep-ph]].

\bibitem{Bahl:2022jnx}
H.~Bahl, J.~Braathen and G.~Weiglein,
Phys. Rev. Lett. \textbf{129} (2022) no.23, 23
[arXiv:2202.03453 [hep-ph]].

\bibitem{Bahl:2022gqg}
H.~Bahl, W.~H.~Chiu, C.~Gao, L.~T.~Wang and Y.~M.~Zhong,
Eur. Phys. J. C \textbf{82} (2022) no.10, 944
[arXiv:2207.04059 [hep-ph]].

\bibitem{Falaki:2023tyd}
J.~E.~Falaki,
Phys. Lett. B \textbf{840} (2023), 137879
[arXiv:2301.13773 [hep-ph]].

\bibitem{Bahl:2023eau}
H.~Bahl, J.~Braathen, M.~Gabelmann and G.~Weiglein,
Eur. Phys. J. C \textbf{83} (2023) no.12, 1156
[erratum: Eur. Phys. J. C \textbf{84} (2024) no.5, 498]
[arXiv:2305.03015 [hep-ph]].

\bibitem{Aiko:2023nqj}
M.~Aiko, J.~Braathen and S.~Kanemura,
Eur. Phys. J. C \textbf{85} (2025) no.5, 489
[arXiv:2307.14976 [hep-ph]].

\bibitem{Cherchiglia:2024abx}
A.~Cherchiglia and L.~J.~Ferreira Leite,
[arXiv:2411.00094 [hep-ph]].

\bibitem{Basler:2024aaf}
P.~Basler, L.~Biermann, M.~M{\"u}hlleitner, J.~M{\"u}ller, R.~Santos and J.~Viana,
Comput. Phys. Commun. \textbf{316} (2025), 109766
[arXiv:2404.19037 [hep-ph]].

\bibitem{Bahl:2025wzj}
H.~Bahl, J.~Braathen, M.~Gabelmann and S.~Pa\ss{}ehr,
[arXiv:2503.15645 [hep-ph]].

\bibitem{Braathen:2025qxf}
J.~Braathen, S.~Heinemeyer, A.~Parra Arnay and A.~Verduras Schaeidt,
[arXiv:2507.02569 [hep-ph]].

\bibitem{ATLAS:2022jtk}
G.~Aad \textit{et al.} [ATLAS],
Phys. Lett. B \textbf{843} (2023), 137745
[arXiv:2211.01216 [hep-ex]].

\bibitem{ATLAS:2022kbf}
 [ATLAS],
ATLAS-CONF-2022-050.

\bibitem{CMS:2022dwd}
A.~Tumasyan \textit{et al.} [CMS],
Nature \textbf{607} (2022) no.7917, 60-68
[erratum: Nature \textbf{623} (2023) no.7985, E4]
[arXiv:2207.00043 [hep-ex]].

\bibitem{ATLAS:2024ish}
G.~Aad \textit{et al.} [ATLAS],
Phys. Rev. Lett. \textbf{133} (2024) no.10, 101801
[arXiv:2406.09971 [hep-ex]].

\bibitem{CMS:2024awa}
A.~Hayrapetyan \textit{et al.} [CMS],
Phys. Lett. B \textbf{861} (2025), 139210
[arXiv:2407.13554 [hep-ex]].

\bibitem{ATLAS:2025lbo}
 [ATLAS],
ATLAS-CONF-2025-005.

\bibitem{CMS:2025hfp}
 [ATLAS and CMS], ATL-PHYS-PUB-2025-018, CMS-HIG-25-002.
[arXiv:2504.00672 [hep-ex]].

\bibitem{Arco:2025pgx}
F.~Arco, S.~Heinemeyer and M.~M\"uhlleitner,
[arXiv:2505.02947 [hep-ph]].

\bibitem{anyHH}
H.~Bahl, J.~Braathen, M.~Gabelmann, K.~Radchenko and G.~Weiglein, DESY-2026-010,
\textit{in preparation}. See: (https://gitlab.com/anybsm/anybsm).

\bibitem{Arco:2025nii}
F.~Arco, S.~Heinemeyer, M.~M{\"u}hlleitner, A.~P.~Arnay, N.~R.~Gonz{\'a}lez and A.~V.~Schaeidt,
JHEP \textbf{06} (2025), 211
[arXiv:2502.03878 [hep-ph]].

\bibitem{Grojean:2006bp}
C.~Grojean and G.~Servant,
Phys. Rev. D \textbf{75} (2007), 043507
[arXiv:hep-ph/0607107 [hep-ph]].

\bibitem{Ashoorioon:2009nf}
A.~Ashoorioon and T.~Konstandin,
JHEP \textbf{07} (2009), 086
[arXiv:0904.0353 [hep-ph]].

\bibitem{No:2011fi}
J.~M.~No,
Phys. Rev. D \textbf{84} (2011), 124025
[arXiv:1103.2159 [hep-ph]].

\bibitem{Huber:2015znp}
S.~J.~Huber, T.~Konstandin, G.~Nardini and I.~Rues,
JCAP \textbf{03} (2016), 036
[arXiv:1512.06357 [hep-ph]].

\bibitem{Dorsch:2016nrg}
G.~C.~Dorsch, S.~J.~Huber, T.~Konstandin and J.~M.~No,
JCAP \textbf{05} (2017), 052
[arXiv:1611.05874 [hep-ph]].

\bibitem{Kang:2017mkl}
Z.~Kang, P.~Ko and T.~Matsui,
JHEP \textbf{02} (2018), 115
[arXiv:1706.09721 [hep-ph]].

\bibitem{Bruggisser:2018mrt}
S.~Bruggisser, B.~Von Harling, O.~Matsedonskyi and G.~Servant,
JHEP \textbf{12} (2018), 099
[arXiv:1804.07314 [hep-ph]].

\bibitem{Chala:2018ari}
M.~Chala, C.~Krause and G.~Nardini,
JHEP \textbf{07} (2018), 062
[arXiv:1802.02168 [hep-ph]].

\bibitem{Morais:2018uou}
A.~P.~Morais, R.~Pasechnik and T.~Vieu,
PoS \textbf{EPS-HEP2019} (2020), 054
[arXiv:1802.10109 [hep-ph]].

\bibitem{Hashino:2018zsi}
K.~Hashino, M.~Kakizaki, S.~Kanemura, P.~Ko and T.~Matsui,
JHEP \textbf{06} (2018), 088
[arXiv:1802.02947 [hep-ph]].

\bibitem{Hashino:2018wee}
K.~Hashino, R.~Jinno, M.~Kakizaki, S.~Kanemura, T.~Takahashi and M.~Takimoto,
Phys. Rev. D \textbf{99} (2019) no.7, 075011
[arXiv:1809.04994 [hep-ph]].

\bibitem{Goncalves:2021egx}
D.~Gon{\c{c}}alves, A.~Kaladharan and Y.~Wu,
Phys. Rev. D \textbf{105} (2022) no.9, 095041
[arXiv:2108.05356 [hep-ph]].

\bibitem{Lewicki:2024ghw}
M.~Lewicki, P.~Toczek and V.~Vaskonen,
Phys. Rev. Lett. \textbf{133} (2024) no.22, 221003
[arXiv:2402.04158 [astro-ph.CO]].

\bibitem{Athron:2023xlk}
P.~Athron, C.~Bal{\'a}zs, A.~Fowlie, L.~Morris and L.~Wu,
Prog. Part. Nucl. Phys. \textbf{135} (2024), 104094
[arXiv:2305.02357 [hep-ph]].

\bibitem{Kodama:1982sf}
H.~Kodama, M.~Sasaki and K.~Sato,
Prog. Theor. Phys. \textbf{68} (1982), 1979.

\bibitem{Liu:2021svg}
J.~Liu, L.~Bian, R.~G.~Cai, Z.~K.~Guo and S.~J.~Wang,
Phys. Rev. D \textbf{105} (2022) no.2, L021303
[arXiv:2106.05637 [astro-ph.CO]].

\bibitem{Hashino:2021qoq}
K.~Hashino, S.~Kanemura and T.~Takahashi,
Phys. Lett. B \textbf{833} (2022), 137261
[arXiv:2111.13099 [hep-ph]].

\bibitem{Jung:2021mku}
T.~H.~Jung and T.~Okui,
Phys. Rev. D \textbf{110} (2024) no.11, 115014
[arXiv:2110.04271 [hep-ph]].

\bibitem{Kawana:2022olo}
K.~Kawana, T.~Kim and P.~Lu,
Phys. Rev. D \textbf{108} (2023) no.10, 103531
[arXiv:2212.14037 [astro-ph.CO]].

\bibitem{Lewicki:2023ioy}
M.~Lewicki, P.~Toczek and V.~Vaskonen,
JHEP \textbf{09} (2023), 092
[arXiv:2305.04924 [astro-ph.CO]].

\bibitem{Gouttenoire:2023naa}
Y.~Gouttenoire and T.~Volansky,
Phys. Rev. D \textbf{110} (2024) no.4, 4
[arXiv:2305.04942 [hep-ph]].

\bibitem{Baldes:2023rqv}
I.~Baldes and M.~O.~Olea-Romacho,
JHEP \textbf{01} (2024), 133
[arXiv:2307.11639 [hep-ph]].

\bibitem{Flores:2024lng}
M.~M.~Flores, A.~Kusenko and M.~Sasaki,
Phys. Rev. D \textbf{110} (2024) no.1, 015005
[arXiv:2402.13341 [hep-ph]].

\bibitem{Kanemura:2024pae}
S.~Kanemura, M.~Tanaka and K.~P.~Xie,
JHEP \textbf{06} (2024), 036
[arXiv:2404.00646 [hep-ph]].

\bibitem{Hashino:2025fse}
K.~Hashino, S.~Kanemura, T.~Takahashi, M.~Tanaka and C.~M.~Yoo,
JCAP \textbf{09} (2025), 006
[arXiv:2501.11040 [hep-ph]].

\bibitem{Franciolini:2025ztf}
G.~Franciolini, Y.~Gouttenoire and R.~Jinno,
[arXiv:2503.01962 [hep-ph]].

\bibitem{Kierkla:2025vwp}
M.~Kierkla, N.~Ramberg, P.~Schicho and D.~Schmitt,
[arXiv:2506.15496 [hep-ph]].

\bibitem{Vachaspati:1991nm}
T.~Vachaspati,
Phys. Lett. B \textbf{265} (1991), 258-261

\bibitem{Ellis:2019tjf}
J.~Ellis, M.~Fairbairn, M.~Lewicki, V.~Vaskonen and A.~Wickens,
JCAP \textbf{09} (2019), 019
[arXiv:1907.04315 [astro-ph.CO]].

\bibitem{Olea-Romacho:2023rhh}
M.~O.~Olea-Romacho,
Phys. Rev. D \textbf{109} (2024) no.1, 015023
[arXiv:2310.19948 [hep-ph]].

\bibitem{Caprini:2019egz}
C.~Caprini, M.~Chala, G.~C.~Dorsch, M.~Hindmarsh, S.~J.~Huber, T.~Konstandin, J.~Kozaczuk, G.~Nardini, J.~M.~No and K.~Rummukainen, \textit{et al.}
JCAP \textbf{03} (2020), 024
[arXiv:1910.13125 [astro-ph.CO]].

\bibitem{LISACosmologyWorkingGroup:2022jok}
P.~Auclair \textit{et al.} [LISA Cosmology Working Group],
Living Rev. Rel. \textbf{26} (2023) no.1, 5
[arXiv:2204.05434 [astro-ph.CO]].

\bibitem{Kawamura:2011zz}
S.~Kawamura, M.~Ando, N.~Seto, S.~Sato, T.~Nakamura, K.~Tsubono, N.~Kanda, T.~Tanaka, J.~Yokoyama and I.~Funaki, \textit{et al.}
Class. Quant. Grav. \textbf{28} (2011), 094011

\bibitem{Corbin:2005ny}
V.~Corbin and N.~J.~Cornish,
Class. Quant. Grav. \textbf{23} (2006), 2435-2446
[arXiv:gr-qc/0512039 [gr-qc]].

\bibitem{Shelton:2010ta}
J.~Shelton and K.~M.~Zurek,
Phys. Rev. D \textbf{82} (2010), 123512
[arXiv:1008.1997 [hep-ph]].

\bibitem{Espinosa:2011eu}
J.~R.~Espinosa, B.~Gripaios, T.~Konstandin and F.~Riva,
JCAP \textbf{01} (2012), 012
[arXiv:1110.2876 [hep-ph]].

\bibitem{Azevedo:2018fmj}
D.~Azevedo, P.~M.~Ferreira, M.~M.~Muhlleitner, S.~Patel, R.~Santos and J.~Wittbrodt,
JHEP \textbf{11} (2018), 091
[arXiv:1807.10322 [hep-ph]].

\bibitem{Hall:2019ank}
E.~Hall, T.~Konstandin, R.~McGehee, H.~Murayama and G.~Servant,
JHEP \textbf{04} (2020), 042
[arXiv:1910.08068 [hep-ph]].

\bibitem{Biermann:2022meg}
L.~Biermann, M.~M{\"u}hlleitner and J.~M{\"u}ller,
Eur. Phys. J. C \textbf{83} (2023) no.5, 439
[arXiv:2204.13425 [hep-ph]].

\bibitem{Hooper:2025fda}
D.~Hooper, G.~Krnjaic, D.~Rocha and S.~Roy,
[arXiv:2507.22975 [hep-ph]].

\bibitem{Roy:2025zvo}
S.~Roy,
[arXiv:2509.19982 [hep-ph]].

\bibitem{Huang:2016cjm}
P.~Huang, A.~J.~Long and L.~T.~Wang,
Phys. Rev. D \textbf{94} (2016) no.7, 075008
[arXiv:1608.06619 [hep-ph]].

\bibitem{Alves:2018jsw}
A.~Alves, T.~Ghosh, H.~K.~Guo, K.~Sinha and D.~Vagie,
JHEP \textbf{04} (2019), 052
[arXiv:1812.09333 [hep-ph]].

\bibitem{Liu:2021jyc}
W.~Liu and K.~P.~Xie,
JHEP \textbf{04} (2021), 015
[arXiv:2101.10469 [hep-ph]].

\bibitem{Ellis:2022lft}
J.~Ellis, M.~Lewicki, M.~Merchand, J.~M.~No and M.~Zych,
JHEP \textbf{01} (2023), 093
[arXiv:2210.16305 [hep-ph]].

\bibitem{Blasi:2023rqi}
S.~Blasi, R.~Jinno, T.~Konstandin, H.~Rubira and I.~Stomberg,
JCAP \textbf{10} (2023), 051
[arXiv:2302.06952 [astro-ph.CO]].

\bibitem{Ramsey-Musolf:2024ykk}
M.~J.~Ramsey-Musolf, T.~V.~I.~Tenkanen and V.~Q.~Tran,
[arXiv:2409.17554 [hep-ph]].

\bibitem{Niemi:2024vzw}
L.~Niemi and T.~V.~I.~Tenkanen,
Phys. Rev. D \textbf{111} (2025) no.7, 075034
doi:10.1103/PhysRevD.111.075034
[arXiv:2408.15912 [hep-ph]].

\bibitem{Gould:2024jjt}
O.~Gould and P.~M.~Saffin,
JHEP \textbf{03} (2025), 105
[arXiv:2411.08951 [hep-ph]].

\bibitem{Niemi:2024axp}
L.~Niemi, M.~J.~Ramsey-Musolf and G.~Xia,
Phys. Rev. D \textbf{110} (2024) no.11, 115016
[arXiv:2405.01191 [hep-ph]].

\bibitem{Ghosh:2022fzp}
P.~Ghosh, T.~Ghosh and S.~Roy,
JHEP \textbf{10} (2023), 057
[arXiv:2211.15640 [hep-ph]].

\bibitem{Roy:2022gop}
S.~Roy,
Phys. Rev. D \textbf{111} (2025) no.1, 015037
[arXiv:2212.11230 [hep-ph]].

\bibitem{Goncalves:2024vkj}
D.~Gon{\c{c}}alves, A.~Kaladharan and Y.~Wu,
Phys. Rev. D \textbf{111} (2025) no.3, 035009
[arXiv:2406.07622 [hep-ph]].

\bibitem{Feuerstake:2024uxs}
F.~Feuerstake, E.~Fuchs, T.~Robens and D.~Winterbottom,
JHEP \textbf{04} (2025), 094
[arXiv:2409.06651 [hep-ph]].

\bibitem{Lewis:2024yvj}
I.~M.~Lewis, J.~Scott, M.~A.~S.~Alcaraz and M.~Sullivan,
[arXiv:2410.08275 [hep-ph]].

\bibitem{Aboudonia:2024frg}
M.~Aboudonia, C.~Balazs, A.~Papaefstathiou and G.~White,
JHEP \textbf{04} (2025), 093
[arXiv:2410.22700 [hep-ph]].

\bibitem{Li:2019tfd}
H.~L.~Li, M.~Ramsey-Musolf, S.~Willocq
Phys.Rev.D \textbf{100} (2019) 7, 075035,
[1906.05289 [hep-ph]].

\bibitem{Zhang:2023jvh}
W.~Zhang, H.~L.~Li, K.~Liu, M.~J.~Ramsey-Musolf, Y.~Zeng and S.~Arunasalam,
JHEP \textbf{12} (2023), 018
[arXiv:2303.03612 [hep-ph]].

\bibitem{Palit:2023dvs}
P.~Palit and S.~Shil,
J. Phys. G \textbf{51} (2024) no.9, 095005
[arXiv:2302.04191 [hep-ph]].

\bibitem{Lerner:2009xg}
R.~N.~Lerner, J.~McDonald,
 Phys.Rev.D \textbf{80} (2009), 123507,
 [arXiv:0909.0520].

\bibitem{Gonderinger:2009jp}
M.~Gonderinger, Y.~Li, H.~Patel, M.~J.~Ramsey-Musolf,
JHEP \textbf{01} (2010) 053,
[0910.3167 [hep-ph]].

\bibitem{Braathen:2017jvs}
J.~Braathen, M.~D.~Goodsell, M.~E.~Krauss, T.~Opferkuch and F.~Staub,
Phys. Rev. D \textbf{97} (2018) no.1, 015011
[arXiv:1711.08460 [hep-ph]].

 \bibitem{Bechtle:2008jh}
  P.~Bechtle, O.~Brein, S.~Heinemeyer, G.~Weiglein and K.~E.~Williams,
  Comput.\ Phys.\ Commun.\  \textbf{ 181} (2010) 138
  [arXiv:0811.4169 [hep-ph]].

\bibitem{Bechtle:2011sb}
  P.~Bechtle, O.~Brein, S.~Heinemeyer, G.~Weiglein and K.~E.~Williams,
  Comput.\ Phys.\ Commun.\  \textbf{182} (2011) 2605
  [arXiv:1102.1898 [hep-ph]].

\bibitem{Bechtle:2013wla}
  P.~Bechtle, O.~Brein, S.~Heinemeyer, O.~St{\aa}l, T.~Stefaniak,
  G.~Weiglein and K.~E.~Williams, 
  Eur.\ Phys.\ J.\ C \textbf{74} (2014) no.3,  2693
  [arXiv:1311.0055 [hep-ph]].

\bibitem{Bechtle:2015pma}
  P.~Bechtle, S.~Heinemeyer, O.~St{\aa}l, T.~Stefaniak and G.~Weiglein,
  Eur.\ Phys.\ J.\ C \textbf{75} (2015) no.9,  421
  [arXiv:1507.06706 [hep-ph]].

\bibitem{Bechtle:2020pkv}
P.~Bechtle, D.~Dercks, S.~Heinemeyer, T.~Klingl, T.~Stefaniak,
G.~Weiglein and J.~Wittbrodt,
Eur. Phys. J. C \textbf{80} (2020) no.12, 1211
[arXiv:2006.06007 [hep-ph]].

\bibitem{Bahl:2022igd}
H.~Bahl, T.~Biek\"otter, S.~Heinemeyer, C.~Li, S.~Paasch, G.~Weiglein and J.~Wittbrodt,
Comput. Phys. Commun. \textbf{291} (2023), 108803
[arXiv:2210.09332 [hep-ph]].

\bibitem{Bechtle:2013xfa}
  P.~Bechtle, S.~Heinemeyer, O.~St{\aa}l, T.~Stefaniak and G.~Weiglein,
  Eur.\ Phys.\ J.\ C \textbf{74} (2014) no.2,  2711
  [arXiv:1305.1933 [hep-ph]].

\bibitem{Bechtle:2014ewa}
  P.~Bechtle, S.~Heinemeyer, O.~St{\aa}l, T.~Stefaniak and G.~Weiglein,
  JHEP \textbf{1411} (2014) 039
  [arXiv:1403.1582 [hep-ph]].

\bibitem{Bechtle:2020uwn}
 P.~Bechtle \textit{et al.},
  Eur.Phys.J.C \textbf{81} (2021) 2, 145
  [2012.09197 [hep-ph]].

\bibitem{cw}
S.R.~Coleman, E.J.~Weinberg,,
Phys. Rev. D \textbf{7} (1973) 1888.

\bibitem{supertrace}
R.~Jackiw,
Phys. Rev. D9 (1974) 1686.

\bibitem{efft1}
M.~Quiros, 
hep-ph/9901312.

\bibitem{efft2}
L.~Dolan, R.~Jackiw,
Phys. Rev. D \textbf{9} (1974) 3320,
hep-ph/9211211.

\bibitem{Arnold:1992rz}
P.~B.~Arnold and O.~Espinosa,
Phys. Rev. D \textbf{47} (1993), 3546
[erratum: Phys. Rev. D \textbf{50} (1994), 6662]
[arXiv:hep-ph/9212235 [hep-ph]].

\bibitem{Klinkhamer:1984di}
F.~R.~Klinkhamer and N.~S.~Manton,
Phys. Rev. D \textbf{30} (1984), 2212.

\bibitem{tunrate}
T.~Banks, C.~M.~Bender, T.~T.~Wu,
Phys. Rev. D \textbf{8}, 3346, (1973),
[https://doi.org/10.1103/PhysRevD.8.3346].

\bibitem{s4}
S.~R.~Coleman,
Phys. Rev. D \textbf{15} (1977), 2929-2936
[erratum: Phys. Rev. D \textbf{16} (1977), 1248]

\bibitem{Linde:1980tt}
A.~D.~Linde,
Phys.Lett.B\textbf{ 100} (1981), 37-40.

\bibitem{Broadbent:1957rm}
S.R.~Broadbent, J.M.~Hammersley,
Math.Proc.Cambridge Phil.Soc. \textbf{53} (2008) 3, 629-641, Proc.Cambridge Phil.Soc. \textbf{53} (1957) 629-641.

\bibitem{Branchina:2024rva}
C.~Branchina, A.~Conaci, S.~De Curtis, L.~Delle~Rose, A.~Guiggiani,
EPJ Web Conf. \textbf{314} (2024) 00031,
[arXiv:[2410.00766 [hep-ph]].

\bibitem{DeCurtis:2024hvh}
S.~De~Curtis, L.~Delle~Rose, A.~Guiggiani, Á.~Gil~Muyor, G~Panico,
JHEP \textbf{05} (2024) 009,
[arXiv:2401.13522 [hep-ph]].

\bibitem{DeCurtis:2023hil}
S.~De~Curtis, L.~Delle~Rose, A.~Guiggiani, Á.~Gil~Muyor, G~Panico,
JHEP \textbf{05 }(2023) 194,
[arXiv: 2303.05846 [hep-ph]].

\bibitem{DeCurtis:2022hlx}
S.~De~Curtis, L.~Delle~Rose, A.~Guiggiani, Á.~Gil~Muyor, G~Panico,
JHEP \textbf{03} (2022) 163,
[arXiv:2201.08220 [hep-ph]].

\bibitem{Branchina:2025adj}
C.~Branchina, A.~Conaci, S.~De Curtis and L.~Delle Rose,
[arXiv:2510.21942 [hep-ph]].

\bibitem{Biekotter:2025npc}
T.~Biek{\"o}tter, A.~Dashko, M.~L{\"o}schner and G.~Weiglein,
[arXiv:2511.14831 [hep-ph]].

\bibitem{Ai:2023see}
W.Y.Ai, Benoit~Laurent, J.~van~de~Vis,
JCAP \textbf{07} (2023) 002,
[2303.10171 [astro-ph.CO]].

\bibitem{Ekstedt:2024fyq}
A.~Ekstedt, O.~Gould, J.~Hirvonen, B.~Laurent, L.~Niemi, P.~Schicho and J.~van de Vis,
JHEP \textbf{04} (2025), 101
[arXiv:2411.04970 [hep-ph]].

\bibitem{Caprini:2015zlo}
C.~Caprini, M.~Hindmarsh, S.~Huber, T.~Konstandin, J.~Kozaczuk, G.~Nardini, J.~M.~No, A.~Petiteau, P.~Schwaller and G.~Servant, \textit{et al.}
JCAP \textbf{04} (2016), 001
[arXiv:1512.06239 [astro-ph.CO]].


\bibitem{Bodeker:2009qy}
D.~Bodeker, G.~D.~Moore,
JCAP\textbf{ 05 }(2009) 009
[arXiv:0903.4099 [hep-ph]].

\bibitem{Ai:2024shx}
W.Y.~Ai, X.~Nagels, M.~Vanvlasselaer
JCAP \textbf{03} (2024) 037
[arXiv: 2401.05911 [hep-ph]].





\bibitem{ATLAS:2020tlo}
G.~Aad \textit{et al.} [ATLAS],
Eur. Phys. J. C \textbf{81} (2021) no.4, 332
[arXiv:2009.14791 [hep-ex]].

\bibitem{ATLAS:2018sbw}
M.~Aaboud \textit{et al.} [ATLAS],
Phys. Rev. D \textbf{98} (2018) no.5, 052008
[arXiv:1808.02380 [hep-ex]].

\bibitem{CMS:2021yci}
A.~Tumasyan \textit{et al.} [CMS],
JHEP \textbf{11} (2021), 057
[arXiv:2106.10361 [hep-ex]].

\bibitem{Guth:1979bh}
A.H. Guth, S.H.H.~Tye,
Phys.Rev.Lett. \textbf{44} (1980) 631, Phys.Rev.Lett. \textbf{44} (1980) 963 (erratum).

\bibitem{Sperling:2013eva}
M.~Sperling, D.~St{\"o}ckinger and A.~Voigt,
JHEP \textbf{07} (2013), 132
[arXiv:1305.1548 [hep-ph]].

\bibitem{Lewicki:2024xan}
M.~Lewicki, M.~Merchand, L.~Sagunski, P.~Schicho and D.~Schmitt,
Phys. Rev. D \textbf{110} (2024) no.2, 023538
[arXiv:2403.03769 [hep-ph]].

\bibitem{Dawson:1998py}
S.~Dawson, S.~Dittmaier and M.~Spira,
Phys. Rev. D \textbf{58} (1998), 115012
[arXiv:hep-ph/9805244 [hep-ph]].

\bibitem{Nhung:2013lpa}
D.~T.~Nhung, M.~M\"uhlleitner, J.~Streicher and K.~Walz,
JHEP \textbf{11} (2013), 181
[arXiv:1306.3926 [hep-ph]].

\bibitem{Grober:2015cwa}
R.~Gr\"ober, M.~M\"uhlleitner, M.~Spira and J.~Streicher,
JHEP \textbf{09} (2015), 092
[arXiv:1504.06577 [hep-ph]].

\bibitem{Grober:2017gut}
R.~Gr\"ober, M.~M\"uhlleitner, M.~Spira,
Nucl.Phys.B \textbf{925} (2017), 1-27,
[arXiv:1705.05314 [hep-ph]].

\bibitem{Abouabid:2021yvw}
H.~Abouabid {\it et al.}, JHEP \textbf{09} (2022), 011
[arXiv:2112.12515 [hep-ph]].

\bibitem{Arco:2022lai}
F.~Arco, S.~Heinemeyer, M.~M\"uhlleitner and K.~Radchenko,
Eur. Phys. J. C \textbf{83} (2023) no.11, 1019
[arXiv:2212.11242 [hep-ph]].

\bibitem{Baglio:2012np}
J.~Baglio, A.~Djouadi, R.~Gr\"ober, M.~M.~M\"uhlleitner, J.~Quevillon and M.~Spira,
JHEP \textbf{04} (2013), 151
[arXiv:1212.5581 [hep-ph]].

\bibitem{Heinemeyer:2024hxa}
S.~Heinemeyer, M.~M{\"u}hlleitner, K.~Radchenko and G.~Weiglein,
Eur. Phys. J. C \textbf{85} (2025) no.4, 437
[arXiv:2403.14776 [hep-ph]].

\bibitem{Frank:2025zmj}
M.~Frank, S.~Heinemeyer, M.~M{\"u}hlleitner and K.~Radchenko,
[arXiv:2506.18981 [hep-ph]].

\bibitem{Alwall:2014hca}
J.~Alwall, R.~Frederix, S.~Frixione, V.~Hirschi, F.~Maltoni, O.~Mattelaer, H.~S.~Shao, T.~Stelzer, P.~Torrielli and M.~Zaro,
JHEP \textbf{07} (2014), 079
[arXiv:1405.0301 [hep-ph]].

\bibitem{Staub:2008uz}
F.~Staub,
[arXiv:0806.0538 [hep-ph]].

\bibitem{Staub:2009bi}
F.~Staub,
Comput. Phys. Commun. \textbf{181} (2010), 1077-1086
[arXiv:0909.2863 [hep-ph]].

\bibitem{Staub:2010jh}
F.~Staub,
Comput. Phys. Commun. \textbf{182} (2011), 808-833
[arXiv:1002.0840 [hep-ph]].

\bibitem{Staub:2012pb}
F.~Staub,
Comput. Phys. Commun. \textbf{184} (2013), 1792-1809
[arXiv:1207.0906 [hep-ph]].

\bibitem{Staub:2013tta}
F.~Staub,
Comput. Phys. Commun. \textbf{185} (2014), 1773-1790
[arXiv:1309.7223 [hep-ph]].

\bibitem{ILC:2013jhg}
H.~Baer \textit{et al.} [ILC],
[arXiv:1306.6352 [hep-ph]].

\bibitem{Moortgat-Pick:2015lbx}
G.~Moortgat-Pick, H.~Baer, M.~Battaglia, G.~Belanger, K.~Fujii, J.~Kalinowski, S.~Heinemeyer, Y.~Kiyo, K.~Olive and F.~Simon, \textit{et al.}
Eur. Phys. J. C \textbf{75} (2015) no.8, 371
[arXiv:1504.01726 [hep-ph]].

\bibitem{Bambade:2019fyw}
P.~Bambade, T.~Barklow, T.~Behnke, M.~Berggren, J.~Brau, P.~Burrows, D.~Denisov, A.~Faus-Golfe, B.~Foster and K.~Fujii, \textit{et al.}
[arXiv:1903.01629 [hep-ex]].

\bibitem{LinearCollider:2025lya}
A.~Subba \textit{et al.} [Linear Collider],
[arXiv:2503.24049 [hep-ex]].

\bibitem{Catani:1991hj}
S.~Catani, Y.~L.~Dokshitzer, M.~Olsson, G.~Turnock and B.~R.~Webber,
Phys. Lett. B \textbf{269} (1991), 432-438.

\bibitem{Abramowicz:2016zbo}
H.~Abramowicz, A.~Abusleme, K.~Afanaciev, N.~A.~Tehrani, C.~Bal{\'a}zs, Y.~Benhammou, M.~Benoit, B.~Bilki, J.~J.~Blaising and M.~J.~Boland, \textit{et al.}
Eur. Phys. J. C \textbf{77} (2017) no.7, 475
[arXiv:1608.07538 [hep-ex]].

\bibitem{Durig:2016jrs}
C.~F.~D{\"u}rig,
doi:10.3204/PUBDB-2016-04283.

\bibitem{Tian:2013qmi}
J.~Tian,
``Study of Higgs self-coupling at the ILC based on the full detector simulation at {\ensuremath{\sqrt{}}} s = 500 GeV and {\ensuremath{\sqrt{}}} s = 1 TeV'',\\
see: \texttt{https://flc.desy.de/lcnotes/notes/LC-REP-2013-003.pdf}.

\bibitem{BriefingBook2025}
J.~de Blas et al.,
[CERN-ESU-2025-001], see: \texttt{https://cds.cern.ch/record/2944678}.

\end{thebibliography}
\end{document}